\documentclass{article}
\PassOptionsToPackage{numbers, compress}{natbib}
\usepackage[preprint]{neurips_2026}

\usepackage[utf8]{inputenc}
\usepackage[T1]{fontenc}
\usepackage[colorlinks=true,linkcolor=blue!70!black,citecolor=blue!70!black,urlcolor=blue!70!black]{hyperref}
\hypersetup{
  pdftitle={Interpretable Allocation and Model-Predictive Rebalancing of Equity Portfolios, Tested on Real and
    Synthetic Markets},
  pdfauthor={Jeffrey D. Varner}
}
\usepackage{url}
\usepackage{booktabs}
\usepackage{amsfonts}
\usepackage{nicefrac}
\usepackage{microtype}
\usepackage{xcolor}
\usepackage{graphicx}
\usepackage{subcaption}
\usepackage{amsmath}
\usepackage{amssymb}
\usepackage{amsthm}
\usepackage{algorithm}
\usepackage{algorithmic}
\usepackage{bm}
\usepackage{multirow}
\usepackage{lineno}
\usepackage{needspace}

\usepackage[section]{placeins} %
\usepackage{flafter} %

\title{Interpretable Allocation and Model-Predictive Rebalancing of Equity Portfolios, Tested on Real and
  Synthetic Markets}

\author{%
  Jeffrey D.~Varner\thanks{Corresponding author.} \\
  Robert Frederick Smith School of Chemical and Biomolecular Engineering \\
  Cornell University, Ithaca, NY 14850 \\
  \texttt{jdv27@cornell.edu} \\
}

\begin{document}
\raggedbottom
\clubpenalty=10000
\widowpenalty=10000
\displaywidowpenalty=10000

\maketitle

\begin{abstract}
A live equity portfolio poses three problems: which stocks to hold, when to trade, and whether the strategy holds up
beyond the observed market. We built one system for all three and tested it on 373 US equities, calibrated on
2014--2024 and traded January 2025 to September 2026. A selection rule chose stocks by score, by a bandit or at
random, and one of seven weighting rules, four consumer-theory utilities and three standard portfolios, set share
counts under limits on budget, single-stock weight, portfolio volatility and turnover, with each stock's preference
weight a bounded score of its trend, market beta and the market's recent direction. A model-predictive controller
forecast portfolio value with a multi-asset geometric Brownian motion and re-planned at its horizon, when the value
left its forecast band, or at a drawdown stop. To probe deeper declines than the real window held, we also ran the
system on simulated markets from per-asset hidden Markov generators with jumps, linked by market and sector factors.
On real prices, the stocks chosen mattered more to terminal wealth than the weighting rule, and the early re-plans
shrank the worst drawdown in most rule combinations against a fixed calendar without consistently raising wealth. In
the deepest simulated declines, every combination, random picks included, ended above simulated SPY and the
equal-weight universe in the median, even without early re-plans, so the margin came from shared parts of the system,
not from stock selection or re-plan timing. A market hedge that tightened a beta cap as the market weakened cost
wealth on real prices but paid off in the most sustained simulated declines, and the generator's persistent trends
favored score-based selection more than the real market did, which suggests a synthetic test judges stock selection
fairly only if its generator does not build in that persistence.

\end{abstract}

\vspace{0.5em}

\section{Introduction}\label{sec:introduction}
Managing a live equity portfolio raises three linked questions. An allocation rule answers what to hold, given
forecasts and limits on risk, concentration and trading; a rebalancing rule answers when the holdings should be
recomputed; and an evaluation asks whether the two rules together would have worked on markets other than the one
that was observed. Mean-variance optimization~\cite{markowitz1952} and its continuous-time
extensions~\cite{merton1969} give a principled answer to the first question, but optimized weights are sensitive to
errors in the estimated means and covariances~\cite{michaud1989, ledoitWolf2004}, and an equal-weight portfolio
often matches optimized portfolios out of sample~\cite{demiguelGarlappiUppal2009}. Transaction costs turn the second
question into a control problem. With proportional costs, the optimal policy leaves the holdings alone inside a
no-trade region and trades only at its boundary~\cite{magillConstantinides1976, davisNorman1990}, and multi-period
formulations solve a sequence of forecast-based optimizations over a receding horizon~\cite{herzogDondiGeering2007,
boydEtAl2017}. The third question is usually answered with a backtest on the historical record, which is a single
path of the market. Choosing among rules on one path inflates their measured performance~\cite{baileyEtAl2014,
harveyLiu2015}, and a short window may contain few of the market conditions a portfolio must survive.

These lines of work mostly treat the three questions one at a time, although a live portfolio answers them
together and the answers interact. The allocation rule decides which positions the rebalancing rule must track, and
the market's path between rebalances decides when the next trade comes and from which holdings it starts. An
allocation that is easy to interpret helps here, because a change in a holding can then be traced to a change in one
stated number. Utility functions from consumer theory offer such a rule when each stock's preference weight is a
score with a plain meaning~\cite{cobbDouglas1928, arrowEtAl1961, stone1954, geary1950}. Testing the combined system
beyond the historical path requires synthetic markets that reproduce the stylized facts of daily growth rates, in
particular heavy tails and volatility clustering~\cite{mandelbrot1963, cont2001}, while keeping the dependence
between stocks and the market~\cite{sharpe1963}. Per-asset hidden Markov generators with a jump mechanism, combined
through a common market factor with each asset's own draw rescaled so that the market term does not inflate the
asset's total variance, reproduce the heavy tails of individual stocks within a multi-asset
market~\cite{alswaidan2026variance}. Such a market makes it possible to run a complete system, with its calibration
repeated on each simulated history, on futures that did not occur, including futures with deeper declines than the historical
window contains.

In this study, we built a portfolio management system that answers the three questions together and tested it on
real and synthetic markets. A selection rule chose the stocks a portfolio could hold, by ranking a score across the
whole universe or within each sector, by a bandit within each sector, or at random as a control, and two short lists
chosen in advance by the author served as further selections. Each stock's score divided its expected growth rate
from a single-index fit to the market by its absolute beta raised to a power that was positive when the market had
recently been falling and negative when it had been rising, so that, among stocks with positive expected growth,
low-beta stocks scored higher after declines and high-beta stocks after rises, and a hyperbolic tangent kept the score bounded. Utility functions from
consumer theory, with the scores as preference weights, and three standard portfolios (minimum variance, tangency
and long-short) set the holdings under shared limits on the budget, each stock's weight, portfolio
volatility and trading cost. An optional market hedge capped the portfolio's beta more tightly as the market
weakened and added up to five stocks with zero or negative beta to those the portfolio could hold. A model-predictive controller forecast the value of the
holdings with a multi-asset geometric Brownian motion and re-solved the allocation from the current holdings when
its horizon had passed, when the value left a band around the forecast, or when a drawdown limit was reached. We
calibrated the system on a decade of United States equity prices and ran it on the following twenty months, and we
ran it again on simulated price paths of the same period (synthetic futures) built from per-asset hidden Markov
generators with jumps, including futures, selected from many simulated paths, with deeper market declines than the
historical window contained. On real prices, which stocks a portfolio held mattered more for its terminal wealth than which utility
function weighted them, and the forecast band and the drawdown limit brought many re-plans forward of the horizon.
Compared with re-planning on the calendar alone, these early re-plans made the worst drawdown smaller in most
combinations of rules, on real prices and in the deepest simulated declines, but did not raise terminal wealth
consistently.
In the deepest simulated declines, every combination of selection and weighting rule ended ahead of both the
simulated market index and an equal-weight portfolio of all stocks, random selections included, and did so with the
early re-plans switched off as well, so the margin came from parts of the system that all combinations shared, such
as the volatility cap, rather than from the choice of stocks or the re-planning triggers. The market hedge cost
wealth while markets rose and helped when declines persisted, and the synthetic futures favored score-based
selection partly because the generator preserved each stock's historical trend. More broadly, the results suggest
that portfolio systems are best evaluated as a whole, on synthetic as well as historical markets, and that a synthetic
test measures the value of stock selection fairly only when its generator does not build in the trend persistence
that selection rules exploit.

\section{Related Work}\label{sec:related-work}
Portfolio allocation is usually posed as a trade-off between expected growth and risk, following mean-variance
analysis~\cite{markowitz1952} and its dynamic extensions~\cite{merton1969}. Optimized weights amplify errors in the
estimated inputs~\cite{michaud1989}. Practical work therefore shrinks the covariance estimate~\cite{ledoitWolf2004}
or adds limits on weights and trading, and judges the result against simple rules such as equal weighting, which is
difficult to beat out of sample~\cite{demiguelGarlappiUppal2009}. The single-index model~\cite{sharpe1963} reduces
the inputs to each stock's intercept and market loading, and factor models add further common
factors~\cite{famaFrench1993}. The allocation used here keeps the single-index inputs for its risk limits and its
scores, but replaces the mean-variance objective with utility functions from consumer theory: the Cobb--Douglas
form~\cite{cobbDouglas1928}, the constant elasticity of substitution form~\cite{arrowEtAl1961}, applied once to
share counts and once to portfolio weights, and the Stone--Geary form with a minimum holding of each
stock~\cite{stone1954, geary1950}, four utilities in all. In each utility, a stock's preference weight is its score,
a number between $-1$ and $1$ equal to the hyperbolic tangent of the ratio of its expected growth rate from the
single-index fit to a penalty on its market loading whose strength depends on the recent direction of the market. Each
preference weight is thus a single number that states why a stock is held. Stocks with a negative score receive no
preference weight in the long-only utilities. The expected growth rate rewards a stock's trend, and trend-following
signals in prices have a long empirical record~\cite{brockLakonishokLeBaron1992, moskowitzOoiPedersen2012}. The
penalty on large loadings, which is harsher when the market is falling, follows the evidence that portfolios tilted
away from high-beta stocks have earned higher risk-adjusted performance than portfolios tilted toward
them~\cite{frazziniPedersen2014}. Three benchmarks run under the same limits: the minimum-variance portfolio, the
tangency portfolio and a $130/30$ long-short portfolio, which holds $130\%$ of wealth long and $30\%$ short and so
relaxes the long-only constraint of the utility portfolios~\cite{jacobsLevyStarer1998}. Choosing which stocks to
hold before weighting them is a separate decision, and one of the selection rules studied here ranks the stocks of
each sector with a multi-armed bandit~\cite{lattimoreSzepesvari2020}, whose arms are the sector's stocks and whose
rewards are their growth in excess of the market over past 21-day windows.

With proportional trading costs, the optimal rebalancing policy leaves the holdings alone inside a no-trade region
and trades only at its boundary~\cite{magillConstantinides1976, davisNorman1990}. When expected growth rates are predictable, the
optimal policy instead trades a fraction of the gap between the current portfolio and the portfolio the investor
would hold without costs~\cite{garleanuPedersen2013}. Model-predictive control solves a finite-horizon optimization
from the measured state, applies the first decision, and solves again from the next measured
state~\cite{mayneEtAl2000}. Its portfolio applications include stochastic formulations with forecast models of asset
prices~\cite{herzogDondiGeering2007}, linear control policies for multi-period allocation~\cite{calafiore2008}, and
convex multi-period trading with costs and limits~\cite{boydEtAl2017}, all of which re-solve at every period.
Drawdown control has its own literature, in which exposure is adjusted continuously to keep wealth above a fixed
fraction of its running maximum~\cite{grossmanZhou1993}. The controller used here solves a single-period allocation
at each re-plan, forecasts the value of the resulting holdings over a $21$-day horizon, and keeps the holdings
unchanged until a trigger fires. Its default schedule is the horizon itself; a re-plan comes earlier when the
measured portfolio value leaves a band around the forecast, so that a forecast miss brings trading forward, or when
the drawdown reaches a limit. A re-plan re-solves the allocation from the current holdings and starts a new
forecast; it does not by itself reduce exposure.

Evaluating trading rules on the historical record carries a selection risk: when many rules or settings are tried on
the same path, the best measured result overstates the result to expect~\cite{baileyEtAl2014, harveyLiu2015}.
Synthetic data can test rules on paths that were not observed~\cite{assefaEtAl2020, jordonEtAl2022}, and the test is
informative to the extent that the generator reproduces the stylized facts of daily growth rates, in particular
heavy tails and volatility clustering~\cite{mandelbrot1963, cont2001}. Block bootstrap resampling keeps short-range
dependence~\cite{politisRomano1994}. Conditional variance models such as GARCH reproduce volatility
clustering~\cite{engle1982, bollerslev1986}. Regime-switching models, including hidden Markov models, represent
shifts between market states~\cite{hamilton1989}, and neural generators learn the joint distribution
directly~\cite{wieseEtAl2020}. Reviews of time-series synthesis note that a realistic distribution of daily values
does not imply realistic behavior over time~\cite{stengerEtAl2024}. Earlier work built the generator used here, a
hidden Markov model for each asset whose hidden states are bins of the daily growth rate, with heavy-tailed draws in
each state and occasional episodes in which the chain is held in its most extreme states for a random number of
days, producing bursts of large moves~\cite{alswaidan2026variance}. The same work combined the draws of many assets
through a common market factor, rescaling each asset's own draw so that adding the market term does not inflate the
asset's total variance. The present study adds drifting market loadings and sector factors to that combination, and
uses the result only as a market for the complete allocation and rebalancing system to trade in; the controller's
own forecast remains a geometric Brownian motion. The generator reproduces heavy tails more closely than volatility
clustering, and the evaluation carries that limitation.

\section{Methods}\label{sec:method}

We tested a rebalancing strategy built from three parts, a score that ranks stocks by expected growth, a
constrained utility maximization that sets how many shares of the selected stocks to hold, and a
model-predictive controller that decides when to re-plan, in a backtest on recorded prices and in a stress test
on synthetic prices. The recorded prices were the daily opens and closes (USD/share) and volumes (shares/day) of
the constituents of the S\&P~500 index and of the exchange-traded fund SPY, which served as the market and was
never held. They came from Polygon.io in three files covering 2014-01-03 to 2024-12-31, calendar 2025, and
2026-01-02 to 2026-09-04, and the trading days were those of SPY. Each name took one of the $11$ sectors of the
Global Industry Classification Standard (GICS) from a snapshot of the index constituent list. The universe kept
every ticker that appeared in all three files with a sector and had a positive open and close on every trading
day from 2014-01-03 to 2026-09-04, which left $N = 373$ names with between $15$ and $60$ per sector
(Appendix~\ref{app:controller}), and we write $s(i)$ for the sector of name $i$. Because the snapshot lists current
members, the universe holds only names that were index members at the snapshot date and traded on every day of
$2014$--$2026$, each with its current sector. The prices were split-adjusted by the vendor and not adjusted for
dividends, and we used them as given, so neither the portfolios nor SPY received dividends. Each file was
split-adjusted as of its own creation date, so for the seven names that split between the creation dates of two
files we put the earlier rows on the newest file's basis (Appendix~\ref{app:controller}).

The calibration window ran from 2014-01-03 to 2024-12-31 ($2{,}767$ trading days) and the trading window from
2025-01-02 to 2026-09-04 ($420$ trading days). Quantities fitted on the calibration window (the prior of the online
single-index fit, the growth moments of SPY, the volatility cap, the gain of the market state and the covariance
shrinkage) were fitted once and then held fixed. In the trading window every decision was made after the close of
day $t$ from data dated on or before day $t$, and its orders filled at the open of day $t+1$. We measured
performance with continuously compounded growth rates. For a price or wealth series $S$ observed on trading days,
the growth rate over the $k$ trading days that end on day $t$ is
\begin{equation}
  g(t;k) \;\equiv\; \frac{1}{k\,\Delta t}\,\ln\!\left(\frac{S(t)}{S(t-k)}\right),
  \label{eq:growth}
\end{equation}
where $\Delta t = 1/252$~yr is the length of one trading day, so that $g$ has units of $\mathrm{yr}^{-1}$. We write
$g_i(t) = g_i(t;1)$ for the daily growth rate of name $i$ and $g_M(t)$ for that of SPY, and the realized growth rate
of a run is Eq.~\eqref{eq:growth} applied to its wealth over the $k = 419$ daily steps of the trading window.

The risk-free growth rate came from the 3-month Treasury bill rate DTB3~\cite{fredDTB3}, a bank discount rate $d$
(a fraction, on a 360-day year), converted to the continuously compounded growth rate of a $91$-day bill,
\begin{equation}
  g_f(t) \;=\; -\,\frac{365}{91}\,\ln\!\left(1 - \frac{91\,d}{360}\right),
  \label{eq:riskfree}
\end{equation}
in $\mathrm{yr}^{-1}$. Day $t$ used the last observation dated before $t$, because the rate for a day is published
on the following day, and $g_f$ ranged from $0.036$ to $0.044$~$\mathrm{yr}^{-1}$ over the trading window. It set
the growth of cash in the backtest and in the controller's forecast, the excess growth of the tangency
benchmark, and the excess growth behind the Sharpe ratio and the regression on SPY.

Each run started with $W_0 = 100{,}000$~USD in cash at the close of 2025-01-02, and its wealth at the close of day
$t$ was $W(t) = c(t) + \sum_i n_i(t)\,S_i(t)$, with cash $c$ (USD), fractional shares $n_i$ (negative for a short
position) and close prices $S_i$ (USD/share). Positive cash grew by the factor $e^{g_f(t)\,\Delta t}$ each day.
Orders filled at the next open, sells before buys, with all buys scaled down by one common factor when the cash
could not cover them, so cash never went negative, and a target position worth at most $0.1\%$ of wealth was set to
zero shares. A fill of $q$ shares (signed) of name $i$ at the open price $S^{\mathrm{o}}_i$ (USD/share) paid
\begin{equation}
  C(q) \;=\; \bar c\, S^{\mathrm{o}}_i\,|q| \;+\; \kappa\,\frac{|q|}{A_i}\,S^{\mathrm{o}}_i\,|q|
  \label{eq:cost}
\end{equation}
in USD, the half spread $\bar c = 5$~bps of the traded value plus a linear impact cost~\cite{almgrenChriss2001}
with coefficient $\kappa = 0.001$ on the participation $|q|/A_i$, where $A_i$ (shares/day) is the mean daily volume
of name $i$ over the $63$ trading days that end on the decision day. There was no commission, and the same half
spread prices the turnover limit of the allocation problem. Short positions, which only the 130/30 benchmark took,
paid a borrow fee at the rate $b = 0.0075$~$\mathrm{yr}^{-1}$ on their value at each close, charged daily, and their
sale proceeds were held as cash.

\subsection{The score and the selection rules}\label{sec:method-score}

The score combines a single-index fit of each name to the market with the market state. The daily growth rates
$g_i(k)$ of name $i$ and $g_M(k)$ of SPY (Eq.~\ref{eq:growth}, yr$^{-1}$) follow the single-index model
$g_i(k) = \alpha_i + \beta_i\, g_M(k) + \varepsilon_i(k)$~\cite{sharpe1963}, with intercept $\alpha_i$ (yr$^{-1}$),
market loading $\beta_i$ (dimensionless) and residual $\varepsilon_i(k)$ (yr$^{-1}$). We fit the model by ordinary
least squares on 2014--2024 and, during the trading window, updated $(\alpha_i, \beta_i)$ every trading day by
exponentially weighted least squares (EWLS),
\begin{equation}
(\alpha_i, \beta_i)_t = \arg\min_{a,\,b} \Big[ \sum_{k = t_0 + 1}^{t} \delta^{\,t-k}
  \big( g_i(k) - a - b\, g_M(k) \big)^2
  + \delta^{\,t - t_0} N_0\, \mathrm{E}_0 \big[ ( g_i - a - b\, g_M )^2 \big] \Big],
\label{eq:ewls}
\end{equation}
where $t_0$ is the last trading day before the window and $\delta = 2^{-1/63}$ is the daily decay, a half-life of
$63$ trading days. The prior term counts as $N_0 = 63$ observations, with $\mathrm{E}_0$ the expectation under the
2014--2024 fit and SPY's 2014--2024 mean ($0.106$~yr$^{-1}$) and variance ($7.51$~yr$^{-2}$) of daily growth; its
minimizer is the 2014--2024 fit, and its weight decays with the data's. The estimates used on day $t$ include the
growth into day $t$ and nothing later. Two market inputs come from SPY's closes. The market growth average
$\tilde g_M$ (yr$^{-1}$) is the exponential moving average of SPY's daily growth rate with a span of $252$ trading
days (smoothing factor $2/253$). The market state is
\begin{equation}
\xi_t = -G \left( \frac{\mathrm{EMA}_{21}(t)}{\mathrm{EMA}_{63}(t)} - 1 \right),
\label{eq:xi}
\end{equation}
where $\mathrm{EMA}_{21}$ and $\mathrm{EMA}_{63}$ are exponential moving averages of SPY's close with spans of $21$
and $63$ trading days, and the gain $G = 21.23$ is the reciprocal of the $99$th percentile of
$\lvert \mathrm{EMA}_{21}/\mathrm{EMA}_{63} - 1 \rvert$ over 2014--2024, so $\lvert\xi_t\rvert \le 1$ on $99\%$ of
those days. The state $\xi_t$ is positive exactly when the 21-day average lies below the 63-day average, as in a
falling market, and it was positive on $75$ of the $420$ trading days of the window (Fig.~\ref{fig:market-state}).
The score of name $i$ is
\begin{equation}
\gamma_i = \tanh\!\left( \frac{\alpha_i + \beta_i\, \tilde g_M}{\max(\lvert\beta_i\rvert,\, 0.1)^{\xi_t}} \right),
\label{eq:gamma}
\end{equation}
with growth rates in yr$^{-1}$ entering the hyperbolic tangent as plain numbers, so $\gamma_i \in (-1, 1)$. The
numerator is the name's expected growth rate $\mu_{g,i}$ from the single-index fit. It keeps the sign of $\beta_i$,
so a name with a negative loading is charged rather than credited for market growth when $\tilde g_M > 0$. The
denominator, the beta penalty, is positive, so $\gamma_i$ always has the sign of the expected growth rate and the
market state changes only its size. When $\xi_t > 0$ the penalty divides the expected growth of a name with
$\lvert\beta_i\rvert > 1$ by a number above one and that of a name with $\lvert\beta_i\rvert < 1$ by a number below
one, so the score favors names that move less with the market; when $\xi_t < 0$ the effect reverses and the score
favors names that move more; at $\xi_t = 0$ the score is the hyperbolic tangent of expected growth. The floor of
$0.1$ keeps the penalty finite for names with a loading near zero, and at the window's largest $\xi_t$ of $0.90$ it
limits their amplification to a factor of $0.1^{-0.90} \approx 7.9$. The same expected growth rate enters the
tangency rule and the controller's forecast.

A selection rule chose the names of each decision from the $373$ names of the universe in $11$ sectors, using data
dated on or before the day of the pick only, and the study compared ten rules. Three picked $3$ names in each sector,
$33$ in all: \emph{$\gamma$ per sector} kept the $3$ highest $\gamma_i$ in each sector, \emph{bandit per sector} kept
the $3$ names a per-sector bandit ranked highest, and \emph{random per sector} drew $3$ names per sector uniformly at
random. Five picked $K$ names from the whole universe with no sector quota, with $K = 13$ in the main comparison and
other sizes in the size comparison (Fig.~\ref{fig:S-size}). \emph{Global $\gamma$} kept the $K$ highest $\gamma_i$.
\emph{Correlation-aware $\gamma$} walked down the same ranking and accepted a name only if the Pearson correlation of
its daily growth rates with those of every name already accepted, over the trailing $252$ trading days, was at most
$0.7$, and filled any shortfall from the ranking. \emph{Liquid then $\gamma$} kept the $K$ highest $\gamma_i$ among
the $100$ names with the largest median daily dollar volume (close times volume) over the trailing $63$ trading days,
\emph{random liquid 100} drew $K$ names uniformly from that shortlist, and \emph{random whole universe} drew $K$ names
uniformly from the universe. The random rules drew afresh at every pick, from a stream seeded by the run seed and the
date. The last two rules were fixed lists: the hand-picked $13$ (AAPL, MSFT, INTC, MU, AMD, NVDA, GS, BAC, WFC, C, F,
GM, JNJ), which the author chose in advance from sectors the author follows and without 2025 data, and the same list
with WMT and PG added. Each is one list chosen by one person, and a pick never changed it, although the hedge pool
(the stocks with zero or negative beta that join each decision when the market hedge is on) could change.

The bandit rule ran one $\epsilon$-greedy bandit~\cite{lattimoreSzepesvari2020} per sector, retrained from scratch at
each pick, with each of the sector's $A$ names (between $15$ and $60$) as one arm. A pull of arm $j$ drew a window
start $d$ uniformly from the $252$ trading days whose 21-day window ends within the trailing $252$ trading days, so
every window was complete by the day of the pick, and earned the excess growth rate
$x_j(d) = [\ln(S_j(d+21)/S_j(d)) - \ln(S_M(d+21)/S_M(d))]/(21\,\Delta t)$ (yr$^{-1}$) of name $j$ over SPY, with
$S_M$ SPY's close. The bandit made $\min(\max(50A, 500), 5000)$ pulls. The first pull chose an arm uniformly at
random; at pull $k > 1$ the bandit chose a uniformly random arm with probability
$\epsilon_k = \max\{0.05,\, \min[1,\, k^{-1/3} (A \ln k)^{1/3}]\}$, and otherwise the pulled arm with the highest
running-mean reward, ties going to a random rank drawn at the start. With these settings $\epsilon_k$ stayed above
$0.51$ at every pull in every sector, so the floor of $0.05$ never bound and each pull explored with probability
above one half. The rule kept the $3$ arms with the highest mean reward, unpulled arms last
(Algorithm~\ref{alg:S-bandit}). Because every arm can be scored on every window, the bandit approximates a ranking of
each sector's names by their mean 21-day growth rate in excess of SPY over the trailing year, a sector-neutral
momentum screen~\cite{jegadeesh1993}.

\subsection{Allocation by constrained utility maximization}\label{sec:method-allocation}

At each decision, made after the close of trading day $t$, the allocator chose a continuous number of shares
$n_i \ge 0$ for every name $i$ in the decision's name set $\mathcal{N}$, which held the names picked by the selection
rule and, with the market hedge on, a pool of hedge names. With $S_i$ the closing price of name $i$ (USD/share), $W$
the wealth at the close (USD) and $w_i = n_i S_i / W$ the weight of name $i$ (dimensionless), the allocator solved
\begin{equation}
\begin{aligned}
\max_{n \ge 0}\quad & U(n;\gamma) \\
\text{subject to}\quad & \textstyle\sum_{i \in \mathcal{N}} n_i S_i \le W, \qquad
  w_i \le \bar w_i \;\; (i \in \mathcal{N}), \qquad \sqrt{w^{\top} \Sigma_g\, w} \le \sigma_{\mathrm{SPY}}, \\
& \bar c\, \bar S \textstyle\sum_{i \in \mathcal{N}} \lvert n_i - n_i^{\mathrm{prev}} \rvert \le B, \qquad
  \textstyle\sum_{i \in \mathcal{N}} \beta_i w_i \le \bar\beta \;\; \text{(hedge on)},
\end{aligned}
\label{eq:allocation}
\end{equation}
where $U(n;\gamma)$ is the weighting rule's utility and $\gamma_i$ the score of name $i$ (Eq.~\ref{eq:gamma}); wealth
not invested is held as cash, which has a market loading of zero.
The single-name cap was $\bar w_i = \max(0.10,\, 1.5/K)$ for a name picked by the selection rule, with $K$ the number
of picks, and $0.10$ for hedge-pool names and $130/30$ short names, so a pick could hold up to $10\%$ of wealth in the
$33$-name baskets, $11.5\%$ in the $13$-name baskets and $30\%$ at $K = 5$. In the volatility cap, $\Sigma_g$
(yr$^{-2}$) is the covariance of the names' daily growth rates, the single-index market term plus the sample
residual covariance over the trailing $126$ trading days (Eq.~\ref{eq:mpc-cov} with its calibrated shrinkage of
zero), and $\sigma_{\mathrm{SPY}} = 2.741$~yr$^{-1}$ is the sample standard deviation of SPY's daily growth rate over
2014--2024, an annualized volatility of $17.3\%$; the solver imposed the cap as a second-order cone on $L^{\top} w$
with $L L^{\top} = \Sigma_g + 10^{-8} I$. The turnover budget caps the half-spread cost of the share changes from
the holdings $n^{\mathrm{prev}}$ before the decision, priced at the half spread $\bar c$ of the cost model
(Eq.~\ref{eq:cost}) times the mean closing price $\bar S$ (USD/share) of the names in $\mathcal{N}$, with
$\bar c = 5 \times 10^{-4}$ and $B = 0.10\,W$ (USD). For a long-only book this budget can bind only when $\bar S$
exceeds $100$ times the cheapest name's price.
The study set no sector limit. With the hedge on, the portfolio's market loading $\sum_i \beta_i w_i$, computed with
each name's signed online loading (Eq.~\ref{eq:ewls}), was capped at
$\bar\beta = \max(0,\, 1 - \xi_t)$, where $\xi_t$ is the market state (Eq.~\ref{eq:xi}). The cap equals SPY's loading
of one at $\xi_t = 0$, tightens as $\xi_t$ rises in a falling market, reaches zero at $\xi_t \ge 1$ and rises above
one when $\xi_t < 0$; over the trading window $\xi_t$ ranged from $-0.80$ to $0.90$, so $\bar\beta$ ranged from
$1.80$ down to $0.10$ (Fig.~\ref{fig:market-state}).

The hedge pool was refreshed at every pick of the basket. A universe name was eligible when its online market loading
was at most zero and its down-day beta was at most zero, where the down-day beta is the ordinary least squares slope,
with intercept, of the name's daily growth rate on SPY's over the days of the trailing $252$ trading days on which
SPY's growth rate was negative. The pool held up to $H = 5$ eligible names, most negative down-day beta first, and
joined every decision's name set on top of the picks, a name both picked and pooled counting once as a hedge name.
In the rules that use $\gamma$, a pool name with $\gamma_i > 0$ entered the utility like any other name, while a pool
name with $\gamma_i \le 0$ had no utility term and was a free variable above its floor, so the allocator bought it only
when the $\beta$ cap or the volatility cap made it worth its share of the budget. Because the objective is flat along
such a name's weight whenever no limit binds, it subtracted a tie-break cost of $10^{-6}$ per unit of hedge weight,
and a second, lexicographic solve then fixed every other name's shares at their first-stage values and minimized the
total weight of the free hedge names under the same limits. The utility depends only on the fixed names, so the
second solve left it unchanged and held a hedge name above its floor only as far as a binding limit required; if the
second solve failed, the first-stage shares stood. With the hedge off, a decision had neither the $\beta$ cap nor the
pool and was otherwise identical. We added the $\beta$ cap and the pool to the design after inspecting how the online
loadings behaved in the February--April 2025 sell-off; every decision used trailing data only, and every result is
reported with the hedge on and off.

We compared four utilities built on the score. With $\mathcal{P} = \{ i \in \mathcal{N} : \gamma_i > 0 \}$ the names
with a positive score, they are
\begin{equation}
\begin{aligned}
U_{\mathrm{CD}} &= \sum_{i \in \mathcal{P}} \gamma_i \ln n_i, &
U_{\mathrm{CES}} &= \Big( \sum_{i \in \mathcal{P}} \gamma_i\, n_i^{\rho} \Big)^{1/\rho}, \\
U_{\mathrm{SG}} &= \sum_{i \in \mathcal{P}} \gamma_i \ln \left( n_i - b_i \right), &
U_{\mathrm{CESw}} &= \Big( \sum_{i \in \mathcal{P}} \gamma_i\, w_i^{\rho} \Big)^{1/\rho},
\end{aligned}
\label{eq:utilities}
\end{equation}
the Cobb--Douglas (CD)~\cite{cobbDouglas1928}, Stone--Geary (SG)~\cite{geary1950, stone1954} and constant elasticity
of substitution (CES)~\cite{arrowEtAl1961} utilities, the last on share counts (CES) and on weights (CESw). Both CES
rules used the elasticity $\eta = 0.5$, so $\rho = 1 - 1/\eta = -1$, and SG used the baseline $b_i = 0.01\,W/S_i$
shares, a core position worth $1\%$ of wealth that SG holds in every name of $\mathcal{N}$ ($n_i \ge b_i$). Under the
budget alone each utility has a closed-form optimum (Table~\ref{tab:utilities}; derivations in
Appendix~\ref{app:allocation}). CD splits the invested wealth in proportion to $\gamma_i$. CES sets
$n_i \propto (\gamma_i/S_i)^{\eta}$, so its weights $w_i \propto \gamma_i^{\eta} S_i^{1-\eta}$ depend on each name's
share price when $\eta \ne 1$, whereas CESw sets $w_i \propto \gamma_i^{\eta}$ and does not; both equal CD at
$\eta = 1$, approach equal share counts (CES) or equal weights (CESw)~\cite{demiguelGarlappiUppal2009} as
$\eta \to 0$, and concentrate in the name with the largest $\gamma_i/S_i$ (CES) or $\gamma_i$ (CESw) as
$\eta \to \infty$. SG first buys every baseline and then splits the remaining wealth in proportion to $\gamma_i$.
With the limits of Eq.~\ref{eq:allocation} there is no closed form, and each utility becomes a convex cone
program~\cite{boydVandenberghe2004}: CD and SG use exponential cones, the two CES rules use power cones, the
volatility cap is a second-order cone and the other limits are linear. We built each program in JuMP
$1.32.1$~\cite{lubin2023jump} and solved it with the interior-point solver Clarabel $0.11.1$~\cite{goulart2024clarabel}
(absolute and relative duality-gap tolerances $10^{-12}$, feasibility tolerance $10^{-9}$, at most $500$ iterations),
accepting its optimal and almost-optimal statuses; when Clarabel reached neither, the same program went to SCS
$2.7.0$~\cite{odonoghue2016scs} with its default settings. To check the solver against the closed forms, we allocated
a new book of $100{,}000$ USD on 2 January 2025 across the hand-picked $13$-name list, with the hedge off and every
limit except the budget removed (the turnover budget stayed and did not bind). Over $17$ solves (CD, and CES and CESw
at eight values of $\eta$ from $0.05$ to $20$), the largest gap between a solver weight and its closed form was
$1.0 \times 10^{-5}$, below the tolerance of $10^{-3}$ we required.

The rules treat names with $\gamma_i \le 0$ differently. CD, CES and CESw leave them out of the utility and hold them
at a floor of $\varepsilon = 10^{-3}$ shares, which disappears when targets worth at most $0.1\%$ of wealth are set
to zero before trading, so the three rules held only names with $\gamma_i > 0$ and any hedge names the limits
required; when no name in $\mathcal{N}$ had $\gamma_i > 0$, they skipped the solve and moved the book to cash. SG
holds a name with $\gamma_i \le 0$ at its baseline $b_i$; with no positive score it held the baselines, adding the
least hedge weight that met the $\beta$ cap, and it reported the problem infeasible when the baselines exceeded the
budget or broke another limit. The baselines use $1\%$ of wealth per name, so SG cannot hold more than $100$ names.
In the large-book runs every SG decision at $K = 200$ was infeasible and the book stayed in cash, and at $K = 100$ most
SG decisions were infeasible and kept the previous book.

Three benchmarks ran under the same limits. Minimum variance (MinVar) and tangency
(Tangent)~\cite{markowitz1952, tobin1958} do not use $\gamma$ and could hold any name of $\mathcal{N}$, with
$n_i \ge 0$ and no floors. Each followed a two-fund form: it found the largest invested fraction
$s = \min(1, \max \sum_i w_i)$ that the budget, single-name, $\beta$ and turnover limits allowed without the
volatility cap, chose the mix at $\sum_i w_i = s$ that minimized $w^{\top} \Sigma_g w$ (MinVar) or maximized
$(\mu_g - g_f \mathbf{1})^{\top} w / \sqrt{w^{\top} \Sigma_g w}$ (Tangent), and scaled the mix by
$\min(1, \sigma_{\mathrm{SPY}} / \sigma_{\mathrm{mix}})$ onto the volatility cap, holding the rest in cash. Here
$\mu_{g,i} = \alpha_i + \beta_i \tilde g_M$ (yr$^{-1}$) is the expected growth rate of name $i$ from the score's inputs
and $g_f$ (yr$^{-1}$) the risk-free growth rate (Eq.~\ref{eq:riskfree}). Tangent solved its ratio as a quadratic
program after a Charnes--Cooper change of variables~\cite{charnes1962} and fell back to MinVar when no feasible mix
had positive excess growth. The $130/30$ long-short Cobb--Douglas rule (LS)~\cite{jacobsLevyStarer1998} maximized
$\sum_{\gamma_i > 0} \gamma_i \ln n_i^{+} + \sum_{\gamma_i < 0} \lvert\gamma_i\rvert \ln n_i^{-}$ over long shares
$n_i^{+}$ and short shares $n_i^{-}$ ($n_i = n_i^{+} - n_i^{-}$), so it shorted names with $\gamma_i < 0$ weighted by
$\lvert\gamma_i\rvert$. Long weights summed to at most $1.30$, short weights to at most $0.30$ in absolute value and
the net weight to at most $1$, so cash stayed nonnegative; the single-name cap applied to $\lvert w_i \rvert$, the
volatility and $\beta$ caps to the signed weights and the turnover budget to the signed share changes. Its long side
held the picks with $\gamma_i > 0$ and the hedge pool, and its short side held the short candidates, up to $3$ names
per sector with the most negative $\gamma_i < 0$ outside the pool, chosen at each pick of the basket and kept only
while $\gamma_i < 0$. A hedge name was never shorted, since shorting a name with a negative loading raises the
portfolio's loading. With no short candidate LS ran as long-only CD, and with no positive score it moved to cash.

Algorithm~\ref{alg:allocation} lists one allocation decision. A decision whose program a solver certified
infeasible, that neither solver solved, or that produced non-finite shares kept every holding and placed no orders.
Apart from SG with $100$ or more names, this happened in under $1\%$ of the decisions of every grid of real-data runs.

\begin{algorithm}[t]
\caption{One allocation decision after the close of day $t$.}
\label{alg:allocation}
\begin{algorithmic}[1]
  \REQUIRE Closes through day $t$; held shares $n^{\mathrm{prev}}$; wealth $W$ (USD); weighting rule $U$; selection
    rule; hedge on or off; whether a pick of the basket is due (Algorithm~\ref{alg:mpc}).
  \ENSURE Target shares $n^{\star}$ and orders for the open of day $t+1$, or the kept book and no orders.
  \STATE $\gamma_i \gets$ Eq.~\ref{eq:gamma} for every universe name, from the day's $(\alpha_i, \beta_i)$
    (Eq.~\ref{eq:ewls}), $\tilde g_M$ and $\xi_t$ (Eq.~\ref{eq:xi})
  \IF{a pick is due}
    \STATE $\mathcal{K} \gets$ picks of the selection rule; $\mathcal{H} \gets$ hedge pool if the hedge is on, else
      $\emptyset$
    \STATE $\mathcal{S}^{-} \gets$ up to $3$ names per sector with the most negative $\gamma_i < 0$, outside
      $\mathcal{H}$ \hfill \COMMENT{LS only; else $\emptyset$}
  \ENDIF
  \STATE $\mathcal{N} \gets \mathcal{K} \cup \mathcal{H}$; for LS,
    $\mathcal{N} \gets \{ i \in \mathcal{K} : \gamma_i > 0 \} \cup \mathcal{H} \cup
    \{ i \in \mathcal{S}^{-} : \gamma_i < 0 \}$
  \STATE Build $\mu_{g,i} = \alpha_i + \beta_i \tilde g_M$, $\Sigma_g$ (Eq.~\ref{eq:mpc-cov}), $\bar w_i$,
    $\bar c\,\bar S$, $B = 0.10\,W$, and $\bar\beta = \max(0, 1 - \xi_t)$ (hedge on) or $\bar\beta = \infty$
  \IF{$U$ uses $\gamma$ and no $i \in \mathcal{N}$ has $\gamma_i > 0$}
    \STATE $n^{\star} \gets$ floors: $\varepsilon$ (CD, CES, CESw), $0$ (LS), or $b_i$ plus the least hedge weight
      that meets $\bar\beta$ (SG) \hfill \COMMENT{no utility solve}
  \ELSE
    \STATE Solve Eq.~\ref{eq:allocation} for $U$ with Clarabel \hfill \COMMENT{MinVar, Tangent: mix, then scale}
    \IF{Clarabel is neither optimal nor almost optimal}
      \STATE Solve the same program with SCS
    \ENDIF
    \IF{a hedge name has $\gamma_i \le 0$ and $U$ uses $\gamma$}
      \STATE Fix the other names' shares; minimize $\sum_{i \in \mathcal{H}} w_i$ under the same limits
    \ENDIF
  \ENDIF
  \IF{the program is infeasible, both solvers failed, or some $n^{\star}_i$ is not finite}
    \STATE $n^{\star} \gets n^{\mathrm{prev}}$; \textbf{return} no orders \hfill \COMMENT{keep the whole book}
  \ENDIF
  \STATE $n^{\star}_i \gets 0$ where $\lvert n^{\star}_i S_i \rvert \le 0.001\,W$, and for held names outside
    $\mathcal{N}$
  \RETURN orders $n^{\star} - n^{\mathrm{prev}}$, filled at the open of day $t+1$
\end{algorithmic}
\end{algorithm}

\subsection{Model-predictive rebalancing}\label{sec:method-mpc}

The controller decides when to re-plan the book and leaves it alone in between. It follows model-predictive
control in the receding-horizon sense~\cite{garcia1989mpc,mayneEtAl2000}. At each re-plan it measures the
state of the account (the shares $n_i$, the cash $c$, the wealth $W(t)$, the close prices $S_i(t)$ and the online
estimates of that day), solves the constrained utility maximization of Algorithm~\ref{alg:allocation} for target
shares $n^\star$, and predicts the distribution of the wealth of holding $n^\star$ over the next $T = 21$ trading
days with a multi-asset geometric Brownian motion (GBM). It then compares the measured wealth with the prediction
every day and re-plans from the newly measured state when the horizon elapses, when the wealth leaves the
predicted band, or when the book falls a set fraction below its peak. The plan over the horizon is to hold
$n^\star$ without further trades, and the GBM is the controller's only model of prices. The plant is the price
process that the plan meets, the recorded prices in the backtest and synthetic prices from per-asset JumpHMM
generators~\cite{alswaidan2026variance} in the stress test, where the controller still forecasts with the GBM and
never sees the generator. Multi-period trading by convex optimization~\cite{boydEtAl2017} also re-solves from
the measured state, but it plans a sequence of trades over the horizon and executes the first of them every
period. Our controller instead solves a single-period allocation, uses the forecast as a reference for the
realized wealth, and trades only when a trigger fires, so the band around the forecast acts as a no-trade
region~\cite{davisNorman1990} placed on the portfolio value relative to its forecast rather than on the weights.

At a re-plan on day $t$ the drift of name $i$ is its expected growth rate
$\mu_{g,i} = \alpha_i(t) + \beta_i(t)\,\tilde g_M(t)$ ($\mathrm{yr}^{-1}$). Here $\alpha_i(t)$ ($\mathrm{yr}^{-1}$)
and $\beta_i(t)$ (dimensionless) are the online single-index estimates of name $i$ against SPY~\cite{sharpe1963},
fitted by exponentially weighted least squares with a half-life of $63$ trading days from a prior worth $63$
pseudo-observations at the calibration fit (Appendix~\ref{app:controller}), and $\tilde g_M(t)$
($\mathrm{yr}^{-1}$) is the exponential moving average of the daily growth rate of SPY with a span of $252$ trading
days. The covariance $\Sigma_g$ ($\mathrm{yr}^{-2}$) of the daily growth rates of the names in the allocation
problem is
\begin{equation}
  \Sigma_g \;=\; \sigma_M^2(t)\,\beta\beta^{\top} \;+\; (1-\lambda)\,\hat\Omega_\varepsilon
  \;+\; \lambda\,\mathrm{diag}\big(\hat\Omega_\varepsilon\big),
  \label{eq:mpc-cov}
\end{equation}
where $\beta$ is the vector of the loadings $\beta_i(t)$, $\sigma_M(t)$ ($\mathrm{yr}^{-1}$) is the exponentially
weighted standard deviation of the daily growth rate of SPY with the same $63$-day half-life, and
$\hat\Omega_\varepsilon$ ($\mathrm{yr}^{-2}$) is the sample covariance of the residuals
$\varepsilon_i(s) = g_i(s) - \alpha_i(t) - \beta_i(t)\,g_M(s)$ over the $L_\Omega = 126$ daily growth rates that
end on day $t$. The weight $\lambda \in [0, 1]$ shrinks the residual covariance toward its diagonal, from the full
sample covariance at $\lambda = 0$ to the covariance of the single-index model, the target of Ledoit and
Wolf~\cite{ledoitWolf2003}, at $\lambda = 1$. We chose it on the calibration window by the QLIKE
loss~\cite{patton2011} of $21$-day variance forecasts for random portfolios, which selected $\lambda = 0$
(Appendix~\ref{app:controller}). The allocation problem uses the same $\mu_g$ and $\Sigma_g$ in its volatility limit
and in its minimum-variance and tangency benchmarks, so each book is forecast with the model that sized it.

From the target shares $n^\star$ and the close prices $S_i(t)$, the controller simulates $N_p = 1{,}000$ paths over
the next $T = 21$ trading days. On path $p$ and day $\tau = 1, \dots, T$ after the re-plan, the log price of each
name advances by $\Delta t$ times a vector of growth rates drawn from a multivariate normal distribution with mean
$\mu_g$ and covariance $\Sigma_g$, independently across days and paths,
\begin{equation}
  \ln S^{(p)}_i(\tau) \;=\; \ln S^{(p)}_i(\tau-1) \;+\; g^{(p)}_i(\tau)\,\Delta t,
  \qquad g^{(p)}(\tau) \sim \mathcal{N}\big(\mu_g, \Sigma_g\big),
  \qquad S^{(p)}_i(0) = S_i(t),
  \label{eq:mpc-gbm}
\end{equation}
a multi-asset GBM~\cite{glasserman2003} written in growth rates. Because $\mu_g$ is the expected growth rate of the
log price, the drift needs no It\^o correction. The wealth of the plan on path $p$ is
\begin{equation}
  W^{(p)}(\tau) \;=\; \sum_i n^\star_i\, S^{(p)}_i(\tau) \;+\; c^\star\, e^{g_f(t)\,\tau\,\Delta t},
  \qquad c^\star \;=\; W(t) - \sum_i n^\star_i\, S_i(t),
  \label{eq:mpc-value}
\end{equation}
where $c^\star$ (USD) is the cash left after paying for the targets at the close of the decision day. The forecast
is the sample mean $\mu_W(\tau)$ and standard deviation $\sigma_W(\tau)$ (USD) of $W^{(p)}(\tau)$ across the paths.
It omits the trading costs, the overnight price change before the fill and, for a 130/30 book, the borrow fee, and
its paths are seeded with the run seed and the decision date, so a run reproduces its forecasts exactly.

The forecast defines a band $\mu_W(\tau) \pm z\,\sigma_W(\tau)$ with $z = 1.96$, which would contain the wealth on
about $95\%$ of days if the forecast wealth were normal and the model were correct. With $\tau$ the number of trading
days since the last re-plan, the controller marks the wealth $W(t)$ at each close, after that day's fills, costs,
cash growth and fees, updates its running peak $W^{\mathrm{pk}}(t)$, and checks three triggers in turn, acting on
the first that holds. The drawdown trigger fires when $1 - W(t)/W^{\mathrm{pk}}(t) > D_{\max}$ with $D_{\max} = 0.08$, the horizon
trigger when $\tau \ge T$, and the band trigger when $W(t)$ lies outside $\mu_W(\tau) \pm z\,\sigma_W(\tau)$, so the
wealth measured $\tau$ days after a re-plan is compared with the forecast for $\tau$ days ahead on days
$\tau = 1, \dots, T-1$. A drawdown re-plan sets the peak back to the current wealth, so the trigger fires once for
each fresh loss of $D_{\max}$ rather than on every day that the book stays below an old peak.
We adopted this reset after an earlier pass of the Tier 1 grid (the three per-sector selection rules crossed with
the seven weighting rules) without it, in which $88.5\%$ of the decisions were
drawdown re-plans. A re-plan that leaves no position worth more than $0.1\%$ of wealth sets no band, and the next
re-plan then comes from the drawdown trigger or, after $21$ trading days, from a cash revisit.

The selection rule picks the basket on the first day, every $21$ trading days after its previous pick, and at
every band exit. With the hedge on, each pick also refreshes the hedge pool, and for the 130/30 benchmark the short
candidates. When a timer pick changes the set of names on a day when no trigger fired, the controller re-plans as
well. Because band exits depend on each run's own wealth path, the dates of the picks, and with them the
baskets, could differ across weighting rules within a seed. At a re-plan it solves the allocation problem from the shares actually held, so the turnover limit counts
the changes from that book, queues orders to $n^\star$ for the next open, forecasts the new book from the measured
wealth $W(t)$ and resets $\tau$ to $0$. Each forecast therefore starts from the measured wealth rather than from
the previous forecast, and the horizon recedes with every re-plan. If the solve fails, the controller keeps every
holding, queues no orders and forecasts the kept book. One episode of a real run, three consecutive forecasts and
the re-plans that ended them, shows the mechanism (Fig.~\ref{fig:mpc}), and Algorithm~\ref{alg:mpc} gives the daily
loop of a run.

\begin{algorithm}[t]
\caption{Daily loop of one backtest run. A run fixes a selection rule, a weighting rule, the hedge setting and a
seed. Every decision uses data dated on or before the close of its day.}
\label{alg:mpc}
\begin{algorithmic}[1]
\REQUIRE trading days $t_1, \dots, t_{420}$; calibration fit on 2014--2024; $W_0$; $T$, $z$, $D_{\max}$, $N_p$
\ENSURE daily wealth $W(t)$ and the log of re-plans
\STATE $c \gets W_0$, $n \gets 0$, $W^{\mathrm{pk}} \gets W_0$, $\tau \gets 0$; no band, no pending orders
\FOR{each trading day $t = t_1, \dots, t_{420}$}
  \IF{orders are pending}
    \STATE fill them at the open: sells first, then buys scaled by one factor in $[0,1]$ so that $c \ge 0$; pay
      the costs of Eq.~\eqref{eq:cost}
  \ENDIF
  \IF{$t > t_1$}
    \STATE $c \gets c\,e^{g_f(t)\Delta t}$ if $c > 0$; charge the borrow fee on short positions; $\tau \gets \tau + 1$
  \ENDIF
  \STATE mark $W(t) \gets c + \sum_i n_i S_i(t)$ at the close; $W^{\mathrm{pk}} \gets \max(W^{\mathrm{pk}}, W(t))$
  \STATE update $\alpha_i, \beta_i, \sigma_M, \tilde g_M, \xi$ and $\gamma_i$ with the growth rates into day $t$
  \STATE reason $\gets$ the first that holds of: first day; cash revisit (last re-plan went to cash and
    $\tau \ge 21$); drawdown ($1 - W(t)/W^{\mathrm{pk}} > D_{\max}$); horizon ($\tau \ge T$); band exit
    ($|W(t) - \mu_W(\tau)| > z\,\sigma_W(\tau)$); otherwise none
  \IF{first day, or $21$ trading days since the last pick, or reason is band exit (except on the fixed schedule)}
    \STATE pick the basket with the selection rule; refresh the hedge pool and the short candidates
    \IF{reason is none and the set of names changed}
      \STATE reason $\gets$ basket change
    \ENDIF
  \ENDIF
  \IF{reason is not none}
    \STATE build $\mu_g$ and $\Sigma_g$ (Eq.~\eqref{eq:mpc-cov}) for the names of the allocation problem
    \STATE solve the allocation problem (Algorithm~\ref{alg:allocation}) from $n$, $W(t)$, $S(t)$: targets $n^\star$
    \IF{the solve failed}
      \STATE $n^\star \gets n$; queue no orders
    \ELSIF{$t < t_{420}$}
      \STATE queue orders to $n^\star$ for the open of the next day, each target worth at most $0.001\,W(t)$ set
        to zero shares
    \ENDIF
    \STATE simulate $N_p$ paths of Eqs.~\eqref{eq:mpc-gbm} and~\eqref{eq:mpc-value} for the book $n^\star$; store
      $\mu_W(\tau)$, $\sigma_W(\tau)$, $\tau = 1, \dots, T$
    \STATE set the band from the forecast (no band if the book holds no position above $0.001\,W(t)$);
      $\tau \gets 0$
    \IF{reason is drawdown}
      \STATE $W^{\mathrm{pk}} \gets W(t)$
    \ENDIF
  \ENDIF
\ENDFOR
\end{algorithmic}
\end{algorithm}

\subsection{Backtest design}\label{sec:method-backtest}

A backtest run fixes a selection rule, a weighting rule, the hedge setting and a seed, and replays
Algorithm~\ref{alg:mpc} over the $420$ trading days of the trading window. We used the seeds $5660$ to $5689$, $30$
runs per cell of an experiment grid. The seed enters in three places: the random selection rules draw each basket
from a stream seeded with the run seed and the pick date, each sector's bandit seeds its exploration with the run
seed, the pick date and the sector, and the forecast paths of every re-plan are seeded with the run seed and the
decision date. For the deterministic selection rules (the $\gamma$-based rules and the hand-picked lists) the seed
changes only the forecast paths, and through them the band exits and so the dates of the re-plans and picks. Each
run with the hedge on has a twin with the hedge off and the same seed, identical except that the beta cap and the
hedge pool are removed.

We ran nine grids over the full trading window, and all $18{,}238$ distinct runs finished
(Table~\ref{tab:S-grids}). The Tier 1 grid crossed the seven weighting rules with three per-sector selection rules
($3$ names per sector, $33$ names) and the hedge on and off ($854$ runs). The fixed-schedule grid repeated its
hedge-on half with basket picks only on the $21$-day timer ($427$ runs), while band exits and drawdown stops still
triggered re-plans, unlike the calendar-only grid. Because the random and bandit streams depend
only on the seed and the pick date, and the $\gamma$ ranking and the hedge pool only on market data, every weighting
rule in that grid held the same baskets within a seed.
The universe-wide grid applied five selection rules that pick $K$ names from the whole universe, at
$K = 5, 8, 13, 21$ and $34$ ($4{,}410$ runs), and the hand-picked grid ran the two hand-picked lists ($28$ runs).
A seed rerun gave each deterministic rule all $30$ seeds ($4{,}200$ runs, $140$ of which repeat earlier runs), and
the large-book grid extended global $\gamma$ and random picks from the whole universe to $K = 100$ and $200$
($1{,}680$ runs).
A calendar-only grid repeated the $4{,}200$ runs of the main comparison, described next, with the band exit and the
drawdown stop switched off ($z = \infty$ and $D_{\max} = \infty$), so that a run re-planned only when the $21$-day
horizon elapsed, and paired each run with the triggered run of the same selection rule, weighting rule, hedge
setting and seed ($4{,}200$ runs).
Two sensitivity grids varied the CES elasticity $\eta$, one on the Tier 1 grid with the single-name cap on and off
($2{,}562$ runs) and one on the hand-picked $13$-name list under the budget alone ($17$ runs). This paper does not
report results from the fixed-schedule grid or from the elasticity grid on the Tier 1 rules, and the budget-only grid
served as the solver check against the closed forms.

The main comparison crosses ten selection rules with the seven weighting rules, with the hedge on and off. The ten
are $\gamma$, bandit and random per sector ($33$ names each), global $\gamma$, correlation-aware $\gamma$, liquid
then $\gamma$, random from the $100$ most liquid names and random from the whole universe ($K = 13$ each), and the
two hand-picked lists. The $13$-name list (AAPL, MSFT, INTC, MU, AMD, NVDA, GS, BAC, WFC, C, F, GM, JNJ) and its
$15$-name extension with WMT and PG were chosen in advance by the author, from sectors the author follows and
without 2025 data. Each is one list chosen by a person. Each of the $140$ cells holds $30$ runs, and we report the
median over the runs of each cell, never a median pooled across selection rules. For each run we computed the
terminal wealth ratio $W_T/W_0$, with $W_T$ the wealth at the close of 2026-09-04, the realized growth rate
$\ln(W_T/W_0)/(419\,\Delta t)$, the maximum drawdown $\max_t\big(1 - W(t)/\max_{s \le t} W(s)\big)$, and the
annualized Sharpe ratio~\cite{sharpe1966} $\bar y/(\sqrt{\Delta t}\,s_y)$, where $\bar y$ and $s_y$ are the mean and
standard deviation of the daily excess growth $y(t) = g_W(t) - g_f(t)$, with $g_W$ the daily growth rate of the
run's wealth. We also recorded the intercept $\alpha_W$ ($\mathrm{yr}^{-1}$) and slope $\beta_W$ of an ordinary least-squares regression of $y(t)$ on the daily excess growth
of SPY, and the number of re-plans by trigger. Two passive benchmarks received the same metrics. SPY buy-and-hold
puts $W_0$ into SPY at the close of 2025-01-02, and the equal-weight universe puts $W_0/373$ into each name at that
close and never rebalances. Both are valued at the closes and pay no trading costs. The universe-wide selection
rules, their thresholds and the grid of $K$ were designed after the Tier 1 results were known, so their comparison
is exploratory. The $30$-seed reruns and the large-book runs were added afterwards to measure how much the forecast
seed moves a deterministic rule and whether the comparison of global $\gamma$ with random picks holds for larger
books. The calendar-only grids, on real prices and on the synthetic stress sets, were added last, after all other
results were known, to measure what the two early triggers changed.

\subsection{Synthetic markets from per-asset JumpHMM generators}\label{sec:method-synthetic}

The backtest covers one real history of $420$ trading days with one sell-off, so it cannot show how the method
behaves in conditions that did not occur in that window. We therefore built synthetic markets for the whole universe
and ran the complete method on each. Every synthetic series came from JumpHMM, the hidden Markov generator with jumps
of Alswaidan and Varner~\cite{alswaidan2026variance} (Fig.~\ref{fig:jumphmm}). A JumpHMM is a hidden Markov
model~\cite{rabinerJuang1986,hamilton1989} whose states are intervals of the daily growth rate bounded by
equal-probability quantiles of a fitted Laplace distribution, whose transition matrix is estimated by counting, and
whose draw for each day is Student-$t$ with $\nu = 5$ degrees of freedom, located and scaled by the training days in
the selected state. Because the counted chain mixes within a few days, JumpHMM adds jump episodes to reproduce the
slow decay of the autocorrelation of absolute growth rates~\cite{mandelbrot1963,cont2001,bullaBulla2006}. On a day
with no episode under way, an episode starts with probability $\epsilon$ and lasts a Poisson number of days with mean
$\lambda$, and on each of its days the state is drawn uniformly from the $N_{\rm tail}$ most negative states with
probability $p_{\rm neg}$ or from the $N_{\rm tail}$ most positive states otherwise, which makes an episode a sustained
burst of large moves of both signs rather than a single price jump~\cite{merton1976,kou2002}. The pair
$(\epsilon, \lambda)$ minimizes on a grid, over simulated paths that contain an episode, the squared error in the
absolute-growth autocorrelation at lags $1$ to $25$ plus $0.20$ times the squared error in excess kurtosis. We write
$\epsilon$ for the jump probability and $\varepsilon$ for a growth-rate draw.

Every JumpHMM here used $50$ states, $\nu = 5$, $p_{\rm neg} = 0.52$ and $N_{\rm tail} = 5$, with jumps enabled
(Table~\ref{tab:S-generator-params}). Two settings differ from the SPY generator of the JumpHMM paper. That paper used
$100$ states, so its tail sets held the most extreme $5\%$ of states on each side, whereas with $50$ states they hold
$10\%$. Its multi-asset comparisons either disabled jumps or gave every asset the SPY values of $(\epsilon, \lambda)$,
whereas each sector and name model here was tuned on its own series. As for SPY in that paper, the tuned $\epsilon$
sat at the lower edge of its grid, $10^{-4}$ per day, for all $11$ sector models and $370$ of the $373$ name models,
while $\lambda$ spread over its whole grid of $10$ to $120$ trading days. We used JumpHMM.jl version $0.1.0$ at a
pinned revision (commit \texttt{aaed969}) without modification and called only its fitting, tuning and simulation
functions. Its multi-asset models either reorder each asset's draw to the ranks of a copula sample or add
bootstrapped or Gaussian regression residuals to a market factor, and neither keeps each name's own JumpHMM path in
order, so we wrote the composition in the study code.

\begin{figure}[H]
    \centering
    \includegraphics[width=0.85\textwidth]{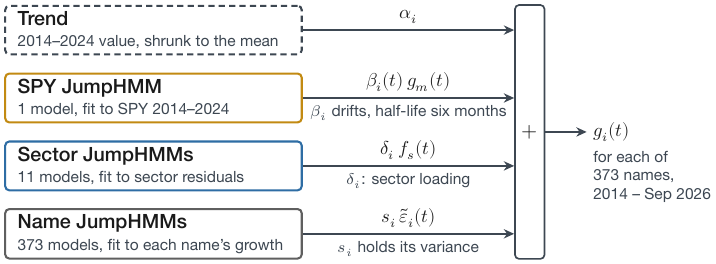}
    \caption{\textbf{Composition of one name's daily growth rate in the synthetic market.} Each name's growth rate
    $g_i(t)$ is the sum of a trend $\alpha_i$ (its 2014--2024 value, shrunk toward the cross-sectional mean), a
    market term $\beta_i(t)\,g_m(t)$ from one JumpHMM (hidden Markov generator with jumps) fit to SPY, a sector term $\delta_i f_s(t)$ from one of $11$
    sector JumpHMMs fit to sector residuals, and a name term $s_i\tilde\varepsilon_i(t)$ from the name's own
    JumpHMM, scaled so that the sum holds the name's variance. The loading $\beta_i(t)$ drifts with a half-life of
    $126$ trading days. The name terms are drawn independently across names, so dependence across names enters only
    through the market and sector terms.}
    \label{fig:multiasset}
\end{figure}

The synthetic market combined three kinds of JumpHMM (Fig.~\ref{fig:multiasset}). The market model, fitted in
earlier work to SPY closing prices over 2014--2024 (the study's series), has $\epsilon = 10^{-4}$ and
$\lambda = 118.9$ trading days. Its draw is the growth rate $g_m(t)$ (yr$^{-1}$) of synthetic SPY on day $t$, with a
flag $J(t)$ that marks the days of a jump episode. For the sector models, we took the residual
$e_i(t) = g_i(t) - \hat\alpha_i - \hat\beta_i\, g_M(t)$ (yr$^{-1}$) of each name from its ordinary least-squares
(OLS) single-index fit to SPY over 2014--2024~\cite{sharpe1963}, where $g_i(t)$ and $g_M(t)$ are the real growth rates
of name $i$ and SPY and $\hat\alpha_i$ (yr$^{-1}$) and $\hat\beta_i$ are the fitted intercept and slope. The sector
factor $f_s(t)$ (yr$^{-1}$) is the equal-weight mean residual of the names in sector $s$, and because JumpHMM.jl fits
price series, each sector model was fitted to the compounded index $\exp(\Delta t \sum_{u \le t} f_s(u))$. As in the
JumpHMM paper, $g_i$ and $f_s$ denote real series when we estimate and synthetic ones when we compose. Finally, one
JumpHMM was fitted to each name's 2014--2024 closing prices. We refitted all $373$ name models on the study's data
rather than reuse models stored from earlier work, whose training file held nine unadjusted stock splits, and each
model's draws had a standard deviation within $0.5$ to $2$ times the real one (median ratio $1.037$).

For name $i$ in sector $s = s(i)$ on day $t$, let $\alpha_i$ (yr$^{-1}$) be its synthetic intercept, $\beta_i(t)$ its
market loading, $\delta_i$ its sector loading, $f_s(t)$ (yr$^{-1}$) the draw of its sector's model,
$\tilde\varepsilon_i(t)$ (yr$^{-1}$) the centered draw of its own model and $s_i$ a scale on that draw. The synthetic
growth rate is
\begin{equation}
  g_i(t) = \alpha_i + \beta_i(t)\, g_m(t) + \delta_i\, f_s(t) + s_i\, \tilde\varepsilon_i(t),
  \qquad t = 1, \dots, 3{,}186,
  \label{eq:compose}
\end{equation}
over every trading day of the real calendar from 2014-01-03 to 2026-09-04, so one continuous path covers both the
calibration and the trading window and every date-based setting of the method applies unchanged. Closes start at each
name's real close on 2014-01-03 and compound as $S_i(t) = S_i(t-1)\exp(g_i(t)\,\Delta t)$, and synthetic SPY
compounds from its real close with $g_m(t)$ (Algorithm~\ref{alg:generator}).

The intercept is the name's real 2014--2024 OLS intercept pulled toward the universe mean $\bar\alpha$,
$\alpha_i = \bar\alpha + \phi(\hat\alpha_i - \bar\alpha)$ with $\phi = 0.6$, constant in time, so names keep their
2014--2024 order at $60\%$ of their real distance from the mean. The pull reflects how little of the real spread
persists. In 2014--2024 the spread of $\hat\alpha_i$ across names was $0.076$~yr$^{-1}$, against a root-mean-square
standard error of $0.075$~yr$^{-1}$, and the intercepts of 2014--2019 and 2020--2024 correlated $0.175$ across names.
We fitted $\phi$ so that OLS intercepts estimated on synthetic paths reproduced both statistics
(Appendix~\ref{app:synthetic}). We dropped a planned drift in $\alpha_i$, of the same form as that of $\beta_i(t)$,
because its fitted size was zero and real intercepts changed between the halves of 2014--2024 by no more than their
estimation noise. We also set aside the raw estimates, which overstate the real spread and persistence, and a common
intercept for every name, which removes them. Because each name keeps a persistent intercept, the synthetic markets
favor selection by $\gamma$ more than real 2025--2026 did, whose intercepts correlated $-0.12$ across names with
real 2014--2024 intercepts, against $0.10$ on synthetic paths (Table~\ref{tab:S-checks}).

The market loading drifts as $\beta_i(t) = \hat\beta_i + x_i(t)$, where $x_i(t)$ is an exponential moving average
(EMA) of independent standard normal shocks $z_i(t)$ with weight $a$, half-life $h_\beta$ (trading days) and scale
$k_i$:
\begin{equation}
  x_i(t) = (1 - a)\, x_i(t-1) + a\, k_i\, z_i(t), \qquad a = 1 - 2^{-1/h_\beta}.
  \label{eq:beta-drift}
\end{equation}
The first value is drawn from the stationary distribution, whose standard deviation is $k_i\sqrt{a/(2-a)}$, so the
synthetic calibration window already contains drift. We calibrated the drift against the method's own online
estimate $\hat\beta_i(t)$ (exponentially weighted least squares with a half-life of $63$ trading days) rather than an
assumed truth. Each $k_i$ reproduces on synthetic 2014--2024 the standard deviation of the name's real
$\hat\beta_i(t)$, and the common half-life reproduces the median lag-$252$ autocorrelation of $\hat\beta_i(t)$,
$0.206$ (Appendix~\ref{app:synthetic}). The fit chose $h_\beta = 126$ trading days, and the stationary standard
deviation of the drift had a median of $0.26$ across names (interquartile range $0.20$ to $0.34$). The synthetic
$\hat\beta_i(t)$ thus carries the estimation noise of the real one, and a name whose $\beta_i(t)$ drifts below zero
becomes a hedge candidate on that path.

The sector loading $\delta_i$ is the OLS slope of the name's real residual $e_i(t)$ on the mean residual of the other
names in its sector, which keeps its own noise out of the regressor in sectors with as few as $15$ names, times one
factor $\kappa_s$ per sector. The slope on this noisy proxy for the common factor is biased toward zero, yet it
multiplies a factor drawn with the variance of the full sector mean, and the unscaled loadings undercounted the shared
co-movement in every sector (in Communication Services, a mean residual correlation of $0.052$ against a real $0.094$).
We chose $\kappa_s$ so that $\delta_i \delta_j \sigma_s^2$, averaged over the pairs $i \ne j$ of the sector, equals
the mean pairwise covariance of the real residuals, where $\sigma_s^2$ (yr$^{-2}$) is the variance of the real sector
factor. The factors ran from $1.02$ in Financials to $1.34$ in Communication Services
(Table~\ref{tab:S-sector-params}), and the sector models were drawn independently of each other and of the market.

The own draw is centered on its model's long-run mean, $\tilde\varepsilon_i(t) = \varepsilon_i(t) - \bar\varepsilon_i$,
where $\varepsilon_i(t)$ is a path of the name's JumpHMM and $\bar\varepsilon_i$ (yr$^{-1}$) the mean of one
$10^6$-day draw with a fixed seed. Let $\sigma_m^2$, $\sigma_s^2$ and $\sigma_i^2$ (yr$^{-2}$) be the variances of the
real 2014--2024 growth rates of SPY, the sector factor and the name, held fixed rather than measured on each path. The
factor share of the name's variance, averaged over the drift in $\beta_i(t)$, is
\begin{equation}
  \bar\rho_i = \frac{\left(\hat\beta_i^2 + k_i^2\, a/(2-a)\right)\sigma_m^2 + \delta_i^2\, \sigma_s^2}{\sigma_i^2},
  \label{eq:rho-bar}
\end{equation}
where the term in parentheses is the long-run mean of $\beta_i(t)^2$. With the JumpHMM paper's floor
$s_{\min}^2 = 0.10$ on the own draw's share, we set $s_i^2 = 1 - \bar\rho_i$ when $\bar\rho_i \le 1 - s_{\min}^2$.
Otherwise we multiplied $\beta_i(t)$ and $\delta_i$ by $\theta_i = \sqrt{(1 - s_{\min}^2)/\bar\rho_i}$ for the whole
path, which keeps their signs, and set $s_i^2 = s_{\min}^2$. This happened for two Utilities names
($\bar\rho_i = 0.917$ and $0.918$), or $0.54\%$ of name-days on every path, and the paper's branch that keeps the real
$R^2$ for names with $R^2 \ge 0.80$ was never used (largest $R^2 = 0.65$). Because the shares are real and each
JumpHMM draws slightly more variance than its training series, the composed variance sat at the level of the name's
own draw, with median ratios of $0.994$ to the draw's variance and $1.066$ to the real variance
(Table~\ref{tab:S-checks}).

We did not reorder the residuals with a copula. The rank reorder that the JumpHMM paper offers as an option imposes a
fitted cross-asset dependence~\cite{patton2006,cherubiniLucianoVecchiato2004}, but it permutes each residual path over
days and so destroys the name's volatility clustering and the timing of its jump episodes. Each name here keeps its
own draw in the order it was produced, and cross-sectional dependence enters only through the shared terms of
Eq.~\eqref{eq:compose}. Names in one sector share $g_m(t)$ and $f_s(t)$, names in different sectors share only
$g_m(t)$, and all dependence is contemporaneous. The mean residual correlation across sectors was therefore $0.0002$
on synthetic paths against $0.031$ in real 2014--2024, and correlations between names rise in a sell-off only through
the market term, whereas real equity correlations rise further in down markets~\cite{longinSolnik2001,angChen2002}.

The composition extends the variance-corrected single-index composition of the JumpHMM
paper~\cite{alswaidan2026variance} and departs from it in six ways. First, the paper centered each draw on its
realized mean, which forces the residuals to sum to zero over the path. On a path that spans both windows, that sum
ties the noise of the trading window to minus the noise of the calibration window, and on $20$ test paths the noise
parts of each name's intercepts in the two windows correlated $-0.48$ across names under path centering against
$0.003$ under centering on the model mean. Second, the paper's variance target was each draw's own sample variance,
whereas the variances in Eq.~\eqref{eq:rho-bar} are fixed, so a market jump burst raises a name's variance through
its market term instead of shrinking its own noise and loadings over the whole path, calibration years included.
Third, the paper composed with the market alone, and Eq.~\eqref{eq:compose} adds the sector term. Fourth, the paper's
loading was constant, whereas $\beta_i(t)$ drifts here and the variance is matched on average over the drift with one
$s_i$ per path. Matching it day by day, as an earlier version did, scaled down $15$ to $24\%$ of name-days in Energy,
Utilities, Financials and Real Estate, weakened the sector co-movement, and scaled $8.8\%$ of all name-days rather
than $0.54\%$. Fifth, when the floor binds, the paper reduced the market loading alone, whereas here the market and
sector loadings shrink by one factor $\theta_i$. Sixth, the paper kept the fitted intercept, and here it is pulled
toward the universe mean. The paper also compared its centered and rescaled full-growth draws with comparators fitted
to regression residuals. Our name terms follow the full-growth approach, whereas our sector terms are residual-fit
models, since each sector JumpHMM is fitted to a mean of regression residuals.

Before running the method on any synthetic path, we compared the generator with the real market on $22$ gated
checks of SPY's growth, the gain from daily rebalancing (a guard against a market that rewards trading more than the
real one), each name's variance, market loading and clustering, the co-movement within sectors, the spread of
$\hat\beta_i(t)$, reproducibility from seeds and the OLS intercepts (Table~\ref{tab:S-checks}). Twenty-one checks
passed. The cross-name standard deviation of OLS intercepts failed, at $0.089$~yr$^{-1}$ against $0.076$~yr$^{-1}$
with a tolerance of $0.01$~yr$^{-1}$. Estimation noise in synthetic intercepts adds spread, and a smaller $\phi$
would push the correlation between the halves of 2014--2024 further below its target, so we reported the failure
rather than tune the generator to pass. Two reported statistics limit what the synthetic market can show. The
composed names clustered as their own draws did but less than real names, with absolute-growth autocorrelations below
the real ones by a median of $0.10$, $0.17$ and $0.10$ at lags $1$, $5$ and $20$. A median of $7$ names per path had
$\hat\beta_i(t) \le 0$ at some point in the trading window, near the median of $5$ over real windows of that length
in 2014--2024 but against $120$ in real 2025--2026, so the hedge had fewer candidates than in the backtest. The
study's test suite, including its look-ahead tests, passed with the synthetic code in place.

\begin{algorithm}[t]
\caption{One synthetic future of the $373$-name market over the full calendar, 2014-01-03 to 2026-09-04.}
\label{alg:generator}
\small
\begin{algorithmic}[1]
  \REQUIRE Path number $p$; JumpHMMs of the market, of each sector $s = 1, \dots, 11$ and of each name
    $i = 1, \dots, N$; real 2014--2024 estimates $\hat\alpha_i$, $\hat\beta_i$, $R_i^2$ and loadings $\delta_i$; real
    variances $\sigma_m^2$, $\sigma_s^2$, $\sigma_i^2$; long-run means $\bar\varepsilon_i$; pull $\phi$; drift
    half-life $h_\beta$ and scales $k_i$; floor $s_{\min}^2$; real closes $S_{\rm SPY}(0)$, $S_i(0)$ on 2014-01-03.
  \ENSURE Closes $S_{\rm SPY}(t)$ and $S_i(t)$ and jump flags $J(t)$ for $t = 1, \dots, 3{,}186$.
  \STATE $a \gets 1 - 2^{-1/h_\beta}$;\quad $\bar\alpha \gets N^{-1}\sum_i \hat\alpha_i$
  \STATE Draw $g_m(t)$ and $J(t)$ from the market JumpHMM with seed $p$
    \hfill \COMMENT{JumpHMM simulation~\cite{alswaidan2026variance}}
  \FOR{$s = 1, \dots, 11$}
    \STATE Draw $f_s(t)$ from the JumpHMM of sector $s$ with a seed set by $(p, s)$
  \ENDFOR
  \FOR{$i = 1, \dots, N$}
    \STATE Draw $\varepsilon_i(t)$ from the JumpHMM of name $i$ with a seed set by $(p, i)$;\quad
      draw $z_i(t) \sim \mathcal{N}(0, 1)$
    \STATE $\alpha_i \gets \bar\alpha + \phi(\hat\alpha_i - \bar\alpha)$
    \STATE $x_i(1) \gets k_i\sqrt{a/(2-a)}\, z_i(1)$;\quad
      $x_i(t) \gets (1-a)\, x_i(t-1) + a\, k_i\, z_i(t)$ for $t \ge 2$;\quad
      $\beta_i(t) \gets \hat\beta_i + x_i(t)$
      \hfill \COMMENT{Eq.~\eqref{eq:beta-drift}}
    \STATE $\bar\rho_i \gets \big[(\hat\beta_i^2 + k_i^2 a/(2-a))\,\sigma_m^2 + \delta_i^2\,\sigma_{s(i)}^2\big]
      / \sigma_i^2$
      \hfill \COMMENT{Eq.~\eqref{eq:rho-bar}}
    \IF{$R_i^2 \ge 0.80$}
      \STATE $\theta_i \gets 1$;\quad $s_i^2 \gets \bar\rho_i (1 - R_i^2)/R_i^2$
        \hfill \COMMENT{keeps the real $R^2$; no name qualified}
    \ELSIF{$\bar\rho_i \le 1 - s_{\min}^2$}
      \STATE $\theta_i \gets 1$;\quad $s_i^2 \gets 1 - \bar\rho_i$
    \ELSE
      \STATE $\theta_i \gets \sqrt{(1 - s_{\min}^2)/\bar\rho_i}$;\quad $s_i^2 \gets s_{\min}^2$
        \hfill \COMMENT{one scale for the whole path}
    \ENDIF
    \STATE $g_i(t) \gets \alpha_i + \theta_i \beta_i(t)\, g_m(t) + \theta_i \delta_i\, f_{s(i)}(t)
      + s_i\big(\varepsilon_i(t) - \bar\varepsilon_i\big)$ for every $t$
      \hfill \COMMENT{Eq.~\eqref{eq:compose}}
  \ENDFOR
  \STATE $S_{\rm SPY}(t) \gets S_{\rm SPY}(0)\exp\big(\Delta t \sum_{u \le t} g_m(u)\big)$;\quad
    $S_i(t) \gets S_i(0)\exp\big(\Delta t \sum_{u \le t} g_i(u)\big)$
  \RETURN $S_{\rm SPY}$, $S_i$ for every name, and $J$
\end{algorithmic}
\end{algorithm}

\subsection{Evaluation on synthetic futures}\label{sec:method-evaluation}

We drew $10{,}000$ market paths over the full calendar, seeding path $p$ with $p$, and ranked them by the maximum
drawdown of synthetic SPY within the trading window, 2025-01-02 to 2026-09-04, with the peak reset on the window's
first day (Algorithm~\ref{alg:evaluation}). We tagged a window as containing a jump burst when the market model's flag
$J(t)$ was set on any of its growth days, which happened in $523$ of the $10{,}000$ windows ($5.2\%$), as the market
model's jump settings imply (Appendix~\ref{app:synthetic}), and in $30$ of the $50$ deepest drawdowns. The typical
set holds $600$ paths drawn at random from all $10{,}000$, the burst-stress set holds the $200$ deepest drawdowns among
windows with a burst, and the bear-stress set holds the $200$ deepest among windows without one. Since $22$ typical
paths also fell in a stress set, the sets hold $978$ distinct paths. Synthetic SPY's median window drawdown was
$0.176$, $0.364$ and $0.406$ in the typical, burst-stress and bear-stress sets, and its median growth rate over the
window was $0.102$, $-0.087$ and $-0.199$~yr$^{-1}$, so a burst window holds months of large moves of both signs with
less net decline than the sustained falls of the bear-stress set. Because the sector, drift and name draws are
independent of the market draw, we drew them only for the chosen paths, from seeds derived from the path number, and
any path can be rebuilt bit for bit.

On each of the $978$ paths, we treated the synthetic history as if it were real and ran the whole method on it. We
first recalibrated every estimated input on synthetic 2014-01-03 to 2024-12-31 with the code of the real run, namely
the OLS single-index fit of each name to synthetic SPY, SPY's growth moments and the volatility cap
$\sigma_{\rm SPY}$, the gain $G$ of the market state $\xi$, and the shrinkage of the residual correlation toward its
diagonal, which the calibration chose anew on every path (zero on $664$ of the $978$ paths, at most $0.2$). The method
thus estimated $\alpha_i$ and $\beta_i(t)$ from synthetic prices without seeing the generator's values. We then traded
the trading window with the controller and settings of the real backtest. The controller still forecast the
portfolio value with a multi-asset geometric Brownian motion (GBM), now calibrated on the synthetic history, and the
plant it acted on was the synthetic closes, so JumpHMM played the market and not the forecast model. The synthetic
data kept the real calendar, sectors and 3-month Treasury bill rate, held each name's daily share volume at its real
2014--2024 median, so that the dollar volume ranked by the liquidity-based rule moves with the synthetic price, and
set opens equal to closes, which the method does not use.

Each path received $532$ runs, a subset of the real grid sized to the compute budget. All seven weighting rules, with
the hedge off and on, were run under the three per-sector selection rules ($\gamma$, bandit and random, each picking
three names in each of the $11$ sectors), under five universe-wide rules at $K = 13$ (global $\gamma$,
correlation-aware $\gamma$, liquid then $\gamma$, random from the whole universe and random from the $100$ most liquid
names), under global $\gamma$ and random selection from the whole universe at $K = 5$ and $K = 34$, and on the two
lists of $13$ and $15$ names that the author chose in advance. The randomized rules, bandit and random, used $5$ seeds
($5660$ to $5664$) per path instead of the $30$ of the real backtest, and the other rules ran once. Each path also
recorded synthetic SPY bought and held and the equal-weight universe bought once and held. The $978$ paths gave
$520{,}296$ runs, all of which finished without error.
On each of the $400$ stress paths, the $364$ runs of the ten main selection rules were repeated with the band exit
and the drawdown stop switched off, as in the calendar-only grid on real prices, and each was paired with the
triggered run of the same selection rule, weighting rule, hedge setting and seed on the same path ($145{,}600$ runs,
all of which finished without error).

For each set and hedge setting, we summarized the terminal wealth ratio $W_T/W_0$, the realized growth rate, the
maximum drawdown and the Sharpe ratio in selection rule by weighting rule grids with the two benchmark rows, never
pooling across selection rules (Fig.~\ref{fig:synthetic}). We took the median over seeds within a path and then,
across paths, the median, the $10$th and $90$th percentiles, and the share of paths on which the cell's terminal
wealth ratio exceeded that of the path's own synthetic SPY. That share is a paired comparison, so a cell can beat SPY
on most paths while its median lies below that of the SPY row. We also formed paired differences across paths between
global $\gamma$ and random selection at $K = 5$, $13$ and $34$, between $\gamma$ per sector and random per sector, and
between hedge on and hedge off, and we located each real backtest result among the $600$ typical futures as the share
of typical paths that ended at or below it (Fig.~\ref{fig:S-percentile}). These summaries describe this generator
combined with this method and are not forecasts.

\begin{algorithm}[t]
\caption{Evaluation of the method on synthetic futures.}
\label{alg:evaluation}
\small
\begin{algorithmic}[1]
  \REQUIRE Generator of Algorithm~\ref{alg:generator}; $P = 10{,}000$ market paths; calibration window
    $\mathcal{C}$ (2014-01-03 to 2024-12-31); trading window $\mathcal{W}$ (2025-01-02 to 2026-09-04); run grid
    $\mathcal{R}$ of $532$ runs; seeds $5660$ to $5664$ for the randomized selection rules.
  \ENSURE Selection rule by weighting rule summaries for the typical, burst-stress and bear-stress sets.
  \FOR{$p = 1, \dots, P$}
    \STATE Draw $g_m(t)$ and $J(t)$ with seed $p$ and compound synthetic SPY closes
    \STATE $d_p \gets$ maximum drawdown of synthetic SPY over $\mathcal{W}$, peak reset at the start of
      $\mathcal{W}$
    \STATE $b_p \gets$ true if $J(t)$ is set on any growth day of $\mathcal{W}$
  \ENDFOR
  \STATE Burst stress $\gets$ the $200$ paths with the largest $d_p$ among those with $b_p$ true
  \STATE Bear stress $\gets$ the $200$ paths with the largest $d_p$ among those with $b_p$ false
  \STATE Typical $\gets$ the first $600$ paths of a seeded random permutation of $1, \dots, P$
  \FOR{each path $p$ in the union of the three sets}
    \STATE Build the synthetic market of path $p$
      \hfill \COMMENT{Algorithm~\ref{alg:generator}}
    \STATE Recalibrate on synthetic $\mathcal{C}$: single-index fits, SPY moments and volatility cap, gain $G$ of
      $\xi$, residual shrinkage
    \FOR{each run in $\mathcal{R}$ (selection rule, weighting rule, hedge setting, seed)}
      \STATE Trade $\mathcal{W}$ with the rebalancing controller, forecasting with GBM calibrated on synthetic
        data and acting on the synthetic closes
      \STATE Record $W_T/W_0$, growth rate, maximum drawdown and Sharpe ratio
    \ENDFOR
    \STATE Record synthetic SPY and the equal-weight universe, each bought and held over $\mathcal{W}$
  \ENDFOR
  \STATE For each set, hedge setting and cell: median over seeds within a path, then the median, the $10$th and
    $90$th percentiles, and the share of paths above the path's own synthetic SPY
\end{algorithmic}
\end{algorithm}

\section{Results}\label{sec:results}

We ran the same allocation and rebalancing system on real prices and on synthetic futures and compared the median
terminal wealth of every cell with two passive benchmarks (Tables~\ref{tab:realrun} and~\ref{tab:synthetic}). A cell
was one selection rule, one weighting rule, and one hedge setting. On real prices, the ten selection rules ($\gamma$
per sector, bandit per sector, random per sector, global $\gamma$, correlation-aware $\gamma$, liquid then $\gamma$,
random liquid 100, random whole universe, and the two hand-picked lists) and the seven weighting rules (CD, CES,
CESw, SG, MinVar, Tangent, and LS) gave $70$ cells for each hedge setting, traded over the $420$ trading days from
2 January 2025 to 4 September 2026.
Each of the $140$ cells held $30$ runs with seeds $5660$--$5689$, for $4{,}200$ real runs, and the seed set the
controller's forecast paths and, for the random and bandit rules, the names picked.
A size comparison set global $\gamma$ against random picks from the whole universe at $K = 5$, $8$, $13$, $21$, $34$,
$100$, and $200$ names, again with $30$ runs per cell.
The synthetic futures were price paths from the JumpHMM generator. We used $600$ typical paths drawn at random from
$10{,}000$ and two stress sets of $200$ paths each, the deepest SPY drawdowns in the trading window among paths with a
jump burst (burst stress) and among paths without one (bear stress), and because $22$ typical paths were also stress
paths, the three sets held $978$ distinct paths. On each path the controller recalibrated on the synthetic 2014--2024
history and traded the synthetic 2025--2026 window, and it still forecast with the multi-asset GBM while JumpHMM
supplied the prices.
Each path carried $532$ runs, from $14$ selection rules (the ten above, plus global $\gamma$ and random whole universe
at $5$ and $34$ names), the seven weighting rules, and both hedge settings, with five seeds ($5660$--$5664$) for the
six random and bandit rules, for $520{,}296$ runs in all, none of which failed. The benchmarks were SPY buy-and-hold
and the equal-weight universe, which put $W_0/373$ into each name at the first close and never rebalanced. We
measured performance by the terminal wealth ratio $W_T/W_0$ and by continuously compounded growth rates in
$\mathrm{yr}^{-1}$. Every reported median of $W_T/W_0$ belongs to one cell and is never pooled across selection
rules, and on synthetic paths the random and bandit rules first took the median over their five seeds within each
path.

\begin{figure}[H]
    \centering
    \includegraphics[width=\textwidth]{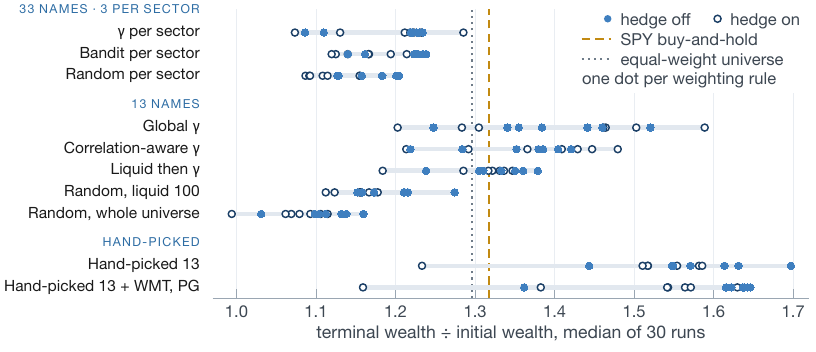}
    \caption{\textbf{Real-data backtest: terminal wealth by selection rule and weighting rule, 2 January 2025 to
    4 September 2026.} Each row is one selection rule, grouped by basket size ($33$ names, three per sector; $13$
    names; and the hand-picked lists of $13$ and $15$ names). Each dot is one of the seven weighting rules
    (Cobb--Douglas, CES on shares, CES on weights, Stone--Geary, minimum variance, tangency, and $130/30$
    long-short), placed at the median over $30$ runs (seeds $5660$--$5689$) of terminal wealth divided by initial
    wealth. Filled dots have the market hedge off and open dots have it on; with the hedge on, a cap on portfolio beta
    tightened as the market weakened and up to five stocks with zero or negative beta could be held. The pale bar
    spans each row's lowest to highest dot. The dashed gold line is SPY buy-and-hold ($1.32$) and the dotted line is the equal-weight universe
    ($1.30$).}
    \label{fig:realrun}
\end{figure}

We compared the seven weighting rules within each selection rule on real prices with the hedge off, and found that
the selection rule moved terminal wealth more than the weighting rule did (Fig.~\ref{fig:realrun};
Table~\ref{tab:realrun}). Within the $\gamma$-per-sector basket, the seven weighting rules ended between $1.086$
(MinVar) and $1.234$ (SG) times initial wealth, and the four utilities lay between $1.219$ and $1.234$, whereas
across all $70$ cells the medians ran from $1.031$ (random whole universe, with LS) to $1.697$ (hand-picked 13, with
Tangent).
The spread within a selection rule, the highest minus the lowest of its seven medians, was $0.078$ to $0.284$, with
a median of $0.14$, and the spread across the ten selection rules within a weighting rule was $0.357$ (MinVar) to
$0.591$ (LS), with a median of $0.518$. Without the two hand-picked lists, the spread across the other eight
selection rules was $0.246$ to $0.422$ (median $0.274$), against $0.078$ to $0.272$ (median $0.135$) within them.
SPY buy-and-hold ended at $1.317$ and the equal-weight universe at $1.296$. Of the $70$ cells, $29$ ended above SPY,
all from five selection rules: global $\gamma$ (six of the seven weighting rules), correlation-aware $\gamma$ (five),
liquid then $\gamma$ (four), and the two hand-picked lists (seven each). None of the three per-sector rules and
neither random $13$-name rule ended above SPY with any weighting rule.
In the $13$-name baskets, global $\gamma$ ended at $1.248$--$1.520$ against $1.031$--$1.160$ for random names from
the whole universe, whereas at three names per sector the $\gamma$ rule ($1.086$--$1.234$) and the bandit
($1.140$--$1.239$) ended close to random picks ($1.127$--$1.205$). The universe-wide rules were designed after the
per-sector results were known, so the $13$-name comparison is exploratory. The hand-picked lists ended highest, at
$1.443$--$1.697$ for the $13$ names and $1.362$--$1.646$ with WMT and PG added. Each is one list chosen by one person,
so it shows what that choice earned in this window and not the result of a repeatable rule.
MinVar or LS gave the lowest median under every selection rule, MinVar under four rules and LS under six.
For the deterministic rules the seed changed only the controller's forecast paths, so their $30$ runs clustered
tightly (hand-picked 13 with CD: interquartile range $1.546$--$1.553$), whereas the runs of random picks spread over
a wider band (random whole universe with CD: $1.048$--$1.220$).

\begin{figure}[H]
    \centering
    \includegraphics[width=\textwidth]{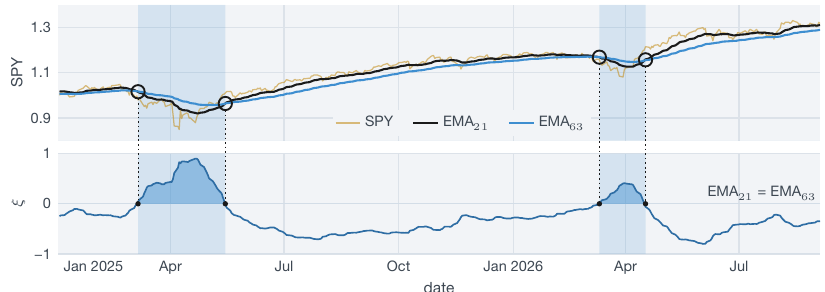}
    \caption{\textbf{SPY and the market state over the trading window, 2 January 2025 to 4 September 2026.}
    Top: SPY's daily close (light gold) with its $21$-day (black) and $63$-day (blue) exponential moving averages
    (EMAs), each divided by SPY's first close in the window. Bottom: the market state
    $\xi = -G(\mathrm{EMA}_{21}/\mathrm{EMA}_{63} - 1)$, where the gain $G = 21.23$ is the reciprocal of the $99$th
    percentile of $\lvert \mathrm{EMA}_{21}/\mathrm{EMA}_{63} - 1 \rvert$ over 2014--2024. Its positive part is filled. Shaded days have $\xi > 0$,
    when the $21$-day EMA lies below the $63$-day EMA; circles mark the crossings of the two EMAs, and dotted lines
    carry them down to the zeros of $\xi$.}
    \label{fig:market-state}
\end{figure}

We repeated every real run with the beta hedge on and compared terminal wealth, portfolio beta, and drawdown with
the hedge-off runs of the same selection and weighting rules, and found that the hedge lowered market exposure but
cost terminal wealth in most cells (Fig.~\ref{fig:realrun}; Table~\ref{tab:S-realrun-on}). At each re-plan with the
hedge on, the allocation obeyed a cap on portfolio $\beta$ that fell as the market state $\xi$ rose, and up to five
names whose online $\beta$ and down-day $\beta$ were at or below zero could enter as a hedge pool. The median
$W_T/W_0$ was lower with the hedge on in $50$ of the $70$ cells, and the median of the $70$ changes was $-0.033$.
Wealth fell in six or seven of the seven cells for the bandit rule, the three random rules, and liquid then
$\gamma$, and in five and six for the two hand-picked lists. Correlation-aware $\gamma$ ended higher in five of its
seven cells, and LS weights ended higher under six of the ten selection rules.
We measured exposure by the portfolio $\beta$, the value-weighted online $\beta$ of the holdings divided by total
wealth, averaged over the days of a run. The hedge lowered this mean portfolio $\beta$ in all $70$ cells (cell
medians $0.388$--$0.903$ with the hedge off and $0.314$--$0.849$ with it on), and hedge-pool names held
$8.4$--$39.8\%$ of wealth on an average day (cell medians). The maximum drawdown nonetheless changed little, by at
most $0.019$ in either direction, with a median change of $+0.001$ across the $70$ cells
(Table~\ref{tab:S-exposure}).
When the hedge-on and hedge-off runs were paired by seed, the hedge-on run had the smaller maximum drawdown in no
more than $14$ of $30$ pairs for any weighting rule under the bandit and random per-sector rules.
The cap explained part of the wealth loss. The state $\xi$ was positive on $75$ of the $420$ trading days
(Fig.~\ref{fig:market-state}), and the portfolio's mean growth rate on the $75$ days that followed a positive $\xi$
fell with the hedge on in $63$ of the $70$ cells (median change $-0.153\,\mathrm{yr}^{-1}$). On those days SPY's mean
daily growth rate was $0.266\,\mathrm{yr}^{-1}$, against $0.144\,\mathrm{yr}^{-1}$ on the other $344$ of the $419$
growth days, so a tighter cap held less of a market that rose on average. The difference of $0.12\,\mathrm{yr}^{-1}$
was small next to its standard error of $0.59\,\mathrm{yr}^{-1}$, so it describes this window and not a property of
the market state.
The real window was one historical path, with a largest SPY drawdown of $0.190$, so we also tested the hedge on
simulated sell-offs.

\begin{figure}[H]
    \centering
    \includegraphics[width=0.9\textwidth]{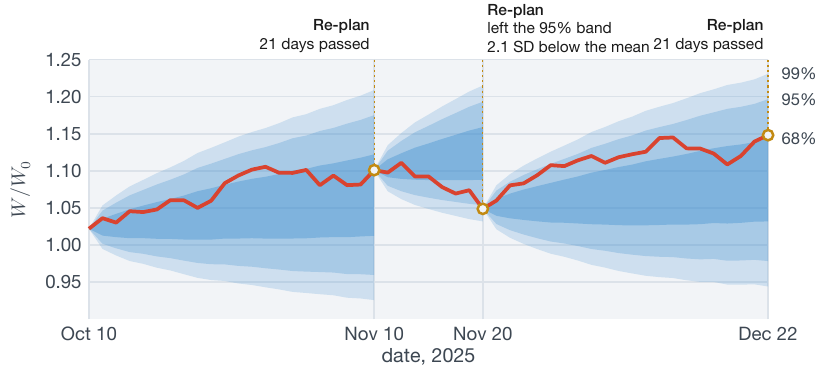}
    \caption{\textbf{One episode of the rebalancing controller on real prices, 10 October to 22 December 2025.}
    The red line is the wealth $W$ of one backtest (the hand-picked $13$ names, Cobb--Douglas weighting, market
    hedge off, seed $5660$) divided by its initial wealth $W_0$. At each re-plan the controller projected $1{,}000$ paths of a
    multi-asset geometric Brownian motion $21$ trading days ahead; the nested cones are the $68$, $95$ and $99\%$
    bands of the projected portfolio value. The first and third re-plans came when $21$ trading days had passed.
    The second came on 20 November, when wealth left the $95\%$ band, $2.1$ standard deviations below the projected
    mean.}
    \label{fig:mpc}
\end{figure}

We followed the controller through one real episode and counted re-plans by trigger across the $4{,}200$
main-comparison runs, and found that the forecast band and the drawdown stop brought forward a share of re-plans
that depended on the selection rule (Fig.~\ref{fig:mpc}; Table~\ref{tab:S-replans}). The episode came from the hand-picked 13 with CD weights,
the hedge off, and seed $5660$, and ran from a re-plan on 10 October 2025 to 22 December 2025. At each re-plan the
controller forecast portfolio value over the next $21$ trading days with the multi-asset GBM and $1{,}000$ paths,
while the portfolio traded real prices. At the horizon re-plan on 10 November the portfolio stood at $1.101$ times
initial wealth, $0.61$ standard deviations (SD) above the mean of the forecast made on 10 October. Eight trading days
into the next forecast, on 20 November, the portfolio had fallen to $1.0489$ against a forecast mean of $1.1235$ with
an SD of $0.0355$, $2.10$~SD below the mean and outside the $1.96$~SD band, and the controller re-planned. The third
forecast ran to its horizon on 22 December, when the portfolio stood at $1.148$, $1.09$~SD above its mean.
In the cell of this run, the median share of re-plans that came before the horizon was $0.37$. Across the $140$
cells, the median share brought forward by a band exit or a drawdown stop ranged from $0.20$ to $0.61$
(Table~\ref{tab:S-replans}). With the hedge off, the share was lowest for the three random rules, at $0.20$ to
$0.40$, and highest for the four $\gamma$ rules, whose highest cells ran from $0.53$ to $0.61$. The other re-plans
came at the horizon or on a basket change, a day on which a timer pick changed the set of names, and no run
re-planned on a cash revisit.

We repeated the main comparison with the band exit and the drawdown stop switched off, and found that the early
re-plans lowered the worst drawdown more consistently than they changed terminal wealth (Table~\ref{tab:S-calendar}).
Each of the $4{,}200$ calendar-only runs kept the selection rule, weighting rule, hedge setting and seed of one
triggered run, so the two runs of a pair differed only in the two triggers, and every calendar-only run re-planned
at the $21$-day horizon alone, $19$ times over the window, whereas in every cell the triggered runs re-planned more
often in the median. We ran this comparison because band exits measure how often the forecast missed, not whether re-planning
early helped the portfolio. With the hedge off, the median maximum drawdown was smaller with the triggers in $55$ of
the $70$ cells, by up to $0.053$ (global $\gamma$), and at most $0.015$ larger in the others. With the
hedge on it was smaller in $61$ of the $70$ cells, by up to $0.054$. Median terminal wealth was higher with the
triggers in $38$ of the $70$ cells with the hedge off and in $26$ with the hedge on, and the differences ran from
$-0.093$ to $0.191$ of initial wealth with the hedge off and from $-0.180$ to $0.183$ with it on. The sign of the
wealth difference did not follow the kind of selection rule: with the hedge off, the triggers raised terminal
wealth under all seven weighting rules for random liquid 100 and under none for random whole universe or liquid then
$\gamma$. Every triggered cell paid more trading cost than its calendar-only twin, $3$ to $33\%$ more with the hedge
off and $4$ to $45\%$ more with it on (Table~\ref{tab:S-replans}), since the triggered runs re-planned more often. On
the two stress sets of synthetic futures, the triggers likewise lowered the worst drawdown in most cells without
consistently raising terminal wealth (Tables~\ref{tab:S-calendar-burst} and~\ref{tab:S-calendar-bear}).

We compared the $K$ highest-$\gamma$ names in the universe with $K$ names drawn at random from the whole universe
for $K$ from $5$ to $200$ on real prices, and found that the advantage of selection by $\gamma$ shrank as the basket
grew (Fig.~\ref{fig:S-size}). For each basket size, weighting rule, and hedge setting, we formed all $900$ pairs of
one of the $30$ global $\gamma$ runs with one of the $30$ random runs, regardless of seed, and counted the share of
pairs in which global $\gamma$ ended with the higher $W_T/W_0$. Like the $13$-name comparison, this size comparison
is exploratory.
Stone--Geary was left out at $100$ and $200$ names, because its baseline holding of $1\%$ of wealth per name used the
whole budget at $100$ names and twice the budget at $200$, so most of its decisions were infeasible.
With the hedge off, global $\gamma$ ended higher in $91.8$--$100\%$ of pairs at $5$ names, $94.7$--$100\%$ at $8$, and
$90.2$--$100\%$ at $13$ across the weighting rules. The share fell to $74.9$--$96.0\%$ at $21$ names and
$56.6$--$96.0\%$ at $34$, and to $33.3$--$83.2\%$ at $100$ names and $34.6$--$82.7\%$ at $200$. Random picks ended
higher in most pairs for MinVar at $100$ names and for CD, CES, and Tangent at $200$.
With CD weights, the global $\gamma$ median fell from $1.568$ at $K = 5$ to $1.520$ at $K = 13$, $1.333$ at
$K = 34$, $1.244$ at $K = 100$, and $1.185$ at $K = 200$, while the random median rose from $1.048$ to $1.098$,
$1.182$, $1.203$, and $1.211$ at the same sizes.
With the hedge on, global $\gamma$ ended higher in $85.9$--$100\%$ of pairs at $5$ to $13$ names, and in
$90.3$--$100\%$ at $21$ and $34$ names for every weighting rule except MinVar, whose share fell to $43.2\%$ at
$21$ names and $26.1\%$ at $34$. At $100$ and $200$ names with the hedge on, the shares were $48.4$--$85.2\%$ and
$30.7$--$81.6\%$. Selection by $\gamma$ thus mattered most for baskets of $5$ to $13$ names, and at $100$ or more of
the $373$ names the CD medians of the two rules differed by less than $0.05$.

\begin{figure}[H]
    \centering
    \includegraphics[width=\textwidth]{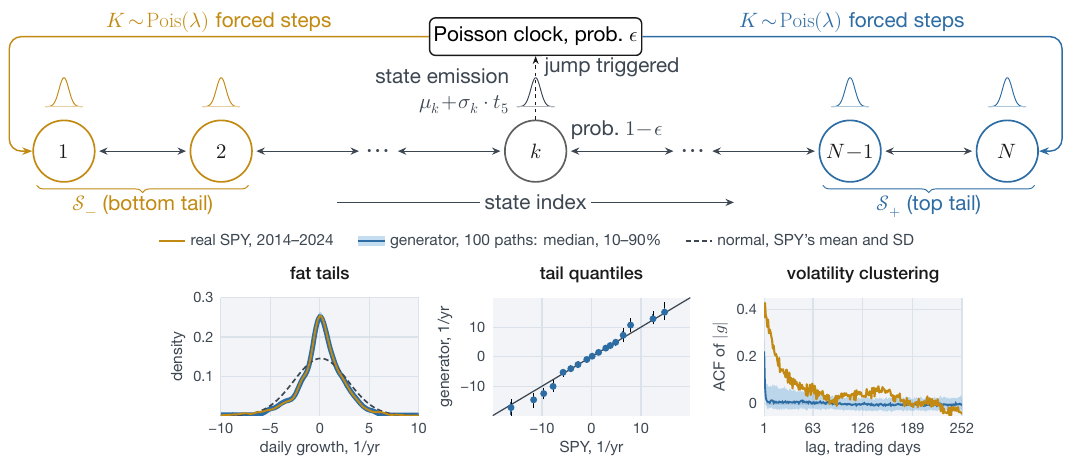}
    \caption{\textbf{The per-asset hidden Markov generator with jumps (JumpHMM) and its stylized facts for SPY.} Top: the generator's hidden
    states $1,\dots,N$ are ordered from the bottom to the top tail of the daily growth rate, and state $k$ emits
    $\mu_k + \sigma_k t_5$, a scaled Student-$t$ draw with five degrees of freedom. With probability $1-\epsilon$
    the chain moves by its fitted transition matrix; with probability $\epsilon$ a jump episode starts and lasts
    $K \sim \mathrm{Pois}(\lambda)$ forced steps, each drawn from the bottom tail set $\mathcal{S}_-$ or the top
    tail set $\mathcal{S}_+$~\citep{alswaidan2026variance}. Bottom: SPY's daily growth rates for 2014--2024 (gold)
    against $100$ generator paths (median line and $10$--$90\%$ band, blue) and a normal distribution with SPY's
    mean and standard deviation (dashed): the density, the generator's quantiles against SPY's, and the
    autocorrelation of the absolute growth rate $|g|$ out to $252$ trading days.}
    \label{fig:jumphmm}
\end{figure}

We compared daily growth rates of the JumpHMM market generator with those of real SPY over 2014--2024 and checked
the composed multi-asset paths against the real universe, and found that the generator reproduced fat tails but not
the persistence of volatility (Fig.~\ref{fig:jumphmm}; Table~\ref{tab:S-checks}). For the comparison with SPY we
used the market factor of the first $100$ typical paths by path number. Across these paths the median kurtosis of
the daily growth rate was $13.7$, against $16.8$ for SPY and $3$ for a normal distribution. In the tails the median
generator quantile lay at or beyond SPY's, at $-17.2$ against $-16.3\,\mathrm{yr}^{-1}$ for the $0.1\%$ quantile,
$-10.1$ against $-7.8$ for the $1\%$ quantile, $7.3$ against $6.4$ for the $99\%$ quantile, and $15.1$ against
$14.8$ for the $99.9\%$ quantile. Both series had small autocorrelation of $g$ at lag 1 ($-0.04$ for the generator
median, $-0.11$ for SPY). The autocorrelation of $|g|$ differed, since for the generator median it fell from $0.22$
at lag 1 to $0.08$ at lag 2, $0.03$ at lag 3, and $0.007$ at lag 21, whereas SPY's was $0.36$ at lag 1 and still
$0.13$ at lag 21, about one trading month.
Of the $22$ gated checks on the composed paths, each with a pass tolerance, $21$ passed. The median synthetic SPY
growth rate over 2014--2024 was $0.105\,\mathrm{yr}^{-1}$ across the $10{,}000$ paths, against
$0.106\,\mathrm{yr}^{-1}$ for real SPY. The median name's variance was $0.994$ of the variance of its own JumpHMM
draw, and $1.066$ of its real variance, a comparison reported without a tolerance. The mean residual correlation
within each of the 11 sectors lay within $0.05$ of the real value (for example $0.535$ against $0.542$ in Energy),
and the SD of the online $\beta$ was $0.985$ of the real one. The failing check was the cross-name SD of $\alpha$,
the intercept of each name's least-squares regression on the market over 2014--2024, which was
$0.089\,\mathrm{yr}^{-1}$ against $0.076\,\mathrm{yr}^{-1}$ for the real universe, outside the tolerance of
$0.01\,\mathrm{yr}^{-1}$.
Comparisons without a tolerance showed what the composition left out. The mean residual correlation across sectors
was $0.0002$ against $0.031$, because cross-sectional dependence entered only through the market and sector factors.
The median name's autocorrelation of $|g|$ was $0.10$ below the real one at lag 1 and $0.17$ below at lag 5, but
within $0.04$ of its own JumpHMM draw, so the shortfall came from the per-name draws and not from the composition.
The synthetic paths also offered the hedge fewer candidates, with a median of seven names per path whose online
$\beta$ reached zero or below in the 2025--2026 window, against $120$ in the real 2025--2026 window and a median of
five in real 2014--2024 windows of the same length (range $0$ to $53$). The synthetic paths thus resembled a typical
stretch of 2014--2024, and the real trading window held more hedge candidates than any earlier stretch of that
length.

We ran all selection and weighting rules on the $600$ typical futures, compared selection by $\gamma$ with random
picks of the same size, and placed each real result among the typical futures of its cell, and found that selection
by $\gamma$ beat random picks in most typical futures (Tables~\ref{tab:synthetic} and~\ref{tab:S-head-to-head};
Fig.~\ref{fig:S-percentile}). In the typical futures the median synthetic SPY ended at $1.185$ and the equal-weight
universe at $1.230$, and the $70$ hedge-off cell medians of the ten main selection rules ran from $1.097$ (random
whole universe, with LS) to $1.244$ (liquid then $\gamma$, with SG).
Of these $70$ cells, $45$ ended above the SPY median and seven above the equal-weight median. The equal-weight
universe ended above its own path's SPY in $77.5\%$ of typical futures, whereas on real prices it ended below SPY
($1.296$ against $1.317$).
With the hedge off, global $\gamma$ ended above $13$ random names from the whole universe in $62.3$--$71.2\%$ of
futures across the weighting rules, with a median difference in $W_T/W_0$ of $0.042$ to $0.101$, and $\gamma$ per
sector ended above random per sector in $60.0$--$73.7\%$. With the hedge on, the $13$-name share was
$63.2$--$71.5\%$.
The hand-picked 13 ended between $1.175$ and $1.242$, within the $1.142$--$1.244$ range of the four $\gamma$ rules
($\gamma$ per sector, global $\gamma$, correlation-aware $\gamma$, and liquid then $\gamma$), whereas on real prices
both hand-picked lists ended highest.
To place the real results, we used the seeds of the synthetic runs, $5660$ for the deterministic rules and the median
over $5660$--$5664$ for the random and bandit rules, rather than the $30$-seed medians, and took the percentile of
each real value among the $600$ typical futures of its cell. Real SPY sat at percentile $70$ of synthetic SPY and the
real equal-weight universe at $60.7$, so the real window was a better-than-median period for the market, and part of
any real margin over the typical futures reflected that period. The real results of the hand-picked 13 sat at
percentiles $79.2$ to $90.7$ of their cells, those of global $\gamma$ at $47.5$ to $81.7$, those of $\gamma$ per
sector at $39.8$ to $56.2$, and those of $13$ random names from the whole universe at $43.3$ to $64.3$. Of the $70$
hedge-off cells, $52$ sat above the median future and $19$ above SPY's percentile of $70$.
The hand-picked lists sat at or above percentile $74$ in all $14$ of their hedge-off cells, above the market's $70$,
whereas the four $\gamma$ rules sat at percentiles $39.8$ to $81.7$ and above $70$ in five of their $28$ cells
(Fig.~\ref{fig:S-percentile}). In the real window the hand-picked lists thus exceeded their typical futures by more
than the market did, and the $\gamma$ rules in most cells did not.

\begin{figure}[H]
    \centering
    \includegraphics[width=\textwidth]{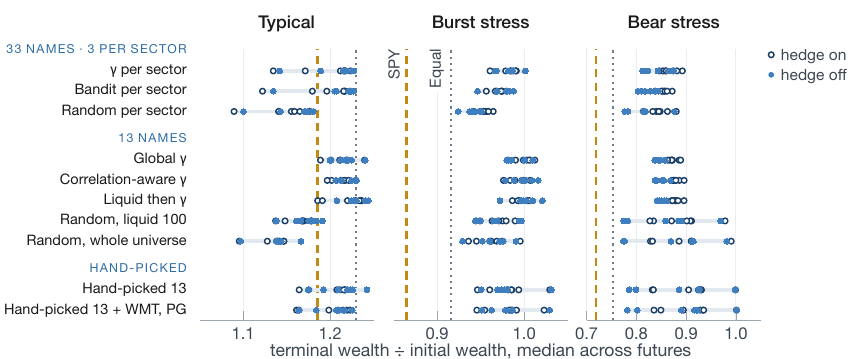}
    \caption{\textbf{Terminal wealth on synthetic futures of the trading window, by selection rule and weighting
    rule.} The three panels are $600$ typical futures and two stress sets drawn from $10{,}000$ futures: the $200$
    with the deepest SPY drawdowns that contain a jump burst (burst stress), and the $200$ deepest without one (bear
    stress). Rows, dots, bars and lines are as in Fig.~\ref{fig:realrun}; each dot is the median across the set's
    futures (after a median over seeds within a future) of terminal wealth divided by initial wealth, and each panel has its own horizontal axis. SPY
    buy-and-hold ended at $1.19$, $0.86$ and $0.72$, and the equal-weight universe at $1.23$, $0.92$ and $0.75$, in
    the typical, burst-stress and bear-stress sets.}
    \label{fig:synthetic}
\end{figure}

We repeated the comparison on the two stress sets of simulated sell-offs, and found that every selection and
weighting median ended above SPY and the equal-weight universe (Fig.~\ref{fig:synthetic};
Table~\ref{tab:synthetic}). The median synthetic SPY drawdown in the 2025--2026 window was $0.364$ in the burst
stress set and $0.406$ in the bear stress set, against $0.176$ in the typical futures, and SPY's median growth rate
was $-0.087$ and $-0.199\,\mathrm{yr}^{-1}$, against $0.102\,\mathrm{yr}^{-1}$.
In both stress sets all $140$ cell medians of the ten main selection rules ended above the SPY median and the
equal-weight median, and so did all $98$ cells per hedge setting of the $14$ synthetic selection rules. Among the
ten main selection rules, the lowest median in the burst stress set was $0.924$ (random per sector, MinVar, hedge off), against $0.865$ for SPY and $0.916$
for the equal-weight universe, and the highest was $1.031$ (hand-picked 13, LS, hedge off). In the bear stress set
the lowest median was $0.774$ (random whole universe, Tangent, hedge off), against $0.718$ for SPY and $0.753$ for
the equal-weight universe, and the highest was $1.001$ (hand-picked 13 with WMT and PG, LS, with either hedge
setting).
Each cell also ended above its own path's SPY in $69.0$--$88.5\%$ of burst futures and $72.5$--$100\%$ of bear
futures. Random picks also ended above SPY, so the margin over SPY did not come from selection by $\gamma$.
The equal-weight universe, which neither caps volatility nor re-plans, also ended above SPY, by $0.051$ in the burst
stress set and $0.035$ in the bear stress set, and the lowest cell medians exceeded it by $0.009$ and $0.021$, so part
of the margin over SPY came from the synthetic market and the rest from parts of the system that every cell shared.
With the hedge off, the cell medians of the mean portfolio $\beta$ were $0.322$ to $0.764$ in the burst stress set and
$0.229$ to $0.850$ in the bear stress set, all below one (Table~\ref{tab:S-synbeta}).
In bear futures selection by $\gamma$ lost its typical-future advantage for five of the seven weighting rules. With
the hedge off, global $\gamma$ ended above $13$ random names in $28.0$--$38.0\%$ of bear futures for CD, CES, CESw,
SG, and LS, and at $5$ names the same five weighting rules gave $38.5$--$45.0\%$. MinVar and Tangent, whose weights do
not use $\gamma$, reversed this pattern, with global $\gamma$ ahead in $70.0\%$ and $71.0\%$ of bear futures. In
burst futures global $\gamma$ ended above $13$ random names in $52.5$--$67.5\%$ of futures with the hedge off.
With the hedge on, the median $W_T/W_0$ ended higher in $65$ of the $70$ cells of the bear stress set (median of the
$70$ cell changes $+0.029$), and the lowest cell median rose from $0.774$ to $0.814$. In the burst stress set it
ended higher in $34$ of the $70$ cells (median change $-0.001$), and in the typical futures in $11$ of the $70$
(median change $-0.006$). The $\beta$ cap and hedge pool thus raised terminal wealth in prolonged declines and
lowered it in typical futures, as they did on real prices.

We repeated the calendar-only comparison on the $400$ stress futures, and found that the early re-plans again lowered
the worst drawdown more consistently than they changed terminal wealth, by about $0.01$ in the median cell as on real
prices, although the stress declines were deeper (Tables~\ref{tab:S-calendar-burst} and~\ref{tab:S-calendar-bear}).
The median maximum drawdown was smaller with the triggers in $65$ of the $70$ cells with the hedge off and in all $70$
with it on in the burst stress set, and in $65$ and $67$ of the $70$ in the bear stress set, by up to $0.022$, and at
most $0.007$ larger in the others. Median terminal wealth was higher with the triggers in $33$ and $28$ of the $70$
cells in the burst stress set and in $51$ and $56$ in the bear stress set, with differences from $-0.031$ to $0.029$
of initial wealth. Every calendar-only run re-planned $19$ times, every $21$ trading days, while the triggered cells
re-planned $25$ to $32$ times in the median. With the two triggers switched off, every cell median still ended above
the SPY median and the equal-weight median in both stress sets, the lowest at $0.930$ in the burst stress set and
$0.772$ in the bear stress set, so the margin over SPY did not depend on the early re-plans.

\section{Discussion}\label{sec:discussion}
The clearest result of the study concerned allocation. On real prices without the market hedge (when on, the hedge
capped the portfolio's beta at a level that fell as the market weakened and let up to five stocks with zero or
negative beta enter), the seven weighting rules applied to one selection of $33$ stocks (three in each of the $11$
sectors) ended within $0.15$ of one another, and all seven ended below SPY. The selections themselves ended further
apart, from $1.03$ to $1.70$ times initial wealth across all combinations, and the wide spread came from the
$13$-stock selections, made across the whole universe of $373$ stocks without a sector quota, and from the two lists the author picked in
advance ($13$ stocks, and the same $13$ plus two). One explanation, consistent with these runs but not tested by them,
is that the shared limits left the weighting rules little room. Every rule, including the three benchmark portfolios (minimum variance, tangency and $130/30$ long-short), faced the
same cap on each stock's weight and the same cap on portfolio volatility, so the rules differed mainly in how they
spread the invested wealth among the selected stocks within those caps. The results do not show that any one
selection rule is reliably better on real data. The hand-picked $13$-stock list ended highest, but it is one list chosen by one
person, and its result cannot be separated from chance. The three score-based $13$-stock rules, which took the $13$
highest scores (the expected growth rate from a single-index fit divided by a power of the stock's market beta that
depended on the market's recent direction) with no screen, a correlation screen or a liquidity screen, ended above
the two random $13$-stock controls, drawn from the whole universe or from its $100$ most liquid stocks, under every
weighting rule, while the three $33$-stock rules (ranked by score, chosen by a
bandit, or drawn at random within each sector) ended in overlapping ranges. The $13$-stock rules were added after the
backtests of the $33$-stock rules on the same trading window had been seen, so their comparison with random
selections is exploratory, and one historical path cannot show whether the difference would
persist~\cite{baileyEtAl2014, harveyLiu2015}.

The rebalancing controller re-planned, re-solving the allocation from the current holdings, on a schedule set partly
by its own forecast errors. Across combinations of selection and weighting rule, the median of each combination's
$30$ runs brought a fifth to three fifths of its re-plans forward of the $21$-day
horizon, through a band exit (the realized portfolio value leaving the range the forecast considered likely) or a
drawdown stop (wealth falling $8\%$ below its peak), and early re-plans were more frequent for stocks selected by
score than for random stocks, a difference the study did not examine. The forecast, a geometric Brownian motion,
represents neither the heavy tails nor the volatility clustering of daily growth rates~\cite{mandelbrot1963,
cont2001}, so band exits are expected to occur more often than they would if the forecast model were correct, and
the band served both as a trigger and as a running check on the forecast. A forecast model that represents
volatility clustering~\cite{engle1982, bollerslev1986} would widen the band after volatile days and narrow it after
calm ones, so its effect on the number of early re-plans cannot be signed in advance, and this was not tested.
Against re-planning every $21$ trading days alone, the early re-plans made the median worst drawdown smaller in most
combinations of selection and weighting rule, with the hedge on or off, and raised trading cost in all of them.
Median terminal wealth, by contrast, rose with the early re-plans in about half of the combinations with the hedge
off and in fewer than half with it on. The same pattern held in the two stress sets of synthetic futures, whose
declines were deeper than the real one: the early re-plans made the median worst drawdown smaller in nearly every
combination, by about as much as on real prices, and raised median terminal wealth in fewer than half of the
combinations in the set with jump bursts and in most of those in the set without (Tables~\ref{tab:S-calendar-burst} and~\ref{tab:S-calendar-bear}).
The triggers therefore limited losses more than they added growth, and the protection they gave did not grow with
the depth of the decline.

Results on the simulated markets describe how this system behaved in one generator's markets and are not forecasts.
In both stress sets, the futures with the deepest declines of simulated SPY, one drawn from futures whose market path
contained a jump episode (a sustained burst of large daily moves) and one from futures without, every combination of
selection and weighting rule ended above simulated SPY bought and held in the median, with the hedge on or off.
Random selections did so as well, so choosing stocks by score or by bandit was not needed for this result, and every
combination also did so with the band exit and the drawdown stop switched off, so the early re-plans were not needed
either (Tables~\ref{tab:S-calendar-burst} and~\ref{tab:S-calendar-bear}). The equal-weight universe,
which has neither limits nor triggers, also ended above simulated SPY in both stress sets, so part of the margin over
SPY belonged to the simulated market itself, and every combination's median exceeded the equal-weight median by as
little as $0.009$ in the set with jump episodes and $0.021$ in the set without. A likely cause of the margin over SPY is
that the portfolios carried less market exposure than the index. With the hedge off, the average market beta of the
portfolios was below one in every combination, at $0.32$ to $0.76$ in the stress set with jump episodes and $0.23$ to
$0.85$ in the set without (medians over futures), as on real prices ($0.39$ to $0.90$), and the market term of a
portfolio with a beta below one falls less than the market when the market factor falls, whatever stocks it holds.
The volatility cap, set at SPY's historical level, is one reason the beta stays below one, but the study did not run
the system without it.

The synthetic market also lacked two features of real sell-offs. Each stock's own simulated residual was generated
independently of the others, so co-movement among stocks came only from the simulated SPY and the sector factors.
Real residual co-movement tends to rise in sell-offs~\cite{longinSolnik2001, angChen2002}, and its absence may
understate how much a portfolio of a few stocks loses relative to the index. The generator's volatility clustering
also decayed within about a day, faster than in the data, so turbulent stretches in the synthetic futures were
shorter than in real markets, which may affect both the depth of drawdowns and how often the triggers fired. The
effect of the market hedge also changed sign across markets. The hedge lowered
market exposure in every combination on real prices but reduced terminal wealth in most of them, and it reduced
terminal wealth in most combinations of the $600$ typical futures, which were drawn at random from $10{,}000$
simulated paths. In the stress set without jump episodes it raised terminal wealth in most combinations, and in the
stress set with jump episodes it raised terminal wealth in about half of them. Since the hedge's beta cap tightens after
the market has started to fall, it pays only when the decline continues.

The comparison of score-based and random selections on synthetic futures needs a further qualification. The
generator gave each synthetic stock a constant intercept, its 2014--2024 single-index intercept shrunk toward the
average across stocks. In the synthetic market, a stock's intercept on the calibration years was therefore positively
correlated with its intercept on the trading window: across stocks, the two correlated at $0.10$, against $-0.12$ in the real market.
The score's trend term is a daily updated estimate of this intercept, so in the synthetic futures it carried
information about future growth by construction, and the share of typical futures in which score-based selections
beat random ones is partly a consequence of that design. The spread of estimated intercepts across stocks was also
wider in the synthetic market than in the real one ($0.089$ against $0.076$ per year), and a wider spread of
persistent intercepts gives a score-based selection more to find. The intercept spread was the one check the
generator failed among the $22$ that had a pass criterion, whereas the intercept correlation and the clustering
shortfall were reported without one. On real prices, the score-based $13$-stock rules still ended above random
selections even though intercepts did not persist from 2014--2024 into 2025--2026. The study does not separate
whether that came from the score's market term, from shorter-lived trends picked up by the daily updated estimate,
or from chance.

The study has further limits. The trading window covered $420$ trading days with one broad sell-off, in the spring
of 2025. The universe held only stocks that were members of the S\&P~500 index at the date of the constituent
snapshot and traded on every day from 2014 to 2026, so it carries survivorship bias~\cite{brownEtAl1992}. Neither
the portfolios nor SPY earned dividends, so every comparison is in prices alone. The simulated fills charged half the
bid-ask spread and a linear market-impact term, the turnover limit, a cap on the half-spread cost of the trades at each re-plan, used the half spread alone, and taxes were not
modeled. A paper-trading deployment of the system is running, but its results are not part of this study. Natural
extensions are a forecast model with heavy tails and volatility clustering for the controller, runs without the volatility cap to measure its share of the stress-set results, residual
co-movement across stocks and less persistent intercepts in the synthetic market, and evaluation of selection rules
over many historical windows.

\section{Conclusion}\label{sec:conclusion}
We built a system that chose stocks, set their share counts by maximizing a utility under limits on weight,
volatility and turnover, and re-planned the holdings with a model-predictive controller, and we tested it on $373$
United States equities over $420$ trading days of 2025--2026 and on simulated markets. On real prices, the choice of
stocks moved terminal wealth more than the choice among seven weighting rules: one $33$-stock selection ended between
$1.09$ and $1.23$ times initial wealth across the weighting rules, while different selections ended between $1.03$
and $1.70$, the highest being a $13$-stock list the author picked in advance. The share of the controller's re-plans that came
before the $21$-day horizon, because the portfolio value left the $95\%$ band of its forecast or fell $8\%$ below its
peak, ranged from a fifth to three fifths across combinations of
selection and weighting rule, highest for stocks selected by their score, the expected growth rate from a
single-index fit divided by a power of the stock's market beta that depended on the market's recent direction.
Compared with re-planning every $21$ trading days alone, these early re-plans made the median worst drawdown smaller
in $55$ of the $70$ combinations of the ten selection rules and seven weighting rules without the market hedge and
in $61$ with it, and raised trading cost in all of them, but raised median terminal wealth in $38$ and $26$ of the
$70$. Among $10{,}000$ simulated price paths (synthetic futures), we kept two sets of $200$ with the
deepest drawdowns of simulated SPY, one from futures whose SPY went through a burst of extreme daily moves and one
from futures without such a burst. In both sets, the median terminal wealth of every combination ended above those
of simulated SPY bought and held and of the equal-weight universe. Random selections did so as well, and so did
every combination with the band exit and the drawdown stop switched off, so the margin came from parts of the system
that every combination shared other than those two triggers, such as the volatility cap, whose separate effect the
study did not measure. In these sets the early re-plans again made the median worst drawdown smaller, in $65$ to
$70$ of the $70$ combinations for each set and hedge setting, without consistently raising median terminal wealth. In $600$ futures drawn at random from the $10{,}000$ (typical futures),
the $13$ stocks with the highest scores ended above $13$ random stocks in $62$ to $71\%$ of futures across the
weighting rules, but the generator kept each stock's 2014--2024 trend constant, and the score's trend term estimates
that trend, whereas in the real market those trends did not carry over into 2025--2026, so
part of this advantage came from the generator's design and may not carry over to real prices. A market hedge, which
capped the portfolio's beta more tightly as the market weakened and added stocks with zero or negative beta, reduced
terminal wealth on real prices and added to it in the set without bursts, whose declines were the most sustained. Deciding which selection rule is
better, and when such a hedge pays, will need more historical windows and generators whose trends persist no more
than real ones do.

\section*{Acknowledgments}
The author thanks Abdulrahman Alswaidan for the JumpHMM generator and its multi-asset composition, which supplied the
synthetic markets used here; Aaron Wheeler for helpful discussions; Evan Vu for an early version of the
Cobb--Douglas allocation and the bandit selection rule; and Alpaca Markets. Computations were run on a single desktop
computer (Apple Mac Studio, M3 Ultra, 32 cores); no shared-resource allocation was required.

\paragraph{Author contributions.}
J.D.V.\ designed the study, wrote the code, ran the experiments, and wrote the manuscript.

\paragraph{Competing interests.}
The author declares no competing interests.

\needspace{12\baselineskip}
\paragraph{Code and data availability.}
The study code, run scripts, result summaries, and the scripts that draw every figure from those summaries are
provided in the study repository (\url{https://github.com/varnerlab/INFORMS-Annual-Meeting-2026-Varnerlab}); the
per-asset generator library is released separately as \texttt{JumpHMM.jl}
(\url{https://github.com/varnerlab/JumpHMM.jl}).

\paragraph{Generative artificial intelligence use.}
During the preparation of this work, the author used Anthropic Claude Code and OpenAI Codex to assist with the
implementation and checking of experiment code, figure preparation, drafting and language editing, consistency
checks, and LaTeX formatting. The author reviewed and edited the assisted work and takes full responsibility for the
content of the publication.

\clearpage
\bibliographystyle{unsrtnat}
\bibliography{References_v1}

\clearpage
\section*{Tables}
\begin{table}[H]
\centering
\small
\caption{\textbf{Weighting rules built on the score.} Each utility sums over the names with $\gamma_i > 0$ in the
decision's name set $\mathcal{N}$; $n_i$ are shares, $S_i$ prices (USD/share), $W$ wealth (USD) and
$w_i = n_i S_i / W$ weights. The optimum under the budget alone pays the floors first and splits the rest among the
names with $\gamma_i > 0$: the proportionality is normalized so that those weights sum to
$\hat w = 1 - \sum_{\gamma_j \le 0} \varepsilon S_j / W$ for CD, CES and CESw, whose names with $\gamma_j \le 0$ sit at
$\varepsilon = 10^{-3}$ shares, and to $1 - 0.01\,\lvert\mathcal{N}\rvert$ for SG, whose baseline is
$b_i = 0.01\,W/S_i$ shares in every name. A hedge-pool name with $\gamma_i \le 0$ is instead free above its floor,
with no utility term. Both CES rules use $\eta = 0.5$ and $\rho = 1 - 1/\eta = -1$; at
$\eta = 1$ they reduce to CD. With the study's limits (Eq.~\ref{eq:allocation}) no closed form exists, and each
utility is maximized numerically as a convex cone program over the cone in the last column, with the volatility cap
as a second-order cone and the other limits as linear constraints. Derivations are in
Appendix~\ref{app:allocation}.}
\label{tab:utilities}
\begin{tabular}{@{}llllc@{}}
\toprule
Rule & Utility $U(n;\gamma)$ & Optimum under the budget alone & Names with $\gamma_i \le 0$ & Cone \\
\midrule
CD   & $\sum \gamma_i \ln n_i$
     & $w_i \propto \gamma_i$
     & $\varepsilon$ shares & exponential \\
CES  & $\big(\sum \gamma_i n_i^{\rho}\big)^{1/\rho}$
     & $n_i \propto (\gamma_i/S_i)^{\eta}$, so $w_i \propto \gamma_i^{\eta} S_i^{1-\eta}$
     & $\varepsilon$ shares & power \\
CESw & $\big(\sum \gamma_i w_i^{\rho}\big)^{1/\rho}$
     & $w_i \propto \gamma_i^{\eta}$
     & $\varepsilon$ shares & power \\
SG   & $\sum \gamma_i \ln (n_i - b_i)$
     & $w_i - 0.01 \propto \gamma_i$
     & $b_i$ shares & exponential \\
\bottomrule
\end{tabular}
\end{table}

\begin{table}[H]
\centering
\small
\setlength{\tabcolsep}{4pt}
\caption{\textbf{Real-data backtest, hedge off.} Terminal wealth ratio $W_T/W_0$ from 2 January 2025 to 4 September 2026, the median
of 30 runs (seeds 5660--5689) for each selection rule (row) and weighting rule (column). Range is the highest minus
the lowest median in a row; the last row gives the same quantity within each column, across the ten selection
rules. SPY buy-and-hold and the equal-weight universe (each of the 373 names bought at the first close and held) do
not depend on the weighting rule. CD, Cobb--Douglas; CES, constant elasticity of substitution (CES) utility on share
counts; CESw, CES utility on portfolio weights; SG, Stone--Geary; MinVar, minimum variance; Tangent, tangency
portfolio; LS, 130/30 long-short Cobb--Douglas. With the market hedge off, no cap on portfolio $\beta$ applied and no
hedge names were held. Hedge-on values are in Table~\ref{tab:S-realrun-on}.}
\label{tab:realrun}
\begin{tabular}{l*{8}{r}}
\toprule
Selection rule & CD & CES & CESw & SG & MinVar & Tangent & LS & Range \\
\midrule
\multicolumn{9}{l}{\emph{33 names, 3 per sector}} \\
$\gamma$ per sector & 1.230 & 1.219 & 1.221 & 1.234 & 1.086 & 1.227 & 1.110 & 0.148 \\
Bandit per sector & 1.226 & 1.234 & 1.223 & 1.239 & 1.162 & 1.229 & 1.140 & 0.099 \\
Random per sector & 1.205 & 1.183 & 1.201 & 1.183 & 1.128 & 1.158 & 1.127 & 0.078 \\
\addlinespace
\multicolumn{9}{l}{\emph{13 names}} \\
Global $\gamma$ & 1.520 & 1.384 & 1.460 & 1.441 & 1.355 & 1.341 & 1.248 & 0.272 \\
Correlation-aware $\gamma$ & 1.352 & 1.386 & 1.421 & 1.380 & 1.284 & 1.404 & 1.219 & 0.202 \\
Liquid then $\gamma$ & 1.333 & 1.310 & 1.379 & 1.350 & 1.238 & 1.361 & 1.305 & 0.141 \\
Random liquid 100 & 1.216 & 1.173 & 1.211 & 1.212 & 1.158 & 1.274 & 1.152 & 0.122 \\
Random whole universe & 1.098 & 1.132 & 1.138 & 1.112 & 1.104 & 1.160 & 1.031 & 0.129 \\
\addlinespace
\multicolumn{9}{l}{\emph{Hand-picked}} \\
Hand-picked 13 & 1.548 & 1.571 & 1.614 & 1.631 & 1.443 & 1.697 & 1.549 & 0.254 \\
Hand-picked 13 + WMT, PG & 1.616 & 1.646 & 1.637 & 1.639 & 1.362 & 1.642 & 1.622 & 0.284 \\
\midrule
SPY buy-and-hold & \multicolumn{7}{c}{1.317} & \\
Equal-weight universe & \multicolumn{7}{c}{1.296} & \\
\midrule
Range across selection rules & 0.518 & 0.514 & 0.499 & 0.527 & 0.357 & 0.539 & 0.591 & \\
\bottomrule
\end{tabular}
\end{table}

\begin{table}[H]
\centering
\small
\caption{\textbf{Synthetic futures, by selection rule.} Lowest and highest of the seven weighting-rule medians of
the terminal wealth ratio $W_T/W_0$ over the synthetic 2 January 2025 to 4 September 2026 window, for each selection
rule, in the 600 typical futures and the 200 futures of each stress set. Typical futures were drawn at random from
$10{,}000$ synthetic markets; the burst and bear stress sets held the 200 markets with the deepest SPY drawdown in the
window among those with a jump burst and among those without one. Each median is taken across futures; for
random and bandit rules, each future first contributes the median of five seeds. SPY buy-and-hold and the
equal-weight universe give the median across futures of each future's own SPY and equal-weight universe. 
With the market hedge on, a cap on portfolio $\beta$ tightened as the market weakened and up to five names with zero or negative $\beta$ could be held. Every selection and weighting cell is shown in
Fig.~\ref{fig:synthetic}.}
\label{tab:synthetic}
\begin{tabular}{lccc}
\toprule
Selection rule & Typical & Burst stress & Bear stress \\
\midrule
\multicolumn{4}{l}{\textbf{Hedge off}} \\
\multicolumn{4}{l}{\emph{33 names, 3 per sector}} \\
$\gamma$ per sector & 1.142--1.227 & 0.967--1.001 & 0.813--0.874 \\
Bandit per sector & 1.135--1.227 & 0.946--0.988 & 0.803--0.844 \\
Random per sector & 1.100--1.180 & 0.924--0.951 & 0.777--0.876 \\
\addlinespace
\multicolumn{4}{l}{\emph{13 names}} \\
Global $\gamma$ & 1.200--1.240 & 0.980--1.010 & 0.837--0.858 \\
Correlation-aware $\gamma$ & 1.206--1.230 & 0.978--1.016 & 0.837--0.868 \\
Liquid then $\gamma$ & 1.208--1.244 & 0.972--1.021 & 0.842--0.861 \\
Random liquid 100 & 1.138--1.191 & 0.944--0.997 & 0.774--0.969 \\
Random whole universe & 1.097--1.166 & 0.929--0.990 & 0.774--0.982 \\
\addlinespace
\multicolumn{4}{l}{\emph{Hand-picked}} \\
Hand-picked 13 & 1.175--1.242 & 0.951--1.031 & 0.785--0.998 \\
Hand-picked 13 + WMT, PG & 1.164--1.225 & 0.951--1.029 & 0.782--1.001 \\
\midrule
\multicolumn{4}{l}{\textbf{Hedge on}} \\
\multicolumn{4}{l}{\emph{33 names, 3 per sector}} \\
$\gamma$ per sector & 1.135--1.223 & 0.961--0.991 & 0.846--0.892 \\
Bandit per sector & 1.122--1.223 & 0.956--0.980 & 0.838--0.872 \\
Random per sector & 1.090--1.176 & 0.942--0.964 & 0.814--0.880 \\
\addlinespace
\multicolumn{4}{l}{\emph{13 names}} \\
Global $\gamma$ & 1.189--1.239 & 0.982--1.012 & 0.863--0.889 \\
Correlation-aware $\gamma$ & 1.197--1.229 & 0.976--1.007 & 0.872--0.896 \\
Liquid then $\gamma$ & 1.186--1.236 & 0.986--1.006 & 0.871--0.896 \\
Random liquid 100 & 1.137--1.178 & 0.945--0.990 & 0.827--0.978 \\
Random whole universe & 1.096--1.147 & 0.936--0.995 & 0.828--0.990 \\
\addlinespace
\multicolumn{4}{l}{\emph{Hand-picked}} \\
Hand-picked 13 & 1.164--1.217 & 0.946--1.029 & 0.832--0.999 \\
Hand-picked 13 + WMT, PG & 1.161--1.223 & 0.946--1.023 & 0.835--1.001 \\
\midrule
SPY buy-and-hold & 1.185 & 0.865 & 0.718 \\
Equal-weight universe & 1.230 & 0.916 & 0.753 \\
\bottomrule
\end{tabular}
\end{table}

\clearpage
\section*{Figures}

\clearpage
\appendix
\section*{Online Appendix}
\renewcommand{\thesection}{S\arabic{section}}
\renewcommand{\thesubsection}{S\arabic{section}.\arabic{subsection}}
\renewcommand{\thefigure}{S\arabic{figure}}
\renewcommand{\thetable}{S\arabic{table}}
\renewcommand{\theequation}{S\arabic{equation}}
\renewcommand{\thealgorithm}{S\arabic{algorithm}}
\renewcommand{\theHsection}{supp.\arabic{section}}
\renewcommand{\theHsubsection}{supp.\arabic{section}.\arabic{subsection}}
\renewcommand{\theHfigure}{supp.\arabic{figure}}
\renewcommand{\theHtable}{supp.\arabic{table}}
\renewcommand{\theHequation}{supp.\arabic{equation}}
\renewcommand{\theHalgorithm}{supp.\arabic{algorithm}}
\setcounter{section}{0}
\setcounter{figure}{0}
\setcounter{table}{0}
\setcounter{equation}{0}
\setcounter{algorithm}{0}

\section{Allocation details}\label{app:allocation}

\paragraph{Closed forms under the budget alone.}
We derive the optima of Table~\ref{tab:utilities} for a decision without hedge names. Let $\mathcal{P}$ be the names
with $\gamma_i > 0$ and $F = W - \sum_{j \notin \mathcal{P}} \varepsilon S_j$ (USD) the wealth left after the floors
of the other names. Each utility increases in every $n_i$ with $i \in \mathcal{P}$, so the budget
$\sum_{i \in \mathcal{P}} S_i n_i \le F$ binds. For CD, stationarity of the Lagrangian gives
$\gamma_i / n_i = \nu S_i$ with multiplier $\nu > 0$ (USD$^{-1}$), so $S_i n_i = \gamma_i / \nu$, and the budget sets
$\nu = \sum_{j \in \mathcal{P}} \gamma_j / F$, hence $w_i = (F/W)\, \gamma_i / \sum_{j \in \mathcal{P}} \gamma_j$.
For CES with $0 < \rho < 1$ ($\eta > 1$) the rule maximizes the concave sum $\sum \gamma_i n_i^{\rho}$; with
$\rho < 0$ ($0 < \eta < 1$) the outer power $1/\rho$ is negative, so the rule minimizes the convex sum
$\sum \gamma_i n_i^{\rho}$. In both cases stationarity gives $\rho\, \gamma_i\, n_i^{\rho - 1} = \nu S_i$ with $\nu$
of the sign of $\rho$, and because $1/(\rho - 1) = -\eta$,
\begin{equation}
n_i = C \left( \frac{\gamma_i}{S_i} \right)^{\eta}, \qquad
C = \frac{F}{\sum_{j \in \mathcal{P}} S_j (\gamma_j / S_j)^{\eta}}, \qquad
w_i = \frac{F}{W}\, \frac{\gamma_i^{\eta} S_i^{1-\eta}}{\sum_{j \in \mathcal{P}} \gamma_j^{\eta} S_j^{1-\eta}}.
\label{eq:S-ces-closed}
\end{equation}
As $\eta \to 0$ every name in $\mathcal{P}$ receives the same share count, a price-weighted book; as
$\eta \to \infty$ the budget goes to the name with the largest $\gamma_i / S_i$. The same steps with $w_i$ in place of
$n_i$ and the budget $\sum_{i \in \mathcal{P}} w_i \le F/W$ give CESw,
$w_i = (F/W)\, \gamma_i^{\eta} / \sum_{j \in \mathcal{P}} \gamma_j^{\eta}$, which tends to equal weights as
$\eta \to 0$ and to the name with the largest $\gamma_i$ as $\eta \to \infty$. For SG every name of $\mathcal{N}$ holds
$n_i \ge b_i$ and the names outside $\mathcal{P}$ hold exactly $b_i$. Writing $e_i = n_i - b_i$ for the excess shares,
SG maximizes $\sum_{i \in \mathcal{P}} \gamma_i \ln e_i$ subject to
$\sum_{i \in \mathcal{P}} S_i e_i \le W - \sum_{j \in \mathcal{N}} S_j b_j = W (1 - 0.01\,\lvert\mathcal{N}\rvert)$,
which is the CD problem on the excess shares, so
\begin{equation}
w_i = 0.01 + \left( 1 - 0.01\,\lvert\mathcal{N}\rvert \right) \frac{\gamma_i}{\sum_{j \in \mathcal{P}} \gamma_j},
\qquad i \in \mathcal{P}.
\label{eq:S-sg-closed}
\end{equation}
A feasible SG book therefore needs $\lvert\mathcal{N}\rvert \le 100$, and a positive excess needs
$\lvert\mathcal{N}\rvert < 100$. The unit tests checked Eq.~\ref{eq:S-sg-closed} against the solver on a three-name
problem to a relative tolerance of $10^{-4}$, and on the hand-picked list the CD, CES and CESw solver weights stayed
within $1.0 \times 10^{-5}$ of their closed forms.

\paragraph{Cone programs.}
We use the cone conventions of MathOptInterface. The exponential cone is the closure of
$\{(x, y, z) : y > 0,\ y\, e^{x/y} \le z\}$, so, with an auxiliary variable $t_i$ per name, $(t_i, 1, n_i)$ in the
cone is $t_i \le \ln n_i$. CD maximizes
$\sum_{i \in \mathcal{P}} \gamma_i t_i$ with $n_i \ge 10^{-8}$, and SG puts $(t_i, 1, n_i - b_i)$ in the cone. The
power cone with exponent $a \in (0, 1)$ is $\{(x, y, z) : x^{a} y^{1-a} \ge \lvert z \rvert,\ x, y \ge 0\}$. The CES
rules act on a scaled variable $u_i = n_i \sum_{j \in \mathcal{P}} S_j / W$ (CES), which equals one at equal share
counts that spend $W$, or $u_i = \lvert\mathcal{P}\rvert\, w_i$ (CESw), which equals one at equal weights; a common
positive scale multiplies $\sum \gamma_i u_i^{\rho}$ by a constant and leaves the optimum unchanged, and it keeps
$u_i$ near one at any wealth and basket size. For $\eta > 1$ the program puts $(u_i, 1, t_i)$ in the power cone with
exponent $\rho$, so $t_i \le u_i^{\rho}$, and maximizes $\sum \gamma_i t_i$. For $\eta < 1$ it puts $(t_i, u_i, 1)$ in
the power cone with exponent $\eta$, so $t_i^{\eta} u_i^{1-\eta} \ge 1$, which is $t_i \ge u_i^{\rho}$, and minimizes
$\sum \gamma_i t_i$. At $\eta = 1$ both CES rules call CD. The volatility cap puts
$(\sigma_{\mathrm{SPY}}, L^{\top} w)$ in the second-order cone. The turnover budget uses slack variables
$v_i \ge \lvert n_i - n_i^{\mathrm{prev}} \rvert$ with $\bar c\, \bar S \sum_i v_i \le B$, added only when the largest
possible share change could break the budget. A name held at its floor enters the limits as a constant, and a limit
row made only of constants is checked directly instead of being sent to the solver, because all-zero rows degraded
Clarabel's accuracy (a share error of $1.3 \times 10^{-5}$ with them against $2 \times 10^{-6}$ without, on a CD
test). LS uses two exponential-cone blocks, $(t_i^{+}, 1, n_i^{+})$ for the long names with $\gamma_i > 0$ and
$(t_i^{-}, 1, n_i^{-})$ for the short names, with $n_i^{\pm} \ge 10^{-8}$, and applies the limits to the signed
weights $w_i = (n_i^{+} - n_i^{-}) S_i / W$. We set Clarabel's duality-gap tolerances to $10^{-12}$ because the
utility is flat at its optimum, so the share error scales with the square root of the gap: on a three-name CD test the
solver's shares missed the closed form by $2.5 \times 10^{-5}$ at a gap tolerance of $10^{-9}$ and by
$6 \times 10^{-9}$ at $10^{-12}$. An exception thrown inside a solver, such as the domain error that Clarabel
$0.11.1$ raised in its power-cone barrier for some CES problems at $\eta \le 0.25$ and $\eta \ge 4$, counted as that
solver not accepting the problem, and the decision moved on to SCS.

\paragraph{The hedge stage.}
Let $\mathcal{H}_f$ be the free hedge names, the pool names with $\gamma_i \le 0$, and $n^{(1)}$ the first-stage
shares. The second stage holds every name with a utility term at $n_i^{(1)}$ as a fixed variable, so its share
change still counts toward the turnover budget, holds every other name outside $\mathcal{H}_f$ at its floor, and
solves
\begin{equation}
\min_{n_i \ge f_i,\ i \in \mathcal{H}_f} \ \sum_{i \in \mathcal{H}_f} w_i
\quad \text{subject to the limits of Eq.~\ref{eq:allocation} (and the 130/30 rows for LS)},
\label{eq:S-hedge}
\end{equation}
with $f_i$ the floor of hedge name $i$ ($\varepsilon$ for CD, CES and CESw, $b_i$ for SG and $10^{-8}$ for LS). The
first stage already subtracts $10^{-6} \sum_{i \in \mathcal{H}_f} w_i$ from the utility, but that cost is below the
solver's dual accuracy and alone would leave hedge holdings of several percent of wealth along the flat direction;
Eq.~\ref{eq:S-hedge} removes them. For SG with no positive score and a finite $\beta$ cap, the second stage runs
with no fixed names and finds the least hedge holding that lets the baselines meet every limit.

\paragraph{Minimum variance and tangency.}
Both benchmarks use long-only shares with no floors and the expected growth rates
$\mu_{g,i} = \alpha_i + \beta_i \tilde g_M$. They first solve
$s = \min\big(1, \max \sum_i w_i\big)$ over the budget, single-name, $\beta$ and turnover limits without the
volatility cap. MinVar then minimizes $w^{\top} \Sigma_g w$ over the same limits with
$\sum_i w_i \ge s - 10^{-7}$. Tangent, with excess growth $e = \mu_g - g_f \mathbf{1}$, first finds the best feasible
excess growth $\bar e = \max e^{\top} w$ over that set; when $\bar e \le 10^{-6}\, s \max_i \lvert e_i \rvert$ no
feasible mix has positive excess growth and the rule falls back to the MinVar portfolio. Otherwise it applies the
Charnes--Cooper change of variables~\cite{charnes1962} $y = \kappa n$, $v = \kappa w$ with $\kappa \ge 0$, and solves
\begin{equation}
\min_{y \ge 0,\ \kappa \ge 0} \ v^{\top} \Sigma_g\, v
\quad \text{subject to} \quad e^{\top} v = \bar e, \qquad \textstyle\sum_i v_i \ge (s - 10^{-7})\,\kappa,
\label{eq:S-tangent}
\end{equation}
with the right-hand side of every limit multiplied by $\kappa$, and sets $w_{\mathrm{mix}} = v / \kappa$. Fixing
$e^{\top} v$ at $\bar e$ instead of one leaves the optimum unchanged, because the Sharpe ratio does not depend on the
scale of $w$, and it keeps $\kappa = \bar e / e^{\top} w_{\mathrm{mix}} \ge 1$, so the division does not amplify solver
error. Each mix is then scaled by $c = \min\big(1, \sigma_{\mathrm{SPY}} / \sqrt{w_{\mathrm{mix}}^{\top} \Sigma_g\,
w_{\mathrm{mix}}}\big)$ with the rest in cash. Scaling down keeps the budget, the single-name caps and the $\beta$ cap,
whose bound is never negative; the turnover budget is checked again, and a scaled book that breaks it keeps the
previous holdings. An earlier version forced $\sum_i w_i$ to the largest investment the volatility cap allowed, which
left a single feasible portfolio whenever the cap bound, so MinVar and Tangent coincided; the two-fund form, a risky
mix scaled against cash~\cite{tobin1958}, avoids this.

\paragraph{The bandit selection rule.}
One sector's bandit at a pick on day $t$ runs as in Algorithm~\ref{alg:S-bandit}. The window starts $d$ run over
$\mathcal{D}_t = \{t - 272, \dots, t - 21\}$, the $252$ starts whose 21-day windows end on days $t - 251$ through $t$.
The random stream of the sector with rank $r$ (sectors in alphabetical order) is seeded with
$(\text{seed} \cdot 1{,}000{,}003 + \text{date}) \cdot 64 + r$, where date is the day count of the pick date, so a
pick depends only on the run seed, the date and the sector.

\begin{algorithm}[tp]
\caption{Bandit selection rule for one sector at a pick on day $t$.}
\label{alg:S-bandit}
\begin{algorithmic}[1]
  \REQUIRE The sector's names $j = 1, \dots, A$; excess growth rates $x_j(d)$ (yr$^{-1}$) for window starts
    $d \in \mathcal{D}_t$; a random stream seeded by the run seed, the date and the sector.
  \ENSURE The $3$ names with the highest mean reward.
  \IF{$A \le 3$}
    \RETURN all $A$ names
  \ENDIF
  \STATE $M \gets \min(\max(50A, 500), 5000)$; $\hat\mu_j \gets 0$ and $c_j \gets 0$ for every $j$; draw a random
    permutation $\pi$ of $1, \dots, A$ as the tie-break rank
  \FOR{$k = 1, \dots, M$}
    \STATE $\epsilon_k \gets 1$ if $k = 1$, else $\max\{0.05,\, \min[1,\, k^{-1/3} (A \ln k)^{1/3}]\}$
    \IF{$\mathrm{Uniform}(0,1) < \epsilon_k$ or no arm has been pulled}
      \STATE $a \gets$ an arm drawn uniformly from $1, \dots, A$ \hfill \COMMENT{explore}
    \ELSE
      \STATE $a \gets \arg\max_{j : c_j > 0} \hat\mu_j$, ties to the smallest $\pi_j$ \hfill \COMMENT{exploit}
    \ENDIF
    \STATE $d \gets$ a start drawn uniformly from $\mathcal{D}_t$; $c_a \gets c_a + 1$;
      $\hat\mu_a \gets \hat\mu_a + (x_a(d) - \hat\mu_a)/c_a$
  \ENDFOR
  \RETURN the first $3$ names ordered by: pulled before unpulled, then $\hat\mu_j$ descending, then $\pi_j$ ascending
\end{algorithmic}
\end{algorithm}

\paragraph{Parameters.}
The settings of the allocation, the score and the selection rules, with the calibrated values from 2014--2024, are
in Table~\ref{tab:S-allocation-params}.

\clearpage
\begin{table}[H]
\centering
\footnotesize
\setlength{\tabcolsep}{4pt}
\caption{\textbf{Settings of the allocation, the score and the selection rules.} Windows and spans are in trading
days. Calibrated values were fitted once on 2014-01-03 to 2024-12-31 and then held fixed.}
\label{tab:S-allocation-params}
\begin{tabular}{@{}lll@{}}
\toprule
Setting & Symbol & Value \\
\midrule
\multicolumn{3}{@{}l}{\emph{Score}} \\
Half-life of the weighted least-squares fit & $h$ ($\delta = 2^{-1/h}$) & $63$ days \\
Prior weight of the 2014--2024 fit & $N_0$ & $63$ observations \\
SPY growth mean; variance (prior, calibrated) & & $0.106$~yr$^{-1}$; $7.51$~yr$^{-2}$ \\
market growth average, EMA span & $\tilde g_M$ & $252$ days \\
market state, EMA spans & $\xi$ & $21$ and $63$ days \\
market state gain (calibrated, $99\%$ rule) & $G$ & $21.23$ \\
floor on $\lvert\beta_i\rvert$ in the score & & $0.1$ \\
\addlinespace
\multicolumn{3}{@{}l}{\emph{Limits}} \\
single-name cap: picks; other names & $\bar w_i$ & $\max(0.10, 1.5/K)$; $0.10$ \\
volatility cap (calibrated) & $\sigma_{\mathrm{SPY}}$ & $2.741$~yr$^{-1}$ ($17.3\%$ annualized) \\
residual covariance window; shrinkage & & $126$ days; $0$ \\
half spread; turnover budget & $\bar c$; $B$ & $5$~bps; $0.10\,W$ \\
sector cap & & none \\
\addlinespace
\multicolumn{3}{@{}l}{\emph{Hedge}} \\
$\beta$ cap & $\bar\beta$ & $\max(0, 1 - \xi_t)$ \\
pool size; down-day beta lookback & $H$ & $5$ names; $252$ days \\
\addlinespace
\multicolumn{3}{@{}l}{\emph{Weighting}} \\
CES elasticity & $\eta$; $\rho$ & $0.5$; $-1$ \\
Stone--Geary baseline & $b_i S_i$ & $0.01\,W$ \\
floor of names with $\gamma_i \le 0$ (CD, CES, CESw) & $\varepsilon$ & $10^{-3}$ shares \\
smallest traded target & & $0.001\,W$ \\
hedge tie-break cost & & $10^{-6}$ per unit weight \\
$130/30$ long; short; net weight & & $1.30$; $0.30$; $1.00$ \\
\addlinespace
\multicolumn{3}{@{}l}{\emph{Selection}} \\
names per sector (per-sector rules) & & $3$ ($33$ in all) \\
names, universe-wide rules (main comparison) & $K$ & $13$ \\
liquid shortlist; dollar-volume window & & $100$ names; $63$ days \\
correlation limit; window & & $0.7$; $252$ days \\
bandit lookback; reward window & & $252$ days; $21$ days \\
bandit exploration floor; pulls & $\epsilon$ & $0.05$; $50A$ within $[500, 5000]$ \\
pick interval; pick at band exit & & $21$ days; yes \\
seeds & & $5660$--$5689$ \\
\addlinespace
\multicolumn{3}{@{}l}{\emph{Solvers}} \\
modeling package; solvers & & JuMP 1.32.1; Clarabel 0.11.1, SCS 2.7.0 \\
Clarabel gap; feasibility tolerance; iterations & & $10^{-12}$; $10^{-9}$; $500$ \\
\bottomrule
\end{tabular}
\end{table}

\section{Controller and estimation details}\label{app:controller}

This appendix gives the preparation of the price data, the estimators behind the controller's forecast and the
details of the simulated fills. The settings of every real-data run are collected in
Table~\ref{tab:S-controller-params} and the experiment grids in Table~\ref{tab:S-grids}.

\subsection{Universe and price files}

The sector snapshot lists $503$ tickers. Of the $418$ tickers that appear in all three price files with a sector,
$373$ have a positive open and close on all $3{,}187$ trading days from 2014-01-03 to 2026-09-04: $15$ in
Communication Services, $39$ in Consumer Discretionary, $31$ in Consumer Staples, $16$ in Energy, $53$ in
Financials, $44$ in Health Care, $60$ in Industrials, $50$ in Information Technology, $17$ in Materials, $22$ in Real
Estate and $26$ in Utilities. Each price file is split-adjusted by the vendor as of its own creation date, so a
split between the creation dates of two files shows as a one-day jump at their boundary.
At each of the two boundaries we flagged a name when its overnight log price move exceeded $0.4$ in magnitude, the
move matched a standard split ratio to within a factor of $e^{0.1}$, its volume moved in the matching direction by
a factor within $3$ of that ratio, and the overnight log move of SPY was within $0.05$ in magnitude. We then divided
the earlier prices of a flagged name by the split ratio and multiplied its earlier volumes by it, which leaves the
growth rates within each file and the dollar volumes unchanged.
This changed seven names, ORLY (ratio $15$), NFLX ($10$), NOW ($5$) and FAST ($2$) at 2025-01-02, and KLAC ($10$),
MNST ($2$) and APH ($2$) at 2026-01-02. Two one-day close moves larger than $0.9$ in log magnitude remain, IR on
2020-03-02 and NI on 2015-07-02, both inside the calibration window, and we left them as given.

\subsection{Online single-index fit}

For each name $i$ the controller updates the intercept $\alpha_i$ ($\mathrm{yr}^{-1}$) and loading $\beta_i$
(dimensionless) of the single-index model $g_i(t) = \alpha_i + \beta_i\,g_M(t) + \varepsilon_i(t)$ once per
trading day by exponentially weighted least squares. With half-life $h$ (trading days) the decay factor is
$\delta = 2^{-1/h}$, and the weighted sums are updated with the growth rates into day $t$ as
\begin{equation}
  \begin{aligned}
    S_w &\leftarrow \delta S_w + 1, &
    S_x &\leftarrow \delta S_x + g_M(t), &
    S_y &\leftarrow \delta S_y + g_i(t),\\
    S_{xx} &\leftarrow \delta S_{xx} + g_M(t)^2, &
    S_{xy} &\leftarrow \delta S_{xy} + g_i(t)\,g_M(t), & &
  \end{aligned}
  \label{eq:ewls-update}
\end{equation}
after which
\begin{equation}
  \beta_i(t) \;=\; \frac{S_w S_{xy} - S_x S_y}{S_w S_{xx} - S_x^2},
  \qquad
  \alpha_i(t) \;=\; \frac{S_y - \beta_i(t)\,S_x}{S_w},
  \label{eq:ewls-solve}
\end{equation}
with the previous estimates kept on a day when the denominator is at most $10^{-12}$ in magnitude. The sums start
from a prior worth $N_0$ pseudo-observations. Let $\alpha_i^{\mathrm{cal}}$ and $\beta_i^{\mathrm{cal}}$ be the
ordinary least-squares fit of name $i$ on SPY over the $2{,}766$ daily growth rates of the calibration window, and
let $m_M$ ($\mathrm{yr}^{-1}$) and $v_M$ ($\mathrm{yr}^{-2}$) be the mean and population variance of SPY's daily
growth rate over the same days. The prior is
\begin{equation}
  \begin{aligned}
    &S_w = N_0, \qquad S_x = N_0\,m_M, \qquad
    S_y = N_0\big(\alpha_i^{\mathrm{cal}} + \beta_i^{\mathrm{cal}} m_M\big),\\
    &S_{xx} = N_0\big(v_M + m_M^2\big), \qquad
    S_{xy} = N_0\big(\alpha_i^{\mathrm{cal}} m_M + \beta_i^{\mathrm{cal}}(v_M + m_M^2)\big),
  \end{aligned}
  \label{eq:ewls-prior}
\end{equation}
the sums that $N_0$ observations would give if SPY's growth had the calibration mean and variance and name $i$
followed its calibration fit in the mean. Before the first update Eq.~\eqref{eq:ewls-solve} gives
$\alpha_i^{\mathrm{cal}}$ and $\beta_i^{\mathrm{cal}}$ exactly, and the weight of the prior halves every $h$ days.
We used $h = 63$ and $N_0 = 63$, with $m_M = 0.1061$~$\mathrm{yr}^{-1}$ and $v_M = 7.511$~$\mathrm{yr}^{-2}$, and
the first update used the growth into 2025-01-02. SPY's standard deviation $\sigma_M(t)$ follows the same decay,
\begin{equation}
  \begin{aligned}
    &Q_w \leftarrow \delta Q_w + 1, \qquad Q_x \leftarrow \delta Q_x + g_M(t), \qquad
    Q_{xx} \leftarrow \delta Q_{xx} + g_M(t)^2,\\
    &\sigma_M^2(t) \;=\; \frac{Q_{xx}}{Q_w} - \left(\frac{Q_x}{Q_w}\right)^2,
  \end{aligned}
  \label{eq:ewvar}
\end{equation}
from the prior $Q_w = N_0$, $Q_x = 0$ and $Q_{xx} = N_0\,\sigma_{\mathrm{SPY}}^2$, where $\sigma_{\mathrm{SPY}} = 2.741$~$\mathrm{yr}^{-1}$ is
the sample standard deviation of SPY's daily growth rate over the calibration window ($0.1727$ when multiplied by
$\sqrt{\Delta t}$ to an annual volatility, the value that also sets the volatility cap).

\subsection{Residual covariance and the choice of the shrinkage weight}

At a re-plan on day $t$ the residual matrix holds $\varepsilon_i(s) = g_i(s) - \alpha_i(t) - \beta_i(t)\,g_M(s)$
for the names of the allocation problem over the $L_\Omega$ daily growth rates that end on day $t$. We write the sample covariance of these residuals as
$\hat\Omega_\varepsilon = D R D$, with $D$ the diagonal matrix of the sample standard deviations of the residual
columns and $R$ their sample correlation matrix, and Eq.~\eqref{eq:mpc-cov} replaces $R$ by $(1-\lambda)R +
\lambda I$. A residual column with zero variance carries no correlation information and gets zero correlation with
the others. We chose $\lambda$ once, on the calibration window, by the accuracy of the residual part of
Eq.~\eqref{eq:mpc-cov} as a forecast of portfolio variance. The anchors were the last trading days of each month
with at least $252$ trading days of the window behind them and $21$ ahead. At each anchor we fitted $\alpha_i$
and $\beta_i$ by ordinary least squares on the trailing $252$ daily growth rates, because the online fit has no
warm-up period inside the calibration window, and formed $\hat\Omega_\varepsilon$ over the trailing
$L_\Omega = 126$ growth rates. We then drew $200$ portfolios per anchor, each of up to $33$ names taken in turn
from the $11$ sectors with weights $w$ drawn from a flat Dirichlet distribution. For each portfolio the realized
variance $r$ is the sample variance of $w^{\top}\varepsilon(s)$ over the $21$ days after the anchor and the
prediction is $p(\lambda) = w^{\top}\big[(1-\lambda)\hat\Omega_\varepsilon +
\lambda\,\mathrm{diag}(\hat\Omega_\varepsilon)\big]w$. The loss of $\lambda$ is the mean over portfolios and
anchors of
\begin{equation}
  \mathcal{L}(\lambda) \;=\; \frac{r}{p(\lambda)} - \ln\frac{r}{p(\lambda)} - 1,
  \label{eq:qlike}
\end{equation}
the QLIKE loss, which ranks variance forecasts consistently when the realized variance is a noisy proxy for the
true one~\cite{patton2011}. Over the grid $\lambda = 0, 0.1, \dots, 1$ we took the $\lambda$ with the smallest mean
loss, breaking ties toward the larger $\lambda$.
The mean loss was $0.353$ at $\lambda = 0$ and rose steadily to $0.619$ at $\lambda = 1$, so the study used
$\lambda = 0$, the full sample residual covariance.

\subsection{Forecast simulation and fills}

The simulation of Eq.~\eqref{eq:mpc-gbm} draws the daily log-price increments as $\Delta t\,\mu_g +
\sqrt{\Delta t}\,L_{\Delta} z$, with $z$ a vector of independent standard normal draws and $L_{\Delta}$ the lower
Cholesky factor of
$\Delta t\,\Sigma_g + 10^{-12} I$. The diagonal term keeps the factorization defined when $\Sigma_g$ is singular
and has no other effect at the scale of the forecast. The value of a 130/30 book on each path counts its short
positions with negative shares. The forecast standard deviation uses the $N_p - 1$ denominator.
The paths of a re-plan come from a Mersenne Twister generator seeded with $1{,}000{,}003$ times the run seed plus
the serial day number of the decision date. The picks of a random rule use the same integer for the pick date
multiplied by $64$, and each sector's bandit adds the rank of its sector to that product, so every run is
reproducible and does not depend on the thread that runs it. At a fill,
the common factor that scales the buys is the largest value in $[0, 1]$ that keeps cash nonnegative after the
buys and their costs, found by $60$ bisection steps, and the average daily volume $A_i$ in Eq.~\eqref{eq:cost} is
floored at one share. Names held before a re-plan that are not part of the new allocation problem are sold at the
next open.

\begin{table}[H]
\centering
\small
\caption{\textbf{Controller and backtest settings.} Settings of every real-data run, from the study
configuration, with the values fitted on the calibration window. Windows, horizons and intervals are in trading
days, and growth rates are in $\mathrm{yr}^{-1}$.}
\label{tab:S-controller-params}
\begin{tabular}{@{}lll@{}}
\toprule
\textbf{Setting} & \textbf{Symbol} & \textbf{Value} \\
\midrule
\multicolumn{3}{@{}l}{\textit{Windows and wealth}} \\[2pt]
Calibration window & & 2014-01-03 to 2024-12-31 ($2{,}767$ days) \\
Trading window & & 2025-01-02 to 2026-09-04 ($420$ days) \\
Length of a trading day & $\Delta t$ & $1/252$~yr \\
Initial wealth & $W_0$ & $100{,}000$~USD \\[4pt]
\multicolumn{3}{@{}l}{\textit{Estimation}} \\[2pt]
Half-life of the online single-index fit & $h$ & $63$ days \\
Prior weight of the online fit & $N_0$ & $63$ pseudo-observations \\
SPY growth mean, 2014--2024 & $m_M$ & $0.1061$~$\mathrm{yr}^{-1}$ \\
SPY growth variance, 2014--2024 & $v_M$ & $7.511$~$\mathrm{yr}^{-2}$ \\
SPY growth SD, 2014--2024 & $\sigma_{\mathrm{SPY}}$ & $2.741$~$\mathrm{yr}^{-1}$ ($0.1727$ annualized) \\
Span of SPY's average growth rate & & $252$ days \\
Residual covariance window & $L_\Omega$ & $126$ days \\
Residual shrinkage weight & $\lambda$ & $0$ (chosen from $0, 0.1, \dots, 1$) \\[4pt]
\multicolumn{3}{@{}l}{\textit{Forecast, triggers and picks}} \\[2pt]
Forecast horizon & $T$ & $21$ days \\
Forecast paths & $N_p$ & $1{,}000$ \\
Band half-width & $z$ & $1.96$ standard deviations \\
Drawdown limit & $D_{\max}$ & $0.08$, peak reset at each drawdown \\
Cash revisit & & $21$ days \\
Pick interval & & $21$ days, and at band exits \\
Smallest position traded & & $0.001\,W$ \\[4pt]
\multicolumn{3}{@{}l}{\textit{Costs and cash}} \\[2pt]
Half spread & $\bar c$ & $5$~bps \\
Impact coefficient & $\kappa$ & $0.001$ \\
Window of the average daily volume & & $63$ days \\
Borrow fee on short positions & $b$ & $0.0075$~$\mathrm{yr}^{-1}$ \\
Commission & & $0$ \\
Cash growth & $g_f$ & DTB3, Eq.~\eqref{eq:riskfree} \\[4pt]
\multicolumn{3}{@{}l}{\textit{Experiment grids}} \\[2pt]
Seeds & & $5660$ to $5689$ ($30$ per cell) \\
Names per sector (per-sector rules) & & $3$ ($33$ names) \\
Basket sizes (universe-wide rules) & $K$ & $5, 8, 13, 21, 34$, and $100$, $200$ \\
\bottomrule
\end{tabular}
\end{table}

\begin{table}[H]
\centering
\small
\caption{\textbf{Experiment grids of the real-data backtest.} Every grid runs the full trading window
(2025-01-02 to 2026-09-04) with the seven weighting rules (CD, CES, CESw, SG, MinVar, Tangent and LS) unless the
row says otherwise. Seeds run from $5660$, and a deterministic selection rule ran with seed $5660$ alone except in
the seed rerun. The seed rerun repeats $140$ runs of the Tier 1, universe-wide and hand-picked grids, and each repeated run counts
once in the total.}
\label{tab:S-grids}
\begin{tabular}{@{}p{2.2cm}p{5.4cm}lll@{}}
\toprule
\textbf{Grid} & \textbf{Selection rules and weighting rules} & \textbf{Hedge} & \textbf{Seeds} &
  \textbf{Runs} \\
\midrule
Tier 1 & $\gamma$, bandit and random per sector ($33$ names) & on, off & $30$ ($\gamma$: $1$) & $854$ \\
Fixed schedule & as Tier 1, picks on the $21$-day timer only & on & $30$ ($\gamma$: $1$) & $427$ \\
Universe-wide & global $\gamma$, liquid then $\gamma$, correlation-aware $\gamma$, random from the $100$ most
  liquid, random from the universe, at $K = 5, 8, 13, 21, 34$ & on, off & $30$ ($\gamma$ rules: $1$) &
  $4{,}410$ \\
Hand-picked & the $13$-name list and its $15$-name extension & on, off & $1$ & $28$ \\
Seed rerun & $\gamma$ per sector; global $\gamma$ at $K = 5, 8, 13, 21, 34$; correlation-aware $\gamma$ and liquid
  then $\gamma$ at $K = 13$; both hand-picked lists & on, off & $30$ & $4{,}200$ \\
Large book & global $\gamma$ and random from the universe at $K = 100, 200$ & on, off & $30$ & $1{,}680$ \\
Calendar only & the ten selection rules of the main comparison, with the band exit and the drawdown stop off &
  on, off & $30$ & $4{,}200$ \\
CES elasticity & as Tier 1; CES and CESw at $\eta = 0.1, 0.25, 0.5, 2, 4, 10$ with the single-name cap on and
  off, CD with the cap off (CES at $\eta = 0.1$ and the cap-on points at $\eta = 0.5$ left out) & on, off &
  $30$ ($\gamma$: $1$) & $2{,}562$ \\
Budget only & hand-picked $13$; CD, and CES and CESw at $\eta = 0.05, 0.1, 0.25, 0.5, 2, 4, 10, 20$, with no
  single-name or volatility cap & off & $1$ & $17$ \\
\midrule
Total & distinct runs & & & $18{,}238$ \\
\bottomrule
\end{tabular}
\end{table}

\clearpage

\section{Synthetic market details}\label{app:synthetic}

The settings and fitted values of the generator are collected in Table~\ref{tab:S-generator-params}, and those of
the $11$ sector models in Table~\ref{tab:S-sector-params}. This appendix gives the calculations behind the sector
loading scale, the variance of a composed name, the fits of the loading drift and the intercept pull, the choice of
centering, and the frequency of jump bursts in the trading window.

For a sector $s$ with $n_s$ names, the real sector factor is $f_s(t) = n_s^{-1}\sum_{i \in s} e_i(t)$, with variance $\sigma_s^2$, so the sum of
the real residual covariances over the ordered pairs $i \ne j$ of the sector is
$n_s^2 \sigma_s^2 - \sum_{i \in s} \mathrm{Var}(e_i)$. In the generator, the shared part of two names in the sector
is $\delta_i f_s(t)$ and $\delta_j f_s(t)$, whose covariance is $\delta_i \delta_j \sigma_s^2$ when the sector draw has
the real factor variance, and the sum of $\delta_i \delta_j$ over the ordered pairs is
$(\sum_i \delta_i)^2 - \sum_i \delta_i^2$. With $\delta_i^{\rm loo}$ the slope of $e_i$ on the mean residual of the other $n_s - 1$ names of the sector,
the scale $\kappa_s$ that makes the generator's mean pairwise covariance equal the real one is
\begin{equation}
  \kappa_s^2 = \frac{n_s^2 \sigma_s^2 - \sum_{i \in s} \mathrm{Var}(e_i)}
    {\left[\left(\sum_{i \in s} \delta_i^{\rm loo}\right)^2 - \sum_{i \in s} \left(\delta_i^{\rm loo}\right)^2\right]
    \sigma_s^2},
  \label{eq:kappa}
\end{equation}
and $\delta_i = \kappa_s \delta_i^{\rm loo}$. The leave-one-out mean contains the common factor plus the average
noise of the other $n_s - 1$ names, so the slope on it is attenuated as in an errors-in-variables regression. For
names with equal loadings and a common residual correlation $r_s$, the slope is attenuated by the factor
$r_s/(r_s + (1 - r_s)/(n_s - 1))$, and to first order
\begin{equation}
  \kappa_s^2 \approx 1 + \frac{1 - r_s}{(n_s - 1)\, r_s},
  \label{eq:kappa-approx}
\end{equation}
so the scale is largest in small sectors with weak residual co-movement. With $15$ names and $r_s = 0.094$,
Eq.~\eqref{eq:kappa-approx} gives $\kappa_s \approx 1.30$ for Communication Services against a fitted $1.34$, and with $17$ names and
$r_s = 0.166$ it gives $1.15$ for Materials, as fitted. In Financials ($53$ names, $r_s = 0.239$), Utilities ($26$,
$0.549$) and Energy ($16$, $0.542$), the fitted scales were $1.02$. With the scaled loadings, every sector's mean
residual correlation on the check paths came within $0.01$ of its real value (Table~\ref{tab:S-checks}).

The draws $g_m$, $f_s$ and $\varepsilon_i$ come from separate models with separate seeds, and the drift shocks
$z_i(t)$ are independent of all three, so the variance of Eq.~\eqref{eq:compose} on a day is approximately
\begin{equation}
  \mathrm{Var}\left(g_i(t)\right) \approx \theta_i^2 \left(\mathrm{E}\left[\beta_i(t)^2\right] v_m
    + \delta_i^2 v_s\right) + s_i^2 v_i,
  \label{eq:composed-variance}
\end{equation}
where $v_m$, $v_s$ and $v_i$ (yr$^{-2}$) are the variances of the market, sector and name draws. The omitted term,
the variance of the drift times the squared mean of $g_m$, is at most $\bar g_m^2/v_m$ times the first term, about
$0.105^2/7.51 \approx 0.0015$. The stationary
variance of the drift in Eq.~\eqref{eq:beta-drift} is $a^2 k_i^2/(1 - (1-a)^2) = k_i^2 a/(2-a)$, so
$\mathrm{E}[\beta_i(t)^2] = \hat\beta_i^2 + k_i^2 a/(2-a)$, the term in Eq.~\eqref{eq:rho-bar}. If each draw
variance were exactly its real counterpart, with $v_m = \sigma_m^2$, $v_s = \sigma_s^2$ and $v_i = \sigma_i^2$, the
rule $s_i^2 = 1 - \bar\rho_i$ would give $\mathrm{Var}(g_i) = \sigma_i^2$. Each JumpHMM instead drew a few percent
more variance than its training series (the median standard deviation ratio of the name models was $1.037$), and when
the market, sector and name models inflate variance by similar ratios, Eq.~\eqref{eq:composed-variance} lands near
$v_i$, the variance of the name's own draw.
The checks found median ratios of $0.994$ to $v_i$ and $1.066$ to $\sigma_i^2$.

The loading drift was fitted on a bank of $20$ synthetic paths, drawn once without drift and reused at every
candidate value. On each bank path, a name was composed with a trial $(h_\beta, k_i)$, the online estimator was
started from the OLS fit of that path's 2014--2024 and the market moments of that path, and the standard deviation and
lag-$252$ autocorrelation of $\hat\beta_i(t)$ over 2014--2024 were averaged over the bank. For each half-life, each
$k_i$ came from $30$ bisection steps on the stationary standard deviation $k_i\sqrt{a/(2-a)}$ between $0$ and $1.5$,
with $k_i = 0$ reserved for a name whose estimation noise alone reached its target, which happened for no name. The
median lag-$252$ autocorrelation then rose with the half-life, from $0.059$ at $63$ trading days to $0.178$ at $126$,
$0.307$ at $252$, $0.397$ at $504$, $0.447$ at $1{,}008$ and $0.481$ at $2{,}016$, and $126$ came closest to the real
$0.206$. The median drift scale at the chosen half-life was $4.97$, and the real median standard deviation of
$\hat\beta_i(t)$ that the fit reproduced was $0.208$.

The intercept pull used the same bank. OLS is linear in the growth path and $\alpha_i$ does not enter $\bar\rho_i$,
so on each bank path and window the synthetic OLS intercept of name $i$ is $\bar\alpha + \phi(\hat\alpha_i -
\bar\alpha) + r_i$, where $r_i$ is the intercept of the same name composed with a zero intercept. We computed $r_i$
once for every name, path and window (2014--2024, 2014--2019 and 2020--2024) and searched $\phi$ from $0$ to $1.5$
in steps of $0.01$, jointly with an intercept drift whose stationary standard deviation ran from $0$ to
$0.15$~yr$^{-1}$ in steps of $0.0025$~yr$^{-1}$. With $\mathrm{spread}$ the cross-name standard deviation of the
2014--2024 OLS intercepts and $\mathrm{corr}$ the cross-name correlation of the 2014--2019 and 2020--2024 intercepts,
each computed on a path and averaged over the bank, the loss was $((\mathrm{spread} - 0.0764)/0.01)^2 +
((\mathrm{corr} - 0.175)/0.05)^2$.
The joint minimum lay at $\phi = 0.6$ with no drift, where the bank gave a spread of $0.088$~yr$^{-1}$ and a
correlation of $0.150$ (loss $1.61$). The real pooled statistic, the mean over names of the squared change in OLS
intercept between 2014--2019 and 2020--2024 minus the mean over names of the sum of its two squared standard
errors, was
$-0.0043$~yr$^{-2}$, so the real changes were no larger than their noise. The synthetic spread contains the
estimation noise of each name's intercept, whose real root-mean-square standard error is $0.075$~yr$^{-1}$, so the
spread could not fall below about $0.075$~yr$^{-1}$ for any $\phi$, close to the real spread of $0.076$~yr$^{-1}$,
while the half-window correlation fell toward zero as $\phi$ shrank. A value of $\phi$ low enough to bring the spread
within its tolerance would therefore have pushed the correlation further below its target, and the fitted $\phi$
left the spread $0.013$~yr$^{-1}$ above the
real value on the check paths, the one failed check.

The composition centers each name's draw on the model's long-run mean $\bar\varepsilon_i$, not on the draw's own mean
over the path. The full calendar holds $2{,}766$ growth days in the calibration window, one growth day from
2024-12-31 to 2025-01-02, and $419$ growth days in the trading window. Centering on the path mean forces
$2{,}766\, m_{\mathcal{C}} + \tilde\varepsilon_i(2{,}767) + 419\, m_{\mathcal{W}} = 0$, where $m_{\mathcal{C}}$ and
$m_{\mathcal{W}}$ are the mean draws in the two windows, so $m_{\mathcal{W}} \approx -6.6\, m_{\mathcal{C}}$. A name
whose own draw raised its calibration intercept by chance would then have its trading-window intercept lowered by
about $6.6$ times that amount. On the bank, the noise parts of the two intercepts correlated $-0.48$ across names under
path centering (from $-0.71$ to $-0.16$ across paths) and $0.003$ when centered on the model's mean. Leaving the draws
uncentered is not an alternative, since the long-run mean of a name model is the name's own mean growth (median
$0.087$~yr$^{-1}$), which would act as a second, persistent intercept and in the same diagnostic made the noise parts
of the intercepts of 2014--2019 and 2020--2024 correlate $0.20$.

A jump episode of the market model overlaps a trading window of $L = 419$ growth days if it starts inside the window
or starts before it and is still under way on its first day. With $\epsilon$ small, the expected number of such
episodes is $\epsilon L$ for the first kind plus $\epsilon\, \mathrm{E}[K] = \epsilon\lambda$ for the second, where
$K$ is the episode length, so the probability that a window contains a burst is approximately
\begin{equation}
  1 - \exp\left(-\epsilon (L + \lambda)\right) = 1 - \exp\left(-10^{-4} (419 + 118.9)\right) = 0.052,
  \label{eq:burst-prob}
\end{equation}
against $523$ of $10{,}000$ windows in the simulation. Over the whole calendar of $3{,}186$ growth days, with no
episode under way before the first day, $1 - \exp(-3{,}186\,\epsilon)$ gives a probability of $0.27$ that a path contains at least one episode, most of which fall in the
calibration window.

Every random stream is seeded from the path number $p$. The market draw uses seed $p$, sector $s$ uses
$10^{12} + 10^4 p + s$, name $j$ uses $2 \times 10^{12} + 10^4 p + j$, and the drift shocks of all names come from
one stream seeded with $3 \times 10^{12} + p$. The long-run means use fixed seeds above $4 \times 10^{12}$, and the
tuning of name $j$ and sector $s$ used seeds $5660 + j$ and $7000 + s$. JumpHMM.jl reseeds the generator at each
simulation call, so the draws do not depend on the number of threads, and building one path twice gave identical
market data (Table~\ref{tab:S-checks}).

\begin{table}[H]
\centering
\caption{\textbf{Synthetic generator settings and fitted values.} JumpHMM settings shared by every model, the tuning
grids, the tuned jump parameters, and the composition parameters of Eq.~\eqref{eq:compose}. Ranges across names are
given as median (interquartile range) or as the smallest and largest value. The fit targets are real 2014--2024
statistics, and the fitted values are averages over the bank of $20$ synthetic paths used for fitting. Growth rates
and intercepts are in yr$^{-1}$, variances in yr$^{-2}$, and half-lives and episode lengths in trading days.}
\label{tab:S-generator-params}
\small
\begin{tabular}{@{}p{5.6cm} l p{4.8cm}@{}}
\toprule
\textbf{Quantity} & \textbf{Symbol} & \textbf{Value} \\
\midrule
\multicolumn{3}{@{}l}{\textit{JumpHMM settings, every model}} \\[2pt]
Number of states & & $50$ \\
Student-$t$ degrees of freedom & $\nu$ & $5$ \\
Probability that a forced step targets the negative tail & $p_{\rm neg}$ & $0.52$ \\
States per tail set & $N_{\rm tail}$ & $5$ \\
Risk-free rate; time step & $r_f$; $\Delta t$ & $0$; $1/252$~yr \\
Tuning objective & & ACF lags $1$ to $25$, kurtosis weight $0.20$ \\[4pt]
\multicolumn{3}{@{}l}{\textit{Tuning grids}} \\[2pt]
Sector and name models, jump probability & $\epsilon$ & $10$ values from $10^{-4}$ to $2.5 \times 10^{-2}$ \\
Sector and name models, mean episode length & $\lambda$ & $10$ values from $10$ to $120$ \\
Sector and name models, paths per grid point & & $100$ \\
Market model (fitted in earlier work) & $\epsilon$, $\lambda$ & $15 \times 15$ grid, $\epsilon$ to $0.03$, $\lambda$
  from $5$ to $150$, $200$ paths \\[4pt]
\multicolumn{3}{@{}l}{\textit{Tuned jump parameters}} \\[2pt]
Market model & $\epsilon$, $\lambda$ & $10^{-4}$, $118.9$ \\
Sector models ($11$) & $\epsilon$, $\lambda$ & $10^{-4}$ for all; $10$ to $120$ (Table~\ref{tab:S-sector-params}) \\
Name models ($373$), jump probability & $\epsilon$ & $10^{-4}$ for $370$; $2.87 \times 10^{-3}$ for $2$;
  $1.12 \times 10^{-2}$ for $1$ \\
Name models, mean episode length & $\lambda$ & $58.9$ ($34.4$ to $83.3$); $10$ to $120$ \\
Name models, draw SD divided by real SD (gate $0.5$ to $2$) & & $1.037$; $1.006$ to $1.251$ \\
Name models, long-run mean of the draw & $\bar\varepsilon_i$ & median $0.087$ \\[4pt]
\multicolumn{3}{@{}l}{\textit{Composition}} \\[2pt]
Real SPY growth variance, 2014--2024 & $\sigma_m^2$ & $7.51$ (SD $2.74$) \\
Intercept pull toward the universe mean & $\phi$ & $0.6$ \\
Universe mean of the real intercepts & $\bar\alpha$ & $-0.014$ \\
Cross-name SD of the synthetic intercepts & $\alpha_i$ & $0.046$ \\
Intercept drift & & none (fitted size zero) \\
Half-life of the loading drift & $h_\beta$ & $126$ \\
Drift scale & $k_i$ & $4.97$ ($3.74$ to $6.45$) \\
Stationary SD of the loading drift & $k_i\sqrt{a/(2-a)}$ & $0.26$ ($0.20$ to $0.34$); $0.03$ to $0.77$ \\
Real market loading, 2014--2024 & $\hat\beta_i$ & $0.98$; $0.33$ to $1.74$ \\
Sector loading & $\delta_i$ & $0.94$; $-0.34$ to $2.70$ \\
Sector scale of the leave-one-out loadings & $\kappa_s$ & $1.02$ to $1.34$ \\
Floor on the own draw's variance share & $s_{\min}^2$ & $0.10$ \\
Real $R^2$ above which $R^2$ is kept & & $0.80$ (largest real $R^2$: $0.645$) \\
Factor share of variance & $\bar\rho_i$ & $0.51$; $0.097$ to $0.918$ \\
Names scaled by $\theta_i < 1$ & & $2$ ($0.54\%$ of name-days) \\[4pt]
\multicolumn{3}{@{}l}{\textit{Fit targets (real) and fitted values (bank)}} \\[2pt]
Cross-name SD of OLS intercepts, 2014--2024 & & $0.0764$; $0.0881$ \\
Correlation of OLS intercepts, 2014--2019 with 2020--2024 & & $0.175$; $0.150$ \\
Median lag-$252$ autocorrelation of $\hat\beta_i(t)$ & & $0.206$; $0.178$ \\
Median SD of $\hat\beta_i(t)$ & & $0.208$; matched per name \\
Pooled change in intercept between halves, net of noise & & $-0.0043$; $-0.0002$~yr$^{-2}$ \\
\bottomrule
\end{tabular}
\end{table}

\begin{table}[H]
\centering
\caption{\textbf{Sector factors and their JumpHMMs.} For each sector, the number of names, the standard deviation of
the real sector factor $f_s$ over 2014--2024 ($\sigma_s$, yr$^{-1}$), the scale $\kappa_s$ applied to the leave-one-out
loadings, the median sector loading $\delta_i$ after scaling, the tuned mean episode length $\lambda$ of the sector
JumpHMM in trading days (every sector model has $\epsilon = 10^{-4}$), and the long-run mean of its draw (yr$^{-1}$),
which enters the composition uncentered.}
\label{tab:S-sector-params}
\small
\begin{tabular}{@{}lrrrrrr@{}}
\toprule
\textbf{Sector} & \textbf{Names} & \textbf{$\sigma_s$} & \textbf{$\kappa_s$} & \textbf{Median $\delta_i$} &
  \textbf{$\lambda$} & \textbf{Mean} \\
\midrule
Communication Services & $15$ & $1.500$ & $1.337$ & $0.784$ & $71.1$ & $-0.0007$ \\
Consumer Discretionary & $39$ & $1.719$ & $1.084$ & $0.834$ & $95.6$ & $0.0010$ \\
Consumer Staples & $31$ & $1.576$ & $1.072$ & $0.941$ & $95.6$ & $0.0013$ \\
Energy & $16$ & $4.215$ & $1.024$ & $0.968$ & $10.0$ & $0.0048$ \\
Financials & $53$ & $1.724$ & $1.020$ & $1.047$ & $107.8$ & $-0.0067$ \\
Health Care & $44$ & $1.400$ & $1.085$ & $0.911$ & $46.7$ & $0.0003$ \\
Industrials & $60$ & $1.350$ & $1.056$ & $1.002$ & $107.8$ & $-0.0010$ \\
Information Technology & $50$ & $1.471$ & $1.069$ & $0.813$ & $71.1$ & $-0.0005$ \\
Materials & $17$ & $2.004$ & $1.151$ & $0.770$ & $95.6$ & $-0.0027$ \\
Real Estate & $22$ & $2.293$ & $1.040$ & $0.952$ & $71.1$ & $-0.0028$ \\
Utilities & $26$ & $2.460$ & $1.021$ & $1.014$ & $120.0$ & $-0.0014$ \\
\bottomrule
\end{tabular}
\end{table}

\begin{table}[H]
\centering
\caption{\textbf{Generator checks against the real market.} A gated check passes when the synthetic value lies within
the tolerance of the reference. References are real values, except for V3 and V4, which compare each composed name
with its own JumpHMM draw, and V7, which builds one path twice. Each of the first $50$ typical paths was recalibrated
on its synthetic 2014--2024 data. A statistic with one value per path is the median over these paths, and one with
a value per name is the median over names of each name's path median. V1 uses all
$10{,}000$ market paths. Statistics cover 2014--2024 unless a row names the trading window (2025-01-02 to
2026-09-04). Twenty-one of the $22$ gated checks passed, and the test suite (V8) passed
separately. Growth rates and intercepts are in yr$^{-1}$. ACF: autocorrelation of absolute growth
rates. Residual correlation: mean pairwise correlation of single-index-model residuals. $\hat\beta_i$: the method's
online estimate of the market loading.}
\label{tab:S-checks}
\footnotesize
\begin{tabular}{@{}l p{6.3cm} r r l l@{}}
\toprule
\textbf{Check} & \textbf{Statistic} & \textbf{Synthetic} & \textbf{Reference} & \textbf{Tolerance} & \textbf{Pass} \\
\midrule
\multicolumn{6}{@{}l}{\textit{Gated checks}} \\
V1 & SPY growth rate & $0.105$ & $0.106$ & $\pm 0.02$ & yes \\
V2 & Equal-weight growth rate, daily rebalancing minus buy and hold & $-0.023$ & $-0.020$ & $\pm 0.02$ & yes \\
V3 & Composed variance divided by the variance of the own draw & $0.994$ & $1$ & $[0.95, 1.05]$ & yes \\
V3 & Absolute gap between the OLS $\beta$ and the mean of $\beta_i(t)$ & $0.021$ & $0$ & $\le 0.05$ & yes \\
V4 & ACF at lag $1$, composed minus own draw & $-0.038$ & $0$ & $\pm 0.05$ & yes \\
V4 & ACF at lag $5$, composed minus own draw & $0.004$ & $0$ & $\pm 0.05$ & yes \\
V4 & ACF at lag $20$, composed minus own draw & $0.006$ & $0$ & $\pm 0.05$ & yes \\
V5 & Residual correlation, Communication Services & $0.091$ & $0.094$ & $\pm 0.05$ & yes \\
V5 & Residual correlation, Consumer Discretionary & $0.108$ & $0.112$ & $\pm 0.05$ & yes \\
V5 & Residual correlation, Consumer Staples & $0.201$ & $0.204$ & $\pm 0.05$ & yes \\
V5 & Residual correlation, Energy & $0.535$ & $0.542$ & $\pm 0.05$ & yes \\
V5 & Residual correlation, Financials & $0.234$ & $0.239$ & $\pm 0.05$ & yes \\
V5 & Residual correlation, Health Care & $0.121$ & $0.128$ & $\pm 0.05$ & yes \\
V5 & Residual correlation, Industrials & $0.134$ & $0.135$ & $\pm 0.05$ & yes \\
V5 & Residual correlation, Information Technology & $0.098$ & $0.102$ & $\pm 0.05$ & yes \\
V5 & Residual correlation, Materials & $0.159$ & $0.166$ & $\pm 0.05$ & yes \\
V5 & Residual correlation, Real Estate & $0.361$ & $0.366$ & $\pm 0.05$ & yes \\
V5 & Residual correlation, Utilities & $0.540$ & $0.549$ & $\pm 0.05$ & yes \\
V6 & Standard deviation of $\hat\beta_i$ divided by the real one & $0.985$ & $1$ & $[0.9, 1.1]$ & yes \\
V7 & Market-data fields that differ when one path is built twice & $0$ & $0$ & $= 0$ & yes \\
V9 & Cross-name standard deviation of the OLS intercepts & $0.089$ & $0.076$ & $\pm 0.01$ & no \\
V9 & Correlation of OLS intercepts, 2014--2019 with 2020--2024 & $0.134$ & $0.175$ & $\pm 0.05$ & yes \\
\midrule
\multicolumn{6}{@{}l}{\textit{Reported, not gated}} \\
V3 & Composed variance divided by the real variance & $1.066$ & $1$ & & \\
V4 & ACF at lag $1$, synthetic minus real & $-0.097$ & $0$ & & \\
V4 & ACF at lag $5$, synthetic minus real & $-0.171$ & $0$ & & \\
V4 & ACF at lag $20$, synthetic minus real & $-0.097$ & $0$ & & \\
V5 & Residual correlation across sectors & $0.0002$ & $0.031$ & & \\
V6 & Lag-$252$ autocorrelation of $\hat\beta_i$ & $0.188$ & $0.206$ & & \\
V6 & Share of paths with a name whose $\hat\beta_i \le 0$ on some day of the trading window (reference: real
  2025--2026, $1$ = yes) & $1$ & $1$ & & \\
V6 & Names with $\hat\beta_i \le 0$ on some day of the trading window (synthetic range $2$ to $11$ over paths;
  reference: real 2025--2026) & $7$ & $120$ & & \\
V6 & As above, reference: median over $112$ real 2014--2024 windows of the same length, ending every $21$ trading
  days up to 2024-12-31 (range $0$ to $53$) & $7$ & $5$ & & \\
V6 & As above, reference: the real window of the same length ending 2024-12-31 & $7$ & $53$ & & \\
V9 & Correlation of OLS intercepts, 2014--2024 with the trading window (reference: real 2025--2026) & $0.100$ &
  $-0.121$ & & \\
\bottomrule
\end{tabular}
\end{table}

\clearpage
\section{Additional results}\label{app:results}

\begin{figure}[H]
    \centering
    \includegraphics[width=\textwidth]{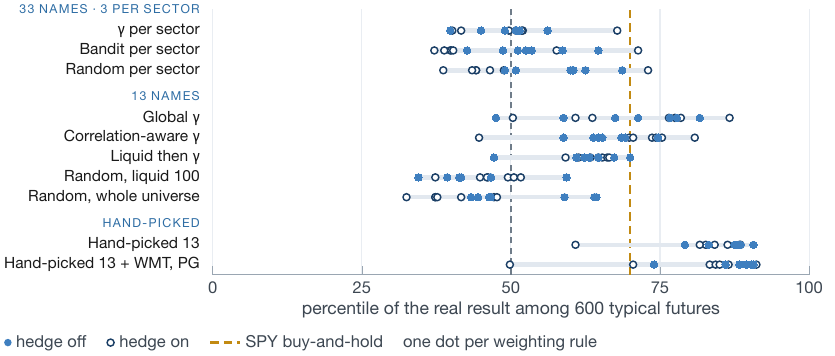}
    \caption{\textbf{Real-data results as percentiles of the typical synthetic futures.} For each selection rule and
    weighting rule, the percentile of the real terminal wealth ratio among the same cell's $600$ typical futures
    Filled dots have the market hedge off and open dots have it on (a $\beta$ cap that tightened as the market
    weakened, plus up to five names with $\beta \le 0$); the pale bar spans each row's lowest to highest dot. At $50$ (dashed) a real
    result equals the median future; the gold line marks real SPY, at the $70$th percentile of synthetic SPY. Real
    results use the seeds of the synthetic runs: the median over seeds $5660$--$5664$ for the random and bandit
    rules, and seed $5660$ for the rest.}
    \label{fig:S-percentile}
\end{figure}

\begin{figure}[H]
    \centering
    \includegraphics[width=0.9\textwidth]{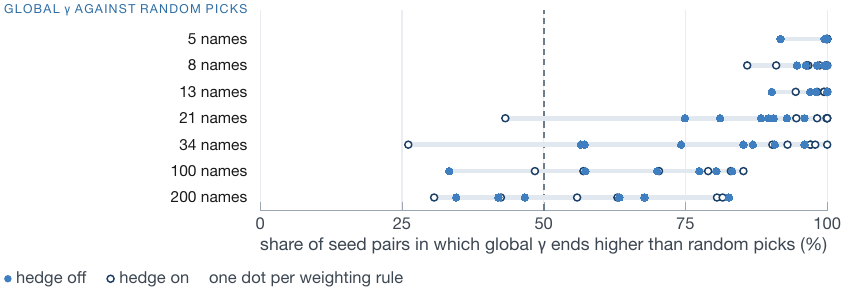}
    \caption{\textbf{Global $\gamma$ against random picks from the whole universe, from $5$ to $200$ names, on real
    prices.} For each basket size and weighting rule, the share of the $900$ pairs of one global $\gamma$ run and one
    random run ($30$ runs each) in which global $\gamma$ ended with the higher terminal wealth ratio (market hedge
    off filled, on open). At $50\%$ (dashed) the two rules end higher equally often. Stone--Geary is left out at
    $100$ and $200$ names, where its baseline holding of $1\%$ of wealth per name uses the whole budget or twice the
    budget.}
    \label{fig:S-size}
\end{figure}

The real-data backtest with the market hedge on, the market exposure of the portfolios on real prices and on the
stress sets, the share of re-plans that the triggers brought forward and the trading cost they added, the comparison
of the re-plan triggers with calendar-only re-planning on real prices and on the two stress sets, the paired
comparisons of selection rules on synthetic futures, the percentile of each real result among the typical synthetic
futures, and the comparison of global $\gamma$ with random picks at basket sizes from $5$ to $200$ names are reported
here (Tables~\ref{tab:S-realrun-on}, \ref{tab:S-exposure}, \ref{tab:S-replans}, \ref{tab:S-calendar},
\ref{tab:S-calendar-burst}, \ref{tab:S-calendar-bear}, \ref{tab:S-head-to-head} and~\ref{tab:S-synbeta};
Figs.~\ref{fig:S-percentile} and~\ref{fig:S-size}).

\begin{table}[H]
\centering
\small
\setlength{\tabcolsep}{4pt}
\caption{\textbf{Real-data backtest, hedge on.} Terminal wealth ratio $W_T/W_0$ from 2 January 2025 to 4 September 2026, the median
of 30 runs (seeds 5660--5689) for each selection rule (row) and weighting rule (column). Range is the highest minus
the lowest median in a row; the last row gives the same quantity within each column, across the ten selection
rules. SPY buy-and-hold and the equal-weight universe (each of the 373 names bought at the first close and held) do
not depend on the weighting rule. CD, Cobb--Douglas; CES, constant elasticity of substitution (CES) utility on share
counts; CESw, CES utility on portfolio weights; SG, Stone--Geary; MinVar, minimum variance; Tangent, tangency
portfolio; LS, 130/30 long-short Cobb--Douglas. With the hedge on, every allocation also obeyed the $\beta$ cap and could hold names from the hedge pool. Hedge-off values are in Table~\ref{tab:realrun}.}
\label{tab:S-realrun-on}
\begin{tabular}{l*{8}{r}}
\toprule
Selection rule & CD & CES & CESw & SG & MinVar & Tangent & LS & Range \\
\midrule
\multicolumn{9}{l}{\emph{33 names, 3 per sector}} \\
$\gamma$ per sector & 1.212 & 1.232 & 1.223 & 1.218 & 1.073 & 1.285 & 1.130 & 0.212 \\
Bandit per sector & 1.125 & 1.140 & 1.167 & 1.194 & 1.166 & 1.214 & 1.120 & 0.094 \\
Random per sector & 1.108 & 1.087 & 1.114 & 1.155 & 1.155 & 1.092 & 1.092 & 0.068 \\
\addlinespace
\multicolumn{9}{l}{\emph{13 names}} \\
Global $\gamma$ & 1.503 & 1.305 & 1.464 & 1.461 & 1.203 & 1.589 & 1.283 & 0.386 \\
Correlation-aware $\gamma$ & 1.409 & 1.429 & 1.479 & 1.447 & 1.213 & 1.366 & 1.292 & 0.266 \\
Liquid then $\gamma$ & 1.322 & 1.285 & 1.347 & 1.331 & 1.184 & 1.317 & 1.336 & 0.163 \\
Random liquid 100 & 1.167 & 1.156 & 1.174 & 1.178 & 1.123 & 1.155 & 1.112 & 0.066 \\
Random whole universe & 1.069 & 1.106 & 1.093 & 1.115 & 1.079 & 1.062 & 0.994 & 0.121 \\
\addlinespace
\multicolumn{9}{l}{\emph{Hand-picked}} \\
Hand-picked 13 & 1.554 & 1.517 & 1.517 & 1.511 & 1.233 & 1.582 & 1.586 & 0.353 \\
Hand-picked 13 + WMT, PG & 1.564 & 1.571 & 1.543 & 1.542 & 1.159 & 1.383 & 1.630 & 0.471 \\
\midrule
SPY buy-and-hold & \multicolumn{7}{c}{1.317} & \\
Equal-weight universe & \multicolumn{7}{c}{1.296} & \\
\midrule
Range across selection rules & 0.495 & 0.484 & 0.450 & 0.427 & 0.160 & 0.527 & 0.636 & \\
\bottomrule
\end{tabular}
\end{table}

\begin{table}[H]
\centering
\small
\setlength{\tabcolsep}{4pt}
\caption{\textbf{Market exposure on real prices.} For each selection rule (row) and weighting rule (column), the
median over the $30$ runs (seeds $5660$--$5689$) of the mean portfolio $\beta$ with the hedge off and on, of the
mean weight of hedge-pool names with the hedge on, and of the maximum drawdown, as the median with the hedge on
minus the median with it off. The portfolio $\beta$ of a day is the value-weighted online $\beta$ of the holdings
divided by total wealth, and its mean is taken over the days of a run. The first three rows of each block hold $33$
names ($3$ per sector), the next five hold $13$, and the last two are the hand-picked lists. With the market hedge
on, a cap on portfolio $\beta$ tightened as the market weakened and up to five names with zero or negative $\beta$
could be held.}
\label{tab:S-exposure}
\begin{tabular}{l*{7}{r}}
\toprule
Selection rule & CD & CES & CESw & SG & MinVar & Tangent & LS \\
\midrule
\multicolumn{8}{l}{\emph{Hedge off, mean portfolio $\beta$}} \\
$\gamma$ per sector & 0.798 & 0.817 & 0.808 & 0.810 & 0.496 & 0.638 & 0.710 \\
Bandit per sector & 0.821 & 0.837 & 0.836 & 0.842 & 0.437 & 0.621 & 0.717 \\
Random per sector & 0.730 & 0.723 & 0.717 & 0.737 & 0.388 & 0.582 & 0.638 \\
\addlinespace[2pt]
Global $\gamma$ & 0.761 & 0.762 & 0.749 & 0.748 & 0.760 & 0.742 & 0.605 \\
Correlation-aware $\gamma$ & 0.774 & 0.787 & 0.783 & 0.779 & 0.761 & 0.758 & 0.639 \\
Liquid then $\gamma$ & 0.784 & 0.784 & 0.787 & 0.785 & 0.782 & 0.759 & 0.621 \\
Random liquid 100 & 0.764 & 0.767 & 0.774 & 0.785 & 0.626 & 0.787 & 0.626 \\
Random whole universe & 0.627 & 0.631 & 0.629 & 0.656 & 0.555 & 0.667 & 0.467 \\
\addlinespace[2pt]
Hand-picked 13 & 0.868 & 0.882 & 0.897 & 0.903 & 0.863 & 0.891 & 0.771 \\
Hand-picked 13 + WMT, PG & 0.883 & 0.894 & 0.887 & 0.899 & 0.784 & 0.899 & 0.765 \\
\addlinespace
\multicolumn{8}{l}{\emph{Hedge on, mean portfolio $\beta$}} \\
$\gamma$ per sector & 0.791 & 0.803 & 0.793 & 0.795 & 0.340 & 0.544 & 0.705 \\
Bandit per sector & 0.793 & 0.800 & 0.799 & 0.813 & 0.338 & 0.536 & 0.695 \\
Random per sector & 0.681 & 0.663 & 0.661 & 0.682 & 0.314 & 0.478 & 0.592 \\
\addlinespace[2pt]
Global $\gamma$ & 0.728 & 0.733 & 0.733 & 0.733 & 0.575 & 0.679 & 0.590 \\
Correlation-aware $\gamma$ & 0.762 & 0.767 & 0.760 & 0.762 & 0.519 & 0.663 & 0.627 \\
Liquid then $\gamma$ & 0.763 & 0.752 & 0.764 & 0.763 & 0.560 & 0.660 & 0.609 \\
Random liquid 100 & 0.702 & 0.696 & 0.706 & 0.715 & 0.383 & 0.563 & 0.559 \\
Random whole universe & 0.564 & 0.549 & 0.557 & 0.578 & 0.342 & 0.470 & 0.379 \\
\addlinespace[2pt]
Hand-picked 13 & 0.838 & 0.846 & 0.848 & 0.849 & 0.449 & 0.692 & 0.746 \\
Hand-picked 13 + WMT, PG & 0.831 & 0.839 & 0.836 & 0.843 & 0.417 & 0.643 & 0.730 \\
\addlinespace
\multicolumn{8}{l}{\emph{Hedge on, hedge-pool weight (\% of wealth)}} \\
$\gamma$ per sector & 8.4 & 11.8 & 8.6 & 9.2 & 34.1 & 19.8 & 13.4 \\
Bandit per sector & 10.1 & 11.5 & 9.5 & 8.4 & 30.2 & 19.2 & 14.6 \\
Random per sector & 11.7 & 12.1 & 11.6 & 10.1 & 25.9 & 20.8 & 14.7 \\
\addlinespace[2pt]
Global $\gamma$ & 27.9 & 28.9 & 29.8 & 28.5 & 39.8 & 30.9 & 33.5 \\
Correlation-aware $\gamma$ & 25.5 & 26.1 & 25.8 & 26.0 & 39.3 & 28.4 & 32.1 \\
Liquid then $\gamma$ & 29.2 & 29.7 & 29.7 & 29.5 & 39.2 & 29.9 & 33.7 \\
Random liquid 100 & 21.8 & 21.7 & 22.0 & 21.4 & 34.0 & 29.7 & 26.9 \\
Random whole universe & 21.1 & 21.4 & 21.4 & 21.0 & 33.0 & 27.6 & 26.0 \\
\addlinespace[2pt]
Hand-picked 13 & 23.2 & 22.9 & 22.4 & 22.6 & 37.6 & 31.4 & 29.6 \\
Hand-picked 13 + WMT, PG & 21.3 & 22.2 & 20.6 & 20.5 & 35.4 & 32.3 & 29.0 \\
\addlinespace
\multicolumn{8}{l}{\emph{Maximum drawdown, hedge on minus hedge off}} \\
$\gamma$ per sector & $0.000$ & $-0.006$ & $-0.001$ & $0.000$ & $0.000$ & $-0.009$ & $0.000$ \\
Bandit per sector & $0.010$ & $0.007$ & $0.002$ & $0.001$ & $0.011$ & $0.007$ & $0.008$ \\
Random per sector & $0.002$ & $0.001$ & $-0.001$ & $-0.003$ & $0.000$ & $0.008$ & $0.002$ \\
\addlinespace[2pt]
Global $\gamma$ & $0.000$ & $0.008$ & $0.007$ & $0.000$ & $0.001$ & $-0.009$ & $0.003$ \\
Correlation-aware $\gamma$ & $0.000$ & $0.000$ & $0.000$ & $0.000$ & $0.004$ & $-0.001$ & $0.005$ \\
Liquid then $\gamma$ & $-0.002$ & $0.008$ & $0.001$ & $0.000$ & $-0.009$ & $0.005$ & $0.005$ \\
Random liquid 100 & $0.003$ & $0.005$ & $0.005$ & $-0.001$ & $0.009$ & $0.019$ & $0.015$ \\
Random whole universe & $0.002$ & $0.005$ & $0.002$ & $0.000$ & $-0.002$ & $-0.018$ & $0.003$ \\
\addlinespace[2pt]
Hand-picked 13 & $0.002$ & $0.000$ & $-0.007$ & $0.003$ & $0.005$ & $0.003$ & $-0.009$ \\
Hand-picked 13 + WMT, PG & $-0.003$ & $-0.003$ & $-0.003$ & $0.004$ & $-0.006$ & $0.000$ & $0.006$ \\
\bottomrule
\end{tabular}
\end{table}

\begin{table}[H]
\centering
\small
\setlength{\tabcolsep}{4pt}
\caption{\textbf{Early re-plans and trading cost on real prices.} For each selection rule (row) and weighting rule
(column), the median over the $30$ runs (seeds $5660$--$5689$) of the share of re-plans after the first allocation
that came before the $21$-day horizon, because the portfolio value left the $95\%$ band of the forecast or fell
$8\%$ below its peak, and the percentage by which the median trading cost of these runs exceeded that of their $30$
calendar-only twins (Table~\ref{tab:S-calendar}). The other re-plans came at the horizon or on a basket change. The
first three rows of each block hold $33$ names ($3$ per sector), the next five hold $13$, and the last two are the
hand-picked lists. With the market hedge on, a cap on portfolio $\beta$ tightened as the market weakened and up to
five names with zero or negative $\beta$ could be held.}
\label{tab:S-replans}
\begin{tabular}{l*{7}{r}}
\toprule
Selection rule & CD & CES & CESw & SG & MinVar & Tangent & LS \\
\midrule
\multicolumn{8}{l}{\emph{Hedge off, share of re-plans brought forward by the band or the drawdown stop}} \\
$\gamma$ per sector & 0.42 & 0.32 & 0.42 & 0.42 & 0.40 & 0.56 & 0.50 \\
Bandit per sector & 0.33 & 0.36 & 0.31 & 0.31 & 0.29 & 0.37 & 0.46 \\
Random per sector & 0.25 & 0.24 & 0.20 & 0.24 & 0.30 & 0.25 & 0.31 \\
\addlinespace[2pt]
Global $\gamma$ & 0.48 & 0.36 & 0.41 & 0.43 & 0.41 & 0.45 & 0.55 \\
Correlation-aware $\gamma$ & 0.50 & 0.38 & 0.41 & 0.47 & 0.50 & 0.52 & 0.61 \\
Liquid then $\gamma$ & 0.29 & 0.29 & 0.28 & 0.29 & 0.38 & 0.53 & 0.52 \\
Random liquid 100 & 0.31 & 0.33 & 0.29 & 0.33 & 0.27 & 0.32 & 0.40 \\
Random whole universe & 0.26 & 0.26 & 0.25 & 0.26 & 0.28 & 0.29 & 0.37 \\
\addlinespace[2pt]
Hand-picked 13 & 0.37 & 0.40 & 0.40 & 0.43 & 0.35 & 0.44 & 0.39 \\
Hand-picked 13 + WMT, PG & 0.41 & 0.37 & 0.37 & 0.37 & 0.37 & 0.40 & 0.39 \\
\addlinespace
\multicolumn{8}{l}{\emph{Hedge on, share of re-plans brought forward by the band or the drawdown stop}} \\
$\gamma$ per sector & 0.42 & 0.35 & 0.42 & 0.42 & 0.39 & 0.52 & 0.46 \\
Bandit per sector & 0.37 & 0.33 & 0.35 & 0.38 & 0.36 & 0.44 & 0.51 \\
Random per sector & 0.27 & 0.29 & 0.25 & 0.25 & 0.39 & 0.35 & 0.35 \\
\addlinespace[2pt]
Global $\gamma$ & 0.47 & 0.41 & 0.50 & 0.48 & 0.45 & 0.47 & 0.56 \\
Correlation-aware $\gamma$ & 0.45 & 0.43 & 0.47 & 0.45 & 0.45 & 0.50 & 0.61 \\
Liquid then $\gamma$ & 0.37 & 0.38 & 0.37 & 0.37 & 0.41 & 0.47 & 0.53 \\
Random liquid 100 & 0.33 & 0.33 & 0.34 & 0.31 & 0.33 & 0.36 & 0.45 \\
Random whole universe & 0.33 & 0.31 & 0.31 & 0.31 & 0.37 & 0.35 & 0.42 \\
\addlinespace[2pt]
Hand-picked 13 & 0.43 & 0.38 & 0.38 & 0.42 & 0.28 & 0.36 & 0.41 \\
Hand-picked 13 + WMT, PG & 0.44 & 0.37 & 0.37 & 0.39 & 0.33 & 0.48 & 0.40 \\
\addlinespace
\multicolumn{8}{l}{\emph{Hedge off, trading cost, \% above the calendar-only twin}} \\
$\gamma$ per sector & 11 & 7 & 9 & 10 & 23 & 21 & 22 \\
Bandit per sector & 26 & 25 & 22 & 24 & 27 & 29 & 31 \\
Random per sector & 13 & 15 & 12 & 14 & 31 & 20 & 21 \\
\addlinespace[2pt]
Global $\gamma$ & 28 & 19 & 18 & 19 & 18 & 24 & 29 \\
Correlation-aware $\gamma$ & 30 & 17 & 15 & 20 & 28 & 22 & 32 \\
Liquid then $\gamma$ & 5 & 3 & 6 & 5 & 10 & 20 & 17 \\
Random liquid 100 & 20 & 22 & 20 & 19 & 17 & 21 & 32 \\
Random whole universe & 18 & 16 & 18 & 16 & 20 & 15 & 33 \\
\addlinespace[2pt]
Hand-picked 13 & 8 & 3 & 6 & 10 & 5 & 9 & 17 \\
Hand-picked 13 + WMT, PG & 10 & 10 & 10 & 10 & 8 & 12 & 22 \\
\addlinespace
\multicolumn{8}{l}{\emph{Hedge on, trading cost, \% above the calendar-only twin}} \\
$\gamma$ per sector & 11 & 6 & 12 & 12 & 26 & 16 & 18 \\
Bandit per sector & 23 & 18 & 21 & 23 & 34 & 29 & 35 \\
Random per sector & 15 & 13 & 13 & 9 & 45 & 24 & 21 \\
\addlinespace[2pt]
Global $\gamma$ & 21 & 24 & 26 & 25 & 21 & 17 & 31 \\
Correlation-aware $\gamma$ & 21 & 17 & 24 & 22 & 25 & 20 & 35 \\
Liquid then $\gamma$ & 24 & 15 & 24 & 25 & 20 & 25 & 22 \\
Random liquid 100 & 21 & 22 & 19 & 18 & 28 & 22 & 27 \\
Random whole universe & 23 & 22 & 21 & 22 & 35 & 23 & 30 \\
\addlinespace[2pt]
Hand-picked 13 & 20 & 4 & 4 & 5 & 19 & 8 & 15 \\
Hand-picked 13 + WMT, PG & 12 & 10 & 10 & 6 & 19 & 7 & 18 \\
\bottomrule
\end{tabular}
\end{table}

\begin{table}[H]
\centering
\small
\setlength{\tabcolsep}{4pt}
\caption{\textbf{Re-plan triggers against calendar-only re-planning on real prices.} Each entry is the median of the
$30$ runs (seeds $5660$--$5689$) with the band exit and the drawdown stop, minus the median of the $30$ runs with
both switched off, for each selection rule (row) and weighting rule (column). Without the two triggers a run
re-planned only at the $21$-day horizon, $19$ times over the window. A positive terminal wealth entry means the
triggered runs ended higher, and a negative maximum drawdown entry means their worst drawdown was smaller. The
first three rows of each block hold $33$ names ($3$ per sector), the next five hold $13$, and the last two are the
hand-picked lists. With the market hedge on, a cap on portfolio $\beta$ tightened as the market weakened and up to
five names with zero or negative $\beta$ could be held. The triggered medians are those of Tables~\ref{tab:realrun}
and~\ref{tab:S-realrun-on}.}
\label{tab:S-calendar}
\begin{tabular}{l*{7}{r}}
\toprule
Selection rule & CD & CES & CESw & SG & MinVar & Tangent & LS \\
\midrule
\multicolumn{8}{l}{\emph{Hedge off, terminal wealth ratio}} \\
$\gamma$ per sector & $0.003$ & $-0.016$ & $-0.002$ & $0.007$ & $-0.019$ & $-0.075$ & $-0.062$ \\
Bandit per sector & $0.013$ & $0.031$ & $0.004$ & $0.010$ & $-0.016$ & $0.003$ & $0.007$ \\
Random per sector & $0.000$ & $-0.006$ & $0.005$ & $-0.014$ & $-0.015$ & $-0.067$ & $-0.058$ \\
\addlinespace[2pt]
Global $\gamma$ & $0.191$ & $0.084$ & $0.104$ & $0.099$ & $0.120$ & $0.002$ & $-0.017$ \\
Correlation-aware $\gamma$ & $-0.006$ & $0.071$ & $0.043$ & $0.013$ & $0.069$ & $0.055$ & $-0.093$ \\
Liquid then $\gamma$ & $-0.063$ & $-0.057$ & $-0.063$ & $-0.064$ & $-0.046$ & $-0.005$ & $-0.022$ \\
Random liquid 100 & $0.013$ & $0.007$ & $0.017$ & $0.008$ & $0.026$ & $0.021$ & $0.004$ \\
Random whole universe & $-0.037$ & $-0.008$ & $-0.006$ & $-0.039$ & $-0.014$ & $-0.054$ & $-0.050$ \\
\addlinespace[2pt]
Hand-picked 13 & $-0.001$ & $0.004$ & $0.034$ & $0.037$ & $0.003$ & $0.009$ & $-0.019$ \\
Hand-picked 13 + WMT, PG & $0.111$ & $0.122$ & $0.112$ & $0.093$ & $0.012$ & $-0.010$ & $0.093$ \\
\addlinespace
\multicolumn{8}{l}{\emph{Hedge off, maximum drawdown}} \\
$\gamma$ per sector & $0.005$ & $-0.015$ & $0.006$ & $0.006$ & $-0.015$ & $-0.007$ & $-0.025$ \\
Bandit per sector & $-0.007$ & $-0.009$ & $-0.005$ & $-0.002$ & $-0.002$ & $0.000$ & $-0.009$ \\
Random per sector & $-0.005$ & $-0.007$ & $-0.008$ & $-0.001$ & $-0.006$ & $0.009$ & $0.002$ \\
\addlinespace[2pt]
Global $\gamma$ & $-0.053$ & $-0.024$ & $-0.052$ & $-0.053$ & $-0.014$ & $-0.021$ & $-0.039$ \\
Correlation-aware $\gamma$ & $-0.013$ & $-0.035$ & $-0.013$ & $-0.013$ & $-0.022$ & $-0.012$ & $-0.031$ \\
Liquid then $\gamma$ & $0.015$ & $0.001$ & $0.010$ & $0.013$ & $-0.026$ & $-0.012$ & $-0.007$ \\
Random liquid 100 & $-0.011$ & $-0.019$ & $-0.023$ & $-0.010$ & $0.000$ & $-0.004$ & $-0.019$ \\
Random whole universe & $-0.010$ & $-0.008$ & $-0.010$ & $-0.009$ & $0.004$ & $0.012$ & $-0.006$ \\
\addlinespace[2pt]
Hand-picked 13 & $-0.020$ & $-0.022$ & $-0.001$ & $-0.008$ & $-0.004$ & $-0.002$ & $0.002$ \\
Hand-picked 13 + WMT, PG & $-0.036$ & $-0.043$ & $-0.038$ & $-0.029$ & $-0.002$ & $0.000$ & $-0.038$ \\
\addlinespace
\multicolumn{8}{l}{\emph{Hedge on, terminal wealth ratio}} \\
$\gamma$ per sector & $-0.006$ & $0.010$ & $0.010$ & $0.002$ & $-0.070$ & $-0.007$ & $-0.045$ \\
Bandit per sector & $-0.042$ & $-0.022$ & $0.002$ & $0.019$ & $-0.046$ & $-0.033$ & $-0.028$ \\
Random per sector & $-0.049$ & $-0.068$ & $-0.047$ & $0.004$ & $-0.016$ & $-0.107$ & $-0.085$ \\
\addlinespace[2pt]
Global $\gamma$ & $0.081$ & $-0.085$ & $0.011$ & $0.025$ & $0.045$ & $0.183$ & $-0.077$ \\
Correlation-aware $\gamma$ & $-0.011$ & $0.051$ & $0.034$ & $0.015$ & $0.060$ & $-0.035$ & $-0.113$ \\
Liquid then $\gamma$ & $-0.165$ & $-0.152$ & $-0.180$ & $-0.174$ & $0.032$ & $-0.069$ & $-0.086$ \\
Random liquid 100 & $-0.006$ & $0.013$ & $0.007$ & $0.022$ & $-0.007$ & $-0.015$ & $-0.042$ \\
Random whole universe & $-0.080$ & $-0.026$ & $-0.049$ & $-0.046$ & $-0.041$ & $-0.055$ & $-0.090$ \\
\addlinespace[2pt]
Hand-picked 13 & $0.036$ & $-0.022$ & $-0.007$ & $-0.040$ & $-0.045$ & $0.025$ & $-0.010$ \\
Hand-picked 13 + WMT, PG & $0.087$ & $0.069$ & $0.063$ & $0.034$ & $-0.058$ & $-0.076$ & $0.064$ \\
\addlinespace
\multicolumn{8}{l}{\emph{Hedge on, maximum drawdown}} \\
$\gamma$ per sector & $0.005$ & $-0.021$ & $0.005$ & $0.005$ & $-0.018$ & $-0.016$ & $-0.025$ \\
Bandit per sector & $-0.003$ & $-0.007$ & $-0.008$ & $-0.008$ & $0.003$ & $-0.002$ & $-0.011$ \\
Random per sector & $-0.009$ & $-0.011$ & $-0.014$ & $-0.009$ & $-0.010$ & $0.011$ & $0.001$ \\
\addlinespace[2pt]
Global $\gamma$ & $-0.053$ & $-0.015$ & $-0.045$ & $-0.054$ & $-0.018$ & $-0.021$ & $-0.036$ \\
Correlation-aware $\gamma$ & $-0.013$ & $-0.035$ & $-0.013$ & $-0.013$ & $-0.025$ & $-0.013$ & $-0.027$ \\
Liquid then $\gamma$ & $0.003$ & $-0.004$ & $0.001$ & $0.003$ & $-0.046$ & $-0.010$ & $-0.014$ \\
Random liquid 100 & $-0.010$ & $-0.015$ & $-0.018$ & $-0.013$ & $-0.002$ & $-0.008$ & $-0.030$ \\
Random whole universe & $-0.021$ & $-0.012$ & $-0.021$ & $-0.015$ & $-0.010$ & $-0.013$ & $-0.027$ \\
\addlinespace[2pt]
Hand-picked 13 & $-0.018$ & $-0.022$ & $-0.008$ & $-0.005$ & $-0.001$ & $-0.001$ & $-0.016$ \\
Hand-picked 13 + WMT, PG & $-0.039$ & $-0.046$ & $-0.040$ & $-0.026$ & $-0.014$ & $-0.003$ & $-0.045$ \\
\bottomrule
\end{tabular}
\end{table}

\begin{table}[H]
\centering
\small
\setlength{\tabcolsep}{4pt}
\caption{\textbf{Re-plan triggers against calendar-only re-planning in the burst stress set.} Each entry is the
median across the $200$ synthetic futures of the set (the deepest SPY drawdowns among futures with a jump burst) of
the runs with the band exit and the drawdown stop, minus the same median of the runs with both switched off, for
each selection rule (row) and weighting rule (column). Within a future, the bandit and random rules contribute the
median over seeds $5660$--$5664$, and the other rules ran once. Without the two triggers a run re-planned every
$21$ trading days, $19$ times over the window, at the horizon or, when its last re-plan had gone to cash, on the
cash revisit. A positive terminal wealth entry means the triggered runs ended higher, and a negative maximum
drawdown entry means their worst drawdown was smaller. Rows are grouped as in Table~\ref{tab:S-calendar}, and the
triggered medians are those of Fig.~\ref{fig:synthetic}.}
\label{tab:S-calendar-burst}
\begin{tabular}{l*{7}{r}}
\toprule
Selection rule & CD & CES & CESw & SG & MinVar & Tangent & LS \\
\midrule
\multicolumn{8}{l}{\emph{Hedge off, terminal wealth ratio}} \\
$\gamma$ per sector & $-0.009$ & $0.008$ & $-0.002$ & $-0.002$ & $0.002$ & $-0.016$ & $0.006$ \\
Bandit per sector & $0.008$ & $-0.007$ & $-0.002$ & $-0.012$ & $0.004$ & $0.015$ & $-0.010$ \\
Random per sector & $0.000$ & $0.003$ & $0.003$ & $0.007$ & $-0.008$ & $0.003$ & $0.002$ \\
\addlinespace[2pt]
Global $\gamma$ & $-0.027$ & $0.001$ & $-0.006$ & $-0.031$ & $-0.022$ & $-0.009$ & $-0.026$ \\
Correlation-aware $\gamma$ & $-0.013$ & $-0.008$ & $-0.013$ & $-0.015$ & $-0.025$ & $-0.013$ & $-0.002$ \\
Liquid then $\gamma$ & $0.007$ & $-0.009$ & $0.002$ & $0.014$ & $0.003$ & $0.005$ & $0.014$ \\
Random liquid 100 & $0.022$ & $-0.009$ & $0.021$ & $-0.006$ & $0.009$ & $0.013$ & $0.016$ \\
Random whole universe & $0.007$ & $0.002$ & $0.004$ & $-0.001$ & $-0.007$ & $-0.011$ & $-0.001$ \\
\addlinespace[2pt]
Hand-picked 13 & $-0.004$ & $-0.007$ & $-0.002$ & $0.000$ & $0.004$ & $0.011$ & $0.024$ \\
Hand-picked 13 + WMT, PG & $0.003$ & $-0.008$ & $-0.001$ & $0.013$ & $0.005$ & $-0.004$ & $0.021$ \\
\addlinespace
\multicolumn{8}{l}{\emph{Hedge off, maximum drawdown}} \\
$\gamma$ per sector & $-0.006$ & $-0.003$ & $-0.007$ & $-0.005$ & $-0.014$ & $-0.001$ & $-0.006$ \\
Bandit per sector & $0.000$ & $-0.001$ & $-0.003$ & $-0.002$ & $-0.002$ & $0.002$ & $-0.008$ \\
Random per sector & $-0.013$ & $-0.013$ & $-0.016$ & $-0.008$ & $-0.007$ & $-0.001$ & $-0.017$ \\
\addlinespace[2pt]
Global $\gamma$ & $0.000$ & $-0.004$ & $-0.009$ & $-0.004$ & $-0.002$ & $-0.005$ & $-0.002$ \\
Correlation-aware $\gamma$ & $-0.002$ & $-0.005$ & $-0.006$ & $0.002$ & $0.001$ & $-0.007$ & $0.001$ \\
Liquid then $\gamma$ & $-0.009$ & $-0.008$ & $-0.009$ & $-0.010$ & $-0.009$ & $-0.003$ & $0.000$ \\
Random liquid 100 & $-0.010$ & $-0.010$ & $-0.013$ & $-0.011$ & $-0.004$ & $-0.003$ & $-0.004$ \\
Random whole universe & $-0.011$ & $-0.010$ & $-0.009$ & $-0.004$ & $-0.002$ & $-0.003$ & $-0.012$ \\
\addlinespace[2pt]
Hand-picked 13 & $-0.022$ & $-0.020$ & $-0.020$ & $-0.012$ & $-0.009$ & $-0.006$ & $-0.022$ \\
Hand-picked 13 + WMT, PG & $-0.020$ & $-0.020$ & $-0.021$ & $-0.015$ & $-0.008$ & $-0.005$ & $-0.019$ \\
\addlinespace
\multicolumn{8}{l}{\emph{Hedge on, terminal wealth ratio}} \\
$\gamma$ per sector & $0.009$ & $0.002$ & $0.002$ & $-0.003$ & $-0.001$ & $0.017$ & $-0.010$ \\
Bandit per sector & $-0.013$ & $-0.019$ & $-0.013$ & $-0.005$ & $-0.002$ & $0.003$ & $-0.007$ \\
Random per sector & $0.012$ & $0.019$ & $0.014$ & $-0.006$ & $0.001$ & $0.000$ & $0.015$ \\
\addlinespace[2pt]
Global $\gamma$ & $-0.003$ & $-0.002$ & $-0.007$ & $0.003$ & $-0.003$ & $-0.010$ & $-0.018$ \\
Correlation-aware $\gamma$ & $-0.012$ & $-0.002$ & $-0.023$ & $-0.016$ & $-0.014$ & $-0.015$ & $-0.026$ \\
Liquid then $\gamma$ & $-0.004$ & $0.004$ & $-0.016$ & $0.002$ & $0.015$ & $0.003$ & $-0.014$ \\
Random liquid 100 & $0.000$ & $-0.010$ & $0.007$ & $0.009$ & $0.015$ & $0.006$ & $-0.002$ \\
Random whole universe & $-0.007$ & $0.001$ & $-0.002$ & $0.013$ & $-0.013$ & $-0.011$ & $0.005$ \\
\addlinespace[2pt]
Hand-picked 13 & $-0.007$ & $-0.017$ & $-0.004$ & $-0.011$ & $0.000$ & $-0.010$ & $0.021$ \\
Hand-picked 13 + WMT, PG & $-0.010$ & $-0.008$ & $-0.007$ & $0.002$ & $-0.007$ & $-0.010$ & $0.010$ \\
\addlinespace
\multicolumn{8}{l}{\emph{Hedge on, maximum drawdown}} \\
$\gamma$ per sector & $-0.007$ & $-0.009$ & $-0.008$ & $-0.010$ & $-0.007$ & $-0.004$ & $-0.011$ \\
Bandit per sector & $-0.005$ & $-0.013$ & $-0.008$ & $-0.013$ & $-0.004$ & $-0.002$ & $-0.011$ \\
Random per sector & $-0.018$ & $-0.015$ & $-0.017$ & $-0.016$ & $-0.009$ & $-0.010$ & $-0.012$ \\
\addlinespace[2pt]
Global $\gamma$ & $-0.012$ & $-0.010$ & $-0.014$ & $-0.012$ & $-0.014$ & $-0.009$ & $-0.006$ \\
Correlation-aware $\gamma$ & $-0.011$ & $-0.005$ & $-0.012$ & $-0.014$ & $-0.002$ & $-0.005$ & $-0.004$ \\
Liquid then $\gamma$ & $-0.009$ & $-0.009$ & $-0.008$ & $-0.012$ & $-0.014$ & $-0.007$ & $-0.009$ \\
Random liquid 100 & $-0.013$ & $-0.013$ & $-0.014$ & $-0.014$ & $-0.007$ & $-0.011$ & $-0.003$ \\
Random whole universe & $-0.014$ & $-0.017$ & $-0.009$ & $-0.009$ & $-0.011$ & $-0.003$ & $-0.017$ \\
\addlinespace[2pt]
Hand-picked 13 & $-0.018$ & $-0.015$ & $-0.021$ & $-0.020$ & $-0.011$ & $-0.007$ & $-0.022$ \\
Hand-picked 13 + WMT, PG & $-0.021$ & $-0.016$ & $-0.020$ & $-0.015$ & $-0.009$ & $0.000$ & $-0.020$ \\
\bottomrule
\end{tabular}
\end{table}

\begin{table}[H]
\centering
\small
\setlength{\tabcolsep}{4pt}
\caption{\textbf{Re-plan triggers against calendar-only re-planning in the bear stress set.} Entries as in
Table~\ref{tab:S-calendar-burst}, across the $200$ synthetic futures of the bear stress set (the deepest SPY
drawdowns among futures without a jump burst).}
\label{tab:S-calendar-bear}
\begin{tabular}{l*{7}{r}}
\toprule
Selection rule & CD & CES & CESw & SG & MinVar & Tangent & LS \\
\midrule
\multicolumn{8}{l}{\emph{Hedge off, terminal wealth ratio}} \\
$\gamma$ per sector & $0.000$ & $0.001$ & $0.005$ & $0.006$ & $0.003$ & $0.007$ & $0.008$ \\
Bandit per sector & $0.001$ & $0.004$ & $-0.005$ & $0.000$ & $-0.010$ & $-0.001$ & $0.004$ \\
Random per sector & $0.004$ & $0.008$ & $0.010$ & $0.000$ & $-0.004$ & $-0.002$ & $0.013$ \\
\addlinespace[2pt]
Global $\gamma$ & $-0.004$ & $0.011$ & $0.004$ & $-0.005$ & $0.010$ & $-0.004$ & $-0.019$ \\
Correlation-aware $\gamma$ & $-0.004$ & $0.004$ & $0.000$ & $-0.007$ & $0.008$ & $-0.003$ & $0.000$ \\
Liquid then $\gamma$ & $0.013$ & $0.018$ & $0.012$ & $0.013$ & $0.012$ & $0.006$ & $-0.009$ \\
Random liquid 100 & $0.011$ & $0.004$ & $0.002$ & $0.005$ & $0.002$ & $0.010$ & $0.017$ \\
Random whole universe & $0.014$ & $0.019$ & $0.017$ & $-0.003$ & $-0.001$ & $0.001$ & $0.008$ \\
\addlinespace[2pt]
Hand-picked 13 & $0.016$ & $0.019$ & $0.020$ & $0.004$ & $0.006$ & $0.002$ & $0.029$ \\
Hand-picked 13 + WMT, PG & $0.004$ & $0.013$ & $0.006$ & $0.015$ & $0.004$ & $0.004$ & $0.024$ \\
\addlinespace
\multicolumn{8}{l}{\emph{Hedge off, maximum drawdown}} \\
$\gamma$ per sector & $-0.010$ & $-0.004$ & $-0.011$ & $-0.006$ & $-0.012$ & $-0.007$ & $-0.009$ \\
Bandit per sector & $-0.006$ & $-0.002$ & $-0.006$ & $-0.005$ & $-0.001$ & $-0.005$ & $-0.011$ \\
Random per sector & $-0.013$ & $-0.009$ & $-0.011$ & $-0.005$ & $0.000$ & $-0.001$ & $-0.014$ \\
\addlinespace[2pt]
Global $\gamma$ & $-0.012$ & $-0.001$ & $-0.008$ & $-0.011$ & $-0.009$ & $0.002$ & $-0.009$ \\
Correlation-aware $\gamma$ & $-0.013$ & $-0.004$ & $-0.015$ & $-0.014$ & $-0.009$ & $0.007$ & $-0.008$ \\
Liquid then $\gamma$ & $-0.009$ & $-0.018$ & $-0.014$ & $-0.010$ & $-0.007$ & $-0.011$ & $-0.007$ \\
Random liquid 100 & $-0.010$ & $-0.014$ & $-0.010$ & $-0.009$ & $-0.004$ & $-0.007$ & $-0.006$ \\
Random whole universe & $-0.013$ & $-0.012$ & $-0.015$ & $-0.007$ & $0.001$ & $-0.003$ & $-0.009$ \\
\addlinespace[2pt]
Hand-picked 13 & $-0.008$ & $-0.009$ & $-0.011$ & $-0.007$ & $-0.006$ & $0.001$ & $-0.013$ \\
Hand-picked 13 + WMT, PG & $-0.011$ & $-0.011$ & $-0.010$ & $-0.008$ & $-0.003$ & $-0.008$ & $-0.014$ \\
\addlinespace
\multicolumn{8}{l}{\emph{Hedge on, terminal wealth ratio}} \\
$\gamma$ per sector & $0.012$ & $0.003$ & $0.014$ & $0.017$ & $0.004$ & $0.006$ & $0.011$ \\
Bandit per sector & $0.012$ & $-0.002$ & $0.010$ & $0.009$ & $-0.005$ & $-0.003$ & $0.002$ \\
Random per sector & $-0.003$ & $-0.001$ & $-0.001$ & $-0.007$ & $-0.001$ & $0.002$ & $0.002$ \\
\addlinespace[2pt]
Global $\gamma$ & $-0.003$ & $0.012$ & $0.002$ & $0.007$ & $0.014$ & $-0.003$ & $0.005$ \\
Correlation-aware $\gamma$ & $-0.001$ & $0.016$ & $0.008$ & $0.001$ & $0.023$ & $0.012$ & $0.014$ \\
Liquid then $\gamma$ & $0.005$ & $0.011$ & $0.006$ & $0.005$ & $0.007$ & $0.002$ & $0.001$ \\
Random liquid 100 & $0.021$ & $0.022$ & $0.014$ & $0.008$ & $-0.002$ & $-0.015$ & $0.020$ \\
Random whole universe & $0.011$ & $0.009$ & $0.012$ & $0.004$ & $0.003$ & $0.009$ & $0.019$ \\
\addlinespace[2pt]
Hand-picked 13 & $-0.002$ & $0.008$ & $0.008$ & $0.009$ & $0.000$ & $0.001$ & $0.022$ \\
Hand-picked 13 + WMT, PG & $0.015$ & $0.011$ & $0.016$ & $0.009$ & $0.003$ & $0.017$ & $0.017$ \\
\addlinespace
\multicolumn{8}{l}{\emph{Hedge on, maximum drawdown}} \\
$\gamma$ per sector & $-0.010$ & $-0.003$ & $-0.008$ & $-0.009$ & $-0.003$ & $-0.014$ & $-0.013$ \\
Bandit per sector & $-0.004$ & $-0.001$ & $-0.008$ & $-0.007$ & $-0.004$ & $0.002$ & $-0.004$ \\
Random per sector & $-0.016$ & $-0.012$ & $-0.012$ & $-0.008$ & $-0.004$ & $-0.006$ & $-0.007$ \\
\addlinespace[2pt]
Global $\gamma$ & $-0.013$ & $-0.013$ & $-0.005$ & $-0.011$ & $-0.012$ & $0.002$ & $-0.016$ \\
Correlation-aware $\gamma$ & $-0.018$ & $-0.014$ & $-0.009$ & $-0.017$ & $-0.006$ & $-0.005$ & $-0.015$ \\
Liquid then $\gamma$ & $-0.007$ & $-0.012$ & $-0.006$ & $-0.007$ & $-0.012$ & $-0.007$ & $-0.012$ \\
Random liquid 100 & $-0.016$ & $-0.015$ & $-0.018$ & $-0.011$ & $-0.001$ & $-0.001$ & $-0.011$ \\
Random whole universe & $-0.014$ & $-0.011$ & $-0.014$ & $-0.010$ & $-0.004$ & $0.004$ & $-0.009$ \\
\addlinespace[2pt]
Hand-picked 13 & $-0.013$ & $-0.013$ & $-0.017$ & $-0.017$ & $0.000$ & $-0.001$ & $-0.015$ \\
Hand-picked 13 + WMT, PG & $-0.015$ & $-0.015$ & $-0.017$ & $-0.013$ & $-0.004$ & $-0.011$ & $-0.015$ \\
\bottomrule
\end{tabular}
\end{table}

\begin{table}[H]
\centering
\small
\setlength{\tabcolsep}{4pt}
\caption{\textbf{Selection by $\gamma$ against random picks on synthetic futures.} Share of futures (\%) in which
the rule that selected by $\gamma$ ended with a higher terminal wealth ratio $W_T/W_0$ than random picks of the
same size, by weighting rule (column), stress set and hedge setting. With the market hedge on, a cap on portfolio $\beta$ tightened as the market weakened and up to five names with zero or negative $\beta$ could be held. Global $\gamma$ held the $K$ highest-$\gamma$
names in the universe and was compared with $K$ names drawn at random from the whole universe; $\gamma$ per sector
was compared with three random names per sector. Each random rule contributed the median of five seeds in each
future. A value of 50 means that each rule ended higher in half of the futures.}
\label{tab:S-head-to-head}
\begin{tabular}{l*{7}{r}}
\toprule
Comparison & CD & CES & CESw & SG & MinVar & Tangent & LS \\
\midrule
\multicolumn{8}{l}{\emph{Typical futures (600), hedge off}} \\
Global $\gamma$ vs.\ random, 5 names & 62.3 & 62.3 & 64.8 & 64.5 & 62.5 & 61.3 & 64.2 \\
Global $\gamma$ vs.\ random, 13 names & 70.7 & 71.2 & 70.5 & 69.7 & 64.5 & 62.3 & 70.0 \\
Global $\gamma$ vs.\ random, 34 names & 70.2 & 68.2 & 71.0 & 70.8 & 67.2 & 58.0 & 69.0 \\
$\gamma$ per sector vs.\ random per sector & 73.2 & 72.7 & 73.2 & 73.3 & 64.8 & 60.0 & 73.7 \\
\addlinespace
\multicolumn{8}{l}{\emph{Typical futures (600), hedge on}} \\
Global $\gamma$ vs.\ random, 5 names & 63.2 & 62.7 & 64.2 & 64.5 & 62.0 & 60.2 & 63.2 \\
Global $\gamma$ vs.\ random, 13 names & 68.8 & 69.8 & 68.7 & 68.8 & 65.7 & 63.2 & 71.5 \\
Global $\gamma$ vs.\ random, 34 names & 67.7 & 68.3 & 70.3 & 71.2 & 65.3 & 60.7 & 68.0 \\
$\gamma$ per sector vs.\ random per sector & 72.3 & 71.7 & 73.7 & 71.5 & 66.7 & 63.5 & 73.2 \\
\addlinespace
\multicolumn{8}{l}{\emph{Burst stress (200), hedge off}} \\
Global $\gamma$ vs.\ random, 5 names & 53.0 & 53.0 & 52.5 & 53.5 & 66.5 & 62.5 & 57.5 \\
Global $\gamma$ vs.\ random, 13 names & 52.5 & 53.0 & 55.0 & 57.5 & 67.0 & 67.5 & 53.5 \\
Global $\gamma$ vs.\ random, 34 names & 58.5 & 63.5 & 61.0 & 65.0 & 60.5 & 55.0 & 56.5 \\
$\gamma$ per sector vs.\ random per sector & 66.0 & 63.0 & 64.5 & 70.0 & 72.0 & 64.5 & 63.0 \\
\addlinespace
\multicolumn{8}{l}{\emph{Burst stress (200), hedge on}} \\
Global $\gamma$ vs.\ random, 5 names & 54.0 & 54.5 & 58.0 & 58.5 & 60.0 & 64.0 & 54.0 \\
Global $\gamma$ vs.\ random, 13 names & 56.0 & 59.0 & 59.0 & 61.0 & 64.5 & 64.0 & 48.5 \\
Global $\gamma$ vs.\ random, 34 names & 53.0 & 60.0 & 66.0 & 69.0 & 61.0 & 54.0 & 57.5 \\
$\gamma$ per sector vs.\ random per sector & 61.5 & 60.0 & 60.0 & 70.0 & 60.5 & 60.0 & 61.0 \\
\addlinespace
\multicolumn{8}{l}{\emph{Bear stress (200), hedge off}} \\
Global $\gamma$ vs.\ random, 5 names & 40.0 & 38.5 & 41.5 & 45.0 & 72.0 & 63.0 & 43.0 \\
Global $\gamma$ vs.\ random, 13 names & 29.0 & 28.5 & 28.0 & 38.0 & 70.0 & 71.0 & 29.5 \\
Global $\gamma$ vs.\ random, 34 names & 53.5 & 46.0 & 47.0 & 65.5 & 69.0 & 72.0 & 35.5 \\
$\gamma$ per sector vs.\ random per sector & 54.0 & 44.0 & 50.5 & 67.0 & 68.5 & 74.5 & 42.5 \\
\addlinespace
\multicolumn{8}{l}{\emph{Bear stress (200), hedge on}} \\
Global $\gamma$ vs.\ random, 5 names & 43.0 & 41.5 & 44.0 & 46.5 & 63.0 & 60.0 & 37.5 \\
Global $\gamma$ vs.\ random, 13 names & 33.5 & 34.0 & 34.5 & 45.0 & 66.5 & 61.0 & 30.5 \\
Global $\gamma$ vs.\ random, 34 names & 48.5 & 50.0 & 49.0 & 61.0 & 59.5 & 63.0 & 40.5 \\
$\gamma$ per sector vs.\ random per sector & 53.5 & 53.5 & 53.0 & 63.5 & 61.0 & 63.0 & 40.0 \\
\bottomrule
\end{tabular}
\end{table}

\begin{table}[H]
\centering
\small
\setlength{\tabcolsep}{4pt}
\caption{\textbf{Market exposure on the stress sets of synthetic futures, hedge off.} For each selection rule (row)
and weighting rule (column), the mean portfolio $\beta$ (Table~\ref{tab:S-exposure}) as the median across the $200$
futures of each stress set, after a median over seeds within a future (seeds $5660$--$5664$ for the random and
bandit rules, $5660$ for the rest). The burst stress set holds the $200$ futures with the deepest SPY drawdowns
among those with a jump burst, and the bear stress set the $200$ deepest among those without one. The first three
rows of each block hold $33$ names ($3$ per sector), the next five hold $13$, and the last two are the hand-picked
lists.}
\label{tab:S-synbeta}
\begin{tabular}{l*{7}{r}}
\toprule
Selection rule & CD & CES & CESw & SG & MinVar & Tangent & LS \\
\midrule
\multicolumn{8}{l}{\emph{Burst stress, mean portfolio $\beta$}} \\
$\gamma$ per sector & 0.739 & 0.747 & 0.749 & 0.751 & 0.542 & 0.637 & 0.690 \\
Bandit per sector & 0.732 & 0.733 & 0.739 & 0.755 & 0.532 & 0.675 & 0.677 \\
Random per sector & 0.659 & 0.667 & 0.663 & 0.728 & 0.557 & 0.695 & 0.552 \\
\addlinespace[2pt]
Global $\gamma$ & 0.690 & 0.695 & 0.693 & 0.692 & 0.652 & 0.662 & 0.625 \\
Correlation-aware $\gamma$ & 0.695 & 0.698 & 0.701 & 0.697 & 0.647 & 0.662 & 0.631 \\
Liquid then $\gamma$ & 0.721 & 0.714 & 0.721 & 0.725 & 0.707 & 0.710 & 0.656 \\
Random liquid 100 & 0.571 & 0.576 & 0.572 & 0.627 & 0.725 & 0.739 & 0.395 \\
Random whole universe & 0.528 & 0.526 & 0.525 & 0.582 & 0.705 & 0.734 & 0.322 \\
\addlinespace[2pt]
Hand-picked 13 & 0.565 & 0.565 & 0.567 & 0.631 & 0.764 & 0.761 & 0.384 \\
Hand-picked 13 + WMT, PG & 0.552 & 0.555 & 0.555 & 0.633 & 0.760 & 0.761 & 0.369 \\
\addlinespace
\multicolumn{8}{l}{\emph{Bear stress, mean portfolio $\beta$}} \\
$\gamma$ per sector & 0.768 & 0.788 & 0.787 & 0.791 & 0.510 & 0.629 & 0.709 \\
Bandit per sector & 0.779 & 0.788 & 0.788 & 0.804 & 0.521 & 0.699 & 0.710 \\
Random per sector & 0.649 & 0.651 & 0.649 & 0.765 & 0.571 & 0.729 & 0.483 \\
\addlinespace[2pt]
Global $\gamma$ & 0.715 & 0.717 & 0.722 & 0.717 & 0.643 & 0.669 & 0.631 \\
Correlation-aware $\gamma$ & 0.724 & 0.725 & 0.729 & 0.726 & 0.647 & 0.670 & 0.642 \\
Liquid then $\gamma$ & 0.776 & 0.768 & 0.778 & 0.782 & 0.739 & 0.750 & 0.693 \\
Random liquid 100 & 0.512 & 0.514 & 0.515 & 0.586 & 0.798 & 0.819 & 0.293 \\
Random whole universe & 0.460 & 0.458 & 0.460 & 0.536 & 0.769 & 0.798 & 0.229 \\
\addlinespace[2pt]
Hand-picked 13 & 0.489 & 0.486 & 0.489 & 0.578 & 0.850 & 0.850 & 0.284 \\
Hand-picked 13 + WMT, PG & 0.484 & 0.487 & 0.485 & 0.584 & 0.847 & 0.850 & 0.250 \\
\bottomrule
\end{tabular}
\end{table}

\end{document}